\documentclass[12pt,a4paper]{article}

\usepackage{amsmath,amssymb}
\usepackage[bookmarks=true]{hyperref}
\usepackage[nosort]{cite}
\usepackage[bulletsep]{collect}
\def\gfxon{\usepackage[final]{graphicx}}

\gfxon

\makeatletter
\let\old@startsection=\@startsection
\renewcommand{\@startsection}[6]{\old@startsection{#1}{#2}{#3}{#4}{#5}{#6\mathversion{bold}}}
\makeatother

\makeatletter \@addtoreset{equation}{section} \makeatother

\makeatletter
\let\old@makecaption=\@makecaption
\def\@makecaption{\small\old@makecaption}
\makeatother

\newcommand{\CN}{\mathcal{N}}
\newcommand{\CL}{\mathcal{L}}

\newcommand{\CM}{\mathcal{M}}

\newcommand{\CR}{\mathcal{R}}
\newcommand{\CW}{\mathcal{W}}
\newcommand{\CZ}{\mathcal{Z}}

\makeatletter
\newcommand{\ellSN}{\mathop{\operator@font sn}\nolimits}
\newcommand{\ellCN}{\mathop{\operator@font cn}\nolimits}
\newcommand{\ellDN}{\mathop{\operator@font dn}\nolimits}
\newcommand{\ellAM}{\mathop{\operator@font am}\nolimits}
\newcommand{\ellK}{\mathop{\smash{\operator@font K}\vphantom{a}}\nolimits}
\newcommand{\ellE}{\mathop{\smash{\operator@font E}\vphantom{a}}\nolimits}
\makeatother

\ifx\genfrac\sdflkaj

\else

\fi
\newcommand{\sfrac}[2]{{\textstyle\frac{#1}{#2}}}
\newcommand{\half}{\sfrac{1}{2}}

\newcommand{\ssN}{\mathcal{N}}

\newcommand{\alg}[1]{\mathfrak{#1}}

\newcommand{\beq}{\begin{equation}}
\newcommand{\eeq}{\end{equation}}

\def\[{\begin{equation}}
\def\]{\end{equation}}
\def\<{\begin{eqnarray}}
\def\>{\end{eqnarray}}

\makeatletter
\def\mr@ignsp#1 {\ifx\:#1\@empty\else #1\expandafter\mr@ignsp\fi}%
\newcommand{\multiref}[1]{\begingroup
\xdef\mr@no@sparg{\expandafter\mr@ignsp#1 \: }%
\def\mr@comma{}%
\@for\mr@refs:=\mr@no@sparg\do{\mr@comma\def\mr@comma{,}\ref{\mr@refs}}%
\endgroup}
\makeatother

\newcommand{\hypref}[2]{\ifx\href\asklfhas #2\else\href{#1}{#2}\fi}
\newcommand{\Secref}[1]{Section~\multiref{#1}}
\newcommand{\secref}[1]{Sec.~\multiref{#1}}

\newcommand{\appref}[1]{App.~\multiref{#1}}

\newcommand{\tabref}[1]{Tab.~\multiref{#1}}

\newcommand{\figref}[1]{Fig.~\multiref{#1}}
\renewcommand{\eqref}[1]{(\multiref{#1})}

\ifx\href\asklfhas\newcommand{\href}[2]{#2}\fi

\begin{document}

\begin{center}

{\Large\textbf{\mathversion{bold}
On Three Dimensional Quiver Gauge Theories \\
and Integrability}
\par}


\vspace{8mm}

\thispagestyle{empty}

\textsc{Davide Gaiotto and Peter Koroteev}

\vspace{7mm}%

\textit{Perimeter Institute for Theoretical Physics\\%
31 Caroline Street North, ON N2L 2Y5, Canada}

\vspace{7mm}

\texttt{dgaiotto@gmail.com, \\
pkoroteev@perimeterinstitute.ca}

\par\vspace{.5cm}

\vfill

\textbf{Abstract}\vspace{5mm}

\begin{minipage}{12.7cm}
In this work we compare different descriptions of the space of vacua of certain three dimensional $\CN=4$ superconformal field theories, compactified on a circle and mass-deformed to $\CN=2$ in a canonical way. The original $\CN=4$ theories are known to admit two distinct mirror descriptions as linear quiver gauge theories, and many more descriptions which involve 
the compactification on a segment of four-dimensional $\CN=4$ super Yang-Mills theory. Each description gives a distinct presentation of the moduli space of vacua. Our main result is to establish the precise dictionary between these presentations. We also study the relationship between this gauge theory problem and integrable systems. 
The space of vacua in the linear quiver gauge theory description is related by Nekrasov-Shatashvili duality to the eigenvalues of quantum integrable spin chain Hamiltonians. 
The space of vacua in the four-dimensional gauge theory description is related to the solution of certain integrable classical many-body problems. 
Thus we obtain numerous dualities between these integrable models.
\end{minipage}

\vspace*{\fill}
\end{center}

\newpage

\tableofcontents

\newpage

\section{Introduction}\label{Sec:Intro}
Three-dimensional $\ssN=4$ superconformal field theories play an important role in our understanding of dualities.
Such theories provided the first examples of three-dimensional mirror symmetry \cite{Intriligator:1996ex, deBoer1997101,deBoer1997148,deBoer1997107}, where the same conformal fixed point can be given 
distinct gauge theory descriptions in the UV. A particularly important aspect of mirror symmetry  is that supersymmetric non-renormalization theorems may allow distinct UV descriptions to compute exactly different properties of the IR SCFT. 
For example, gauge theories with $\ssN=4$ supersymmetry have a Higgs branch of vacua which receives no quantum corrections, and a Coulomb branch of vacua which can be heavily modified by quantum effects in the infrared. The $\ssN=4$ mirror symmetry 
exchanges the two branches of vacua, and thus provides a complete picture of the vacuum structure through a mirror pair of UV descriptions. 

On the other hand, certain properties of the theory can be computed exactly in both descriptions, and matched through the 
mirror symmetry dictionary. An early example is the match of scaling dimensions of protected operators, 
which are built out of elementary fields in one description but become non-perturbative monopole operators in the mirror \cite{Borokhov:2002cg}. 
More recently, we have learned how to compute exactly several more protected probes of a 3d SCFT's properties, 
such as the partition function on the ellipsoid $S^3_b$ \cite{kwy1,2010JHEP...10..013K,hhl} and the superconformal index \cite{1126-6708-2008-02-064,Kim2009241,Iy,sv,2011arXiv1106.2484K,PhysRevD.86.065015}. All of these are intimately connected with a basic object: the effective twisted superpotential \cite{Cecotti:1992rm,Witten:1993yc} of the theory compactified on a circle and mass-deformed in such a way to preserve $(2,2)$ SUSY in two dimensions. This object will be our main tool throughout this paper, but we expect most of our results to admit generalizations involving the other protected probes. 

According to the Nekrasov-Shatashvili duality \cite{Nekrasov:2009uh, Nekrasov:2009ui} the equations which determine the geometry of the space of vacua take a specific form, which resembles the form of Bethe equations for integrable systems such as the XXZ spin chain. 
It is straightforward to engineer a gauge theory whose space of vacua would match the eigenstates of 
a specific spin chain Hamiltonian. If the gauge theory has a mirror, the same space of vacua may be described by different sets of Bethe-like equations, associated to a different integrable system. This leads to {\it bispectral duality} between integrable systems, 
i.e. a correspondence between the eigenstates of two integrable Hamiltonians. 
The first motivation of this paper was to elucidate the details of the bispectral duality induced by the mirror symmetry 
for $\ssN=4$ linear quiver gauge theories. 

String theory allows the identification of a large number of mirror pairs, through brane engineering techniques \cite{Hanany:1996ie}. 
In particular, it provides a mirror for all linear quivers with $\prod_i U(N_i)$ gauge groups and enough fundamental flavors to 
insure a nice RG flow to the IR. The same brane construction can also be used to engineer half-BPS boundary conditions 
and domain walls for $\ssN=4$ SYM gauge theory \cite{Gaiotto:2008sa}. Indeed, each elementary brane intersection descends to a specific domain wall 
in the four-dimensional gauge theory, which can be given a simple field-theoretic description. Both three-dimensional SCFTs and intricate superconformal domain walls and boundary conditions can be assembled in the UV from such elementary building blocks, 
often in many distinct manners. This construction ties together three-dimensional mirror symmetry and four-dimensional S-duality \cite{Gaiotto:2008ak}. 

Clearly, it is natural for us to study these elementary four-dimensional building blocks in the same way as we did for the three-dimensional 
theories. The canonical mass deformation in three dimensions coincides with the standard $\ssN=2^*$ mass deformation of the four-dimensional gauge theory. The moduli space of four-dimensional $U(N)$ $\ssN=4$ gauge theory,
compactified on a circle and mass-deformed to $\ssN=2^*$, is well-known: it coincides with the moduli space $\CM$ of $GL(N)$
flat connections on a one-punctured torus, with minimal semisimple monodromy at the puncture \cite{Donagi:1995cf}. In a given S-duality frame, 
this can be identified \cite{Nekrasov:2011bc} with the phase space of the classical trigonometric Ruijsenaars-Schneider (tRS) model \cite{MR929148, MR1322943}.

Every BPS boundary condition for the four-dimensional theory maps to a specific complex Lagrangian submanifold in $\CM$. 
Domain walls correspond to Lagrangian sub-manifolds in the product of moduli spaces $\CM_L \times \CM_R$ for the theories on the 
two sides of the wall. The moduli space of vacua for the four-dimensional theory on a segment is computed by taking the intersection of 
the appropriate Lagrangian submanifolds in $\CM$. For example, a particularly general construction of the SCFTs which correspond to linear quiver gauge theories with $\prod_i U(N_i)$ gauge groups
consists an $U(N)$ $\ssN=4$ SYM theory on a segment, with a generalized Dirichlet boundary condition, labelled by a Young tableau $\rho$, on one side, and the S-dual of the same type of boundary condition, labelled by a second Young tableau $\rho^\vee$. 
This description was labelled as $T[U(N)]_\rho^{\rho^\vee}$ in \cite{Gaiotto:2008ak}. The moduli space of vacua of the system
is thus described as the intersections of the corresponding Lagrangian sub-manifolds $\CL_\rho \cap \CL^\vee_{\rho^\vee}$. 

We will determine the submanifolds associated to the elementary building blocks of brane constructions, 
through a judicious combination of localization, three-dimensional results and S-duality. In particular, we will characterize the 
$\CL_\rho$ and $\CL^\vee_{\rho^\vee}$ manifolds, and thus provide a novel geometric description of the moduli space of vacua for the three-dimensional SCFTs which are the main subject of this paper. As a bi-product, 
we will derive an intricate network of relations between the XXZ spin chains and the tRS model. 

Finally, we will take the compactification radius to zero to derive some further results about two-dimensional $(4,4)$ gauge theories, XXX spin chains and Gaudin integrable systems. Notice that the three-dimensional problem we study can be connected to mathematical questions concerning the so-called K-theoretic quantum cohomology (see \cite{2012arXiv1211.1287M, 2010arXiv1001.0056B} and references therein), which goes to the standard A-model quantum cohomology in the zero radius limit. 

The paper is organized as follows. In \Secref{Sec:3dN4Quiver} we investigate moduli spaces of vacua $\CL$ of 3d linear quiver gauge theories with $\CN=2^*$ supersymmetry. Following the NS correspondence we regard $\CL$ as a parameter space of solutions of the XXZ chain whose data are given by color and flavor labels of the quiver. We discuss how 3d mirror symmetry acts on the quiver and conclude that in the spin chain language mirror symmetry corresponds to the so-called \textit{bispectral duality}.

\Secref{sec:bcS} is devoted to the study of the moduli space of vacua of the circle compactification of the $\CN=2^*$ four-dimensional SYM theory subject to $1/2$ BPS boundary conditions on a segment. We review the S-duality action on the BPS boundary conditions, thereby refining known results for $\CN=4$ theories \cite{Gaiotto:2008ak}. The corresponding $T[U(N)]_\rho^{\rho^\vee}$ theory is mapped onto its mirror under the S-duality. Then we formulate a geometric description of moduli spaces of vacua of $T[U(N)]_\rho^{\rho^\vee}$ theories and connect them to the solutions of XXZ chain Bethe equations.

After that \Secref{Sec:IntAppl} will explain what the above connection means in terms of integrable systems, such as XXZ chains and trigonometric RS models. Plethora of other integrable models, some of them being dual to each other, appear from the first two by various scaling limits. Conclusions and future directions are given in \Secref{Sec:FutureWork}.

The reader who is more interested in applications of our results to integrable systems may continue reading the paper from 
\Secref{Sec:IntAppl}: The results of the first two sections are briefly summarized in the beginning of \Secref{Sec:IntAppl} and the anticipated dualities between the integrable models we have announced in the abstract are presented. 

During the completion of this paper, we received \cite{Yaakov:2013fza}, which independently proposes and studies in detail the same Seiberg-like duality which we proposed in section \secref{Sec:3dN4Quiver}. 

\section{Quiver Gauge Theories and Mirror Symmetry in Three Dimensions}\label{Sec:3dN4Quiver}
In this section we give the description of the moduli space $\CL$ of a linear quiver gauge theory as a space of solutions of a quantum XXZ spin chain. Following the NS duality strategy, we cast the vacuum equations as Bethe equations for a spin chain Hamiltonian, which we identify with an XXZ $SU(L)$ spin chain with spins transforming in various antisymmetric representations. 
In the process, we identify novel Seiberg-like dualities of the gauge theories, which map to Weyl reflections in the spin chain.  

\subsection{Mass deformations and definition of $\CL$}
We are interested in mass deformations of $\ssN=4$ SCFTs which preserve $\ssN=2$ supersymmetry. It is useful to review first 
the mass-deformations which preserve $\ssN=4$ supersymmetry. 

An $\ssN=4$ SCFT has an $SU(2)_H \times SU(2)_C$ R-symmetry group. The labels $H$ and $C$ 
refer to the fact that the two factors respectively act on the scalar fields which parameterize the 
Higgs and Coulomb branches. Furthermore, an $\ssN=4$ SCFT may have two flavor symmetry groups, $G_H$ and $G_C$, which respectively act on the Higgs and Coulomb branch scalars only. 

The flavor symmetries are associated to canonical $\ssN=4$ mass deformations. The mass parameters can be thought of as the vev of background gauge multiplets coupled to the flavor symmetries.
Higgs branch flavor symmetries correspond to $SU(2)_H$ triplets of parameters $m^A_a$ valued in the Cartan algebra of $G_H$. In a UV Lagrangian description, 
they enter as masses for hypermultiplets. Coulomb branch flavor symmetries correspond to $SU(2)_C$ triplets of parameters $t^I_i$ valued in the Cartan algebra of $G_C$. In a UV Lagrangian description, they enter as FI parameters for Abelian factors in the gauge group \footnote{For certain theories, which we do not consider in this paper, Coulomb branch flavor symmetries are not visible in the UV, and the corresponding FI parameters are absent.}. 

The general strategy to identify $\ssN=4$ massive vacua is to compute the Higgs branch of the theory exactly, keeping the FI parameters into account, and then look at the fixed points under the action of the $m^A_a$ generators in the flavor group $G_H$. The matching of $\ssN=4$ massive vacua in the mirror or four-dimensional descriptions of a theory is already a rather non-trivial, interesting problem, which is related to the mathematical subject 
of symplectic duality. 

We are interested in turning on yet another mass deformation, which breaks SUSY down to $\ssN=2$. There is a simple way to describe the deformation. First, we can pick an $\ssN=2$ sub-algebra of the supersymmetry algebra. From the point of view of the 
$\ssN=2$ sub-algebra, the theory has a flavor group $G_H \times G_C \times U(1)_\epsilon$. Indeed, with no loss of generality 
the $\ssN=2$ $U(1)_R$ R-symmetry generator can be taken to be the diagonal $j^3_H + j^3_C$ Cartan generator of the $\ssN=4$ $SU(2)_H \times SU(2)_C$ R-symmetry algebra. The other Cartan generator $j^3_H - j^3_C$ commutes with the $\ssN=2$ supercharges, and generates the $U(1)_\epsilon$ flavor symmetry.  

We will denote the resulting theory as an 3d $\ssN=2^*$ theory. In a gauge theory description where the hypermultiplets are rotated by $SU(2)_H$, the mass deformation contributes $+\epsilon/2$ to the real mass of the $\ssN=2$ chiral multiplets which come from hypermultiplets, and $-\epsilon$ to the real mass of the $\ssN=2$ chiral multiplets which come from vectormultiplets. In a mirror description where hypermultiplets are rotated by $SU(2)_C$, the charge assignments are opposite. In flat space, given a gauge theory description of a theory, the massive vacua of the $\ssN=2^*$ theory can still be described as fixed points on the Higgs branch, including the effect of $\epsilon$ in the Higgs branch isometry. We will not pursue this characterization here, but it might be interesting to do so. 

Flavor symmetries in $\ssN=2$ SCFTs are also associated to canonical, real mass deformations. The real masses for 
$G_H \times G_C$ correspond to the third components $m_a$ of the $m^A_a$ triplets, and $t_i$ of the $t^I_i$ triplets.
The real mass $\epsilon$ for $U(1)_\epsilon$ is a new deformation parameter, and breaks $\ssN=4$ to $\ssN=2$ explicitly. 
We will sometimes denote all the $\ssN=2$ mass parameters with a uniform notation $u = (m_a, t_i, \epsilon/2)$ (note the factor of $2$ inserted for convenience). The remaining components of $m^A_a$ and $t^I_i$ become superpotential deformation parameters in the $\ssN=2$ language. Unless otherwise specified, we will not turn them on.

Our next step is to compactify the three-dimensional $\ssN=2^*$ on a circle of radius $R$. The three-dimensional mass parameters 
can be naturally combined with the corresponding flavour Wilson lines to give complex mass parameters 
\begin{equation}
u^{2d} = u^{3d} -\frac{i}{R}\oint\limits_{S^1}A^{f} \,,
\end{equation} 
which behave as twisted masses in the language of the $(2,2)$ 2d supersymmetry. We will usually omit the $2d$ superscript. 
As the flavor Wilson lines are periodic, it is often useful to exponentiate the masses to $\nu = e^{2 \pi R u}$, i.e.
\begin{equation}
\mu_a = e^{2 \pi R m_a}\,, \qquad \tau_i = e^{2 \pi R t_i}\,, \qquad \eta = e^{\pi R \epsilon}\,.
\label{eq:nus}
\end{equation}

The mass-deformed, compactified theory has generically a finite set of massive vacua for each value of the mass parameters. Massless directions may open up at codimension two loci. As we vary the mass parameters, the vacua describe a manifold $\CL$. Each massive vacuum is associated to a specific low-energy effective twisted superpotential $\CW(u)$, which is defined up to shifts by constants and integer multiples of the mass parameters. It is very useful to consider the vev of the partial derivatives of $\CW$ with respect to 
the mass parameters of the theory, exponentiated in order to remove the shift ambiguities:
\begin{equation}
p_\nu = \exp \left[2 \pi R \frac{\partial \CW}{\partial u} \right]\,,
\label{eq:Pu}
\end{equation}
in detail:
\begin{equation}
p_\mu^a = \exp \left[2 \pi R \frac{\partial \CW}{\partial m_a} \right]\,, \qquad p_\tau^i = \exp \left[2 \pi R \frac{\partial \CW}{\partial t_i} \right]\,, \qquad p_\eta = \exp \left[4 \pi R \frac{\partial \CW}{\partial \epsilon} \right]\,.
\label{eq:PmuPtauPeta}
\end{equation}
With this definition, the manifold of vacua $\CL$ will sit as a smooth algebraic Lagrangian submanifold in the space of mass parameters and conjugate momenta, equipped with the natural symplectic form $\frac{d\nu}{\nu}\wedge\frac{dp_\nu}{p_\nu}$. 

Crucially, the low-energy effective twisted superpotential $\CW(u)$ is independent from both superpotential 
and gauge couplings of the original 3d $\ssN=2$ theory. Thus it can be computed directly in a UV gauge theory
description of the theory. The general strategy described in \cite{Nekrasov:2009uh} consists of integrating away at first all the 3d chiral multiplets, to obtain a twisted effective superpotential $\CW(u,s)$ for the gauge multiplets, where by $s$ we denote the lowest component of the latter. The vacua of the theory are the extrema of $\CW(u,s)$, i.e. the solutions of 
\begin{equation}
p_\sigma \equiv \exp \left[2 \pi R \frac{\partial \CW}{\partial s} \right] =1\,,
\label{eq:ext}
\end{equation}
and the low-energy effective superpotential $\CW(u)$ in each vacuum is the corresponding extremum of $\CW(u,s)$. 

The contribution to $\CW(u,s)$ of a 3d chiral multiplet of twisted mass $x$ gives an effective superpotential 
\begin{equation}
\CW_{\mathrm{chiral}} = \ell(x)\,,
\label{eq:Wchiralell}
\end{equation}
where $\ell(x)$ is such that $2 \pi R \partial_x \ell(x) = \log 2\sinh \pi R x$. Thus the two chirals inside a 3d $\ssN=4$ hypermultiplet give 
\begin{equation}
\CW_{\mathrm{hyper}}(x) = \ell(x+ \epsilon/2) +  \ell(-x+ \epsilon/2)\,,
\end{equation}
where $\epsilon$ is the $\ssN=4$ twisted mass. The chiral inside a 3d $\ssN=4$ vectormultiplet gives similarly
\begin{equation}
\CW_{\mathrm{vector}}(x) = \ell(x- \epsilon)\,.
\end{equation}

Furthermore, the schematic form of the effective superpotential will be 
\begin{equation}
\CW = \sum_k \CW_{\mathrm{hyper}}(q_k^g \cdot s + q_k^f \cdot m) + \sum_n \CW_{\mathrm{vector}}( e_n^g \cdot s) + t \cdot s\,,
\end{equation}
where $q_k^g$ is the gauge charge of hypermultiplets, $q_k^f$ the flavor charge and $e_n^g$ the gauge charge of the chiral fields in the adjoint vectormultiplets. The last term is the FI coupling. The special form of the effective superpotential means that the conjugate momenta $p_\nu$ and $p_\sigma$ are rational functions of $\nu$ and $\sigma$, and that the manifold of vacua $\CL$ is described by a certain collection of polynomial equations. 

We should address here an important subtlety in the definition of the moduli space $\CL$. The supersymmetric compactification 
of the theory on a circle assumes implicitly a precise choice of a fermion number $(-1)^F$. Such a choice is inherently ambiguous in a generic SCFT, because we can always shift the fermion number by the generator of some other $\mathbb{Z}_2$ flavor symmetry. 
In particular, shifts by a $\mathbb{Z}_2$ subgroup of a continuous $U(1)$ flavor symmetry can be described by a shift by $\pi$ of the corresponding 
flavor Wilson line parameter, i.e. $\nu_a \to - \nu_a$ for some $a$. Equivalently, there is a $\mathbb{Z}_2$ ambiguity in choosing an ``origin'' 
for the flavor Wilson line parameters, $\nu_a = \pm 1$. 

If we are given a Lagrangian UV description of the theory, the chiral matter scalar fields have a canonical fermion number assignment 
of zero, and thus the $\mu_a$ parameters have a natural origin. The situation for the $t_i$ parameters is less clear-cut. 
A simple example is a theory with two chiral fields of opposite charge under some gauge group, which can be lifted 
adding the corresponding mesonic operator to the (standard, not twisted) superpotential. The contribution of such a doublet should naively 
drop out from the twisted superpotential, but we get instead 
\begin{equation}
2 \pi R \partial_s \CW(s) =  \log 2\sinh \pi R s- \log 2\sinh \pi R (-s) = - i \pi
\end{equation} 
In other words, integrating away the doublet shifts the complexified FI parameter as $\tau \to - \tau$. This can be interpreted as 
a shift in the fermion number assignment for monopole operators. 

This is a particular case of a general phenomenon: the canonical fermion number assignment for the elementary fields 
may very well result in an intricate fermion number assignment for monopole operators. Thus the canonical 
fermion number assignment in a mirror description of the theory may correspond in a peculiar choice 
of origin for the FI parameters of the original theory. Tracking down the 
correct sign rule in the mirror symmetry transformation will not be trivial, and will require some careful checks. 
In order to streamline the presentation, we will include some arbitrary shifts of the FI parameters in our 
general formulae, and later fix them in such a way that mirror symmetry acts in a simple way. 
These sign choices will also be important when taking the 2d limits of our formulae in \secref{Sec:IntAppl}. 
 
\subsection{Linear $A_L$ quivers}
\begin{figure}
\begin{center}
\includegraphics[height=3cm, width=8cm]{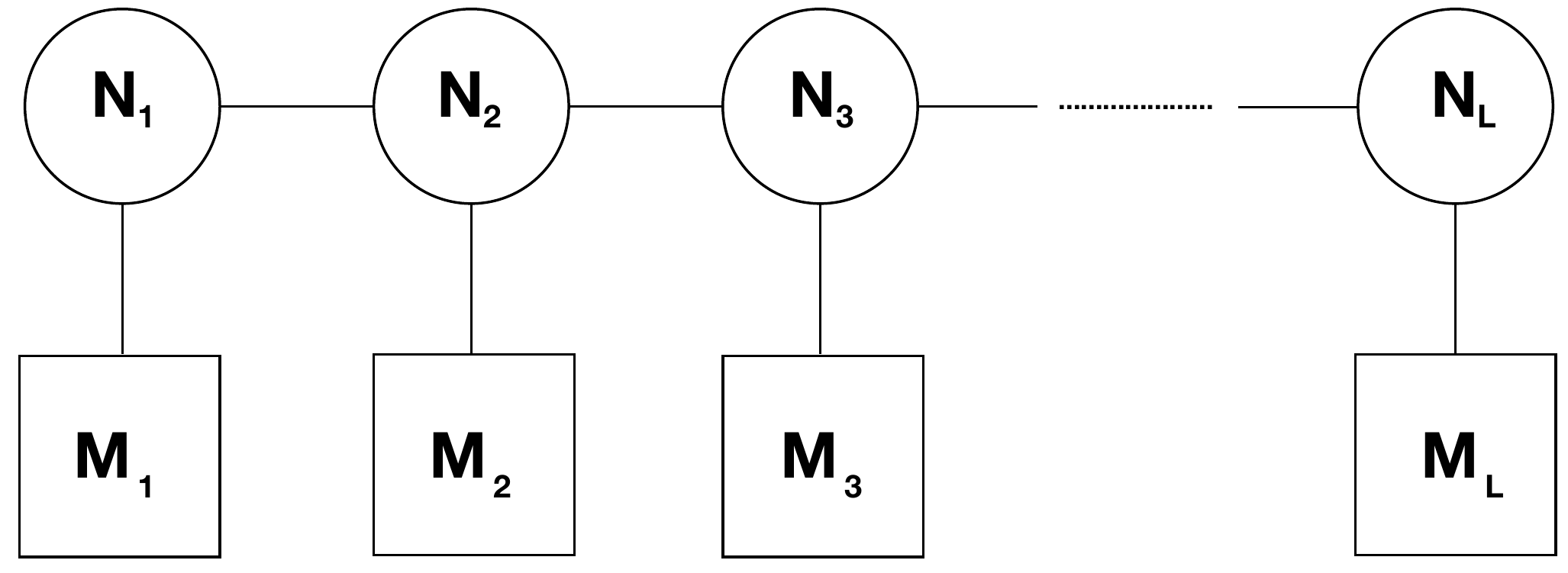}
\caption{$A_L$ linear quiver with labels $(N_1,M_1),\dots, (N_L,M_L)$.}
\label{fig:alquiv}
\end{center}
\end{figure}

Most of our analysis will focus on $A_L$ quivers, i.e. linear quivers with gauge group $\prod_i U(N_i)$ and 
$M_i$ extra fundamental hypermultiplets at each node (see \figref{fig:alquiv}). 
The matter content determines uniquely the $\ssN=4$ Lagrangian, and thus the $A_L$ quivers can be labelled by the ranks $N_i$ and numbers of flavors $M_i$ at each node. Sometimes we shall refer to these numbers as color and flavor labels of each node respectively. 
There is a $S \left(\prod_i U(M_i) \right)$ flavor symmetry acting on the fundamental hypers, and a $U(1)^L$ ``topological'' 
flavor symmetry associated to the $U(1)$ factors in the gauge groups, which can be enhanced in the infrared up to $SU(L+1)$.
We will find it useful to add two extra $U(1)$ flavor symmetries which act trivially on the quiver gauge theory, to complete the hypermultiplet flavor symmetry to $\prod_i U(M_i)$, with real masses $m_a$ acting on the $a$-th fundamental hypermultiplet, and the topological flavor symmetry to $U(1)^{L+1}$ (possibly enhanced up to $U(L+1)$), with real masses $t_i$ defined so that the FI parameter at the $i$-th node is $t_{i+1}-t_i$. The redundant parameters can be eliminated at any time by imposing $\sum_a m_a =0$ and $\sum_i t_i=0$ and adequate constraints on the conjugate momenta. Keeping the parameters in will make formulae simpler and facilitate the comparison with brane constructions and four-dimensional gauge theory. 

The infrared physics of $\ssN=4$ quiver gauge theories, before mass deformation, depends crucially on the quantities 
\begin{equation}
\Delta_i = M_i + N_{i-1} + N_{i+1} - 2 N_i\,, 
\label{eq:deltaqeight}
\end{equation}
which control the R-charge carried by monopole operators. 
If all these integers are non-negative, the quiver gauge theory conjecturally flows to a superconformal field theory with the same R-charge as the UV Lagrangian \cite{Gaiotto:2008ak}. Such ``good'' $A_L$ quivers are the main focus of our paper. 
If some of the $\Delta_i$ are negative, the naive R-charge of monopole operators is inconsistent with the infrared unitarity bounds \cite{Borokhov:2002cg}, and the RG flow must be more complicated, involving the appearance of accidental R-symmetries. 

Our analysis of the manifold of vacua $\CL$ for $A_L$ quivers will also allow us to formulate a simple conjecture
on the infrared descriptions of quivers for which  $2N_i > M_i + N_{i-1} + N_{i+1} \geq N_i$ at some of the nodes: 
up to specific decoupled sectors which arise from monopole operators hitting the unitarity bound, 
the infrared theory has an alternative described in terms of a quiver obtained from the UV quiver by repeated applications of 
the $N_i \to M_i + N_{i-1} + N_{i+1}-N_i$ transformation on the ranks of nodes with negative $\Delta_i$.
Such transformations generate the Weyl group of $A_L$. As long as the ranks remain non-negative, the orbit of the 
original quiver under these Weyl transformations will include a good quiver description.

The conjectural duality between $U(N_c)$ UV theory with $N_f$ flavors and a $U(N_f - N_c)$ IR theory with $N_f$
flavors accompanied by free monopoles is clearly reminiscent of the results of \cite{Aharony:1997bx}. 
It should be possible to derive this relation from the compactification of the corresponding four-dimensional $U(N_c)$ $\ssN=2$ 
gauge theory and its Seiberg-Witten solution \cite{Argyres1996159}.

In what follows we will review the equations which describe the moduli space $\CL$ for $A_L$ quivers. 
It is useful to work our way through a few simple examples at first, which illustrate well several of our general results. 

\subsubsection{The basic Abelian mirror pair}
The simplest non-trivial example is a $U(1)$ theory with a single flavor of hypermultiplets, i.e. $L=N_1=M_1=1$. 
The effective superpotential is simply
\begin{align}
\CW &= \ell(s + \epsilon/2)+\ell(-s+\epsilon/2)+ (t_2-t_1) s+ \ell(-\epsilon)\,,
\label{eq:basic}
\end{align}
The extremum equations become 
\begin{equation}
\frac{\tau_2}{\tau_1} \frac{2\sinh \pi R (s + \epsilon/2)}{2\sinh \pi R (-s+\epsilon/2)} = \frac{\tau_2}{\tau_1} \frac{\sigma\eta - 1}{\eta - \sigma} = 1\,.
\end{equation}
We can simply solve 
\begin{equation}
\sigma = \frac{\eta \tau_1+ \tau_2}{\tau_1 + \eta \tau_2} \,,
\label{eq:sol1hyp}
\end{equation}
and plug back into the momenta
\begin{equation}
p_\tau^2 = (p_\tau^1)^{-1} = \frac{\eta \tau_1+ \tau_2}{\tau_1 + \eta \tau_2}\,,  \qquad p_\eta = \frac{\eta (\sigma \eta -1)(\eta \sigma^{-1} - 1)}{(1-\eta^2)^2} = \frac{\eta \tau_1 \tau_2}{(\tau_1 + \eta \tau_2)(\tau_2 + \eta \tau_1)}\,.
\end{equation}
Thus the low energy effective superpotential evaluated on the solutions \eqref{eq:sol1hyp} reads
\begin{equation}
\CW = \ell(t_1 - t_2 + \frac{i}{2R} - \epsilon/2) + \ell(t_2 - t_1 + \frac{i}{2R} - \epsilon/2)\,,
\end{equation}
and we see clearly the well-known mirror description: a single twisted hypermultiplet. Notice the extra $\frac{i}{2R}$ contribution to the complexified mass parameter. We will combine it with the $\CN=2^*$ mass parameter, 
and refine the expected $\epsilon \to - \epsilon$ action of mirror symmetry to 
\begin{equation}
\epsilon/2 \to \frac{i}{2R}-\epsilon/2\,, \qquad \eta \to - \eta^{-1}\,.
\end{equation}
This choice will work well in more general examples. In all cases, we could alternatively absorb the extra $\frac{i}{2R}$ 
shifts into more complicated mirror transformations for the FI parameter and mass parameters. 

\subsubsection{The $T[U(2)]$ theory}
The next simplest example of $A_1$ quiver is a $U(1)$ theory with two flavors, i.e. $L=N_1=1$ and $M_1=2$.
This gauge theory plays an important role in the action of S-duality on boundary conditions for $U(2)$ $\ssN=4$ SYM in four dimensions. 
It is known to be self-mirror. 
The effective twisted superpotential reads
\begin{align}
\CW &= \ell(s - m_1 + \epsilon/2)+\ell(-s+ m_1+\epsilon/2)+ \ell(s - m_2 + \epsilon/2)+\ell(-s+m_2+\epsilon/2)\notag\\&+ (t_2-t_1) s+ \ell(-\epsilon) + t_1 (m_1 + m_2)  \,,
\label{eq:EfTwSPtsu21}
\end{align}
The last term, which only affects the spurious $U(1)$ symmetry parameters, is added for later convenience. It is also present in the original definition of the $T[U(2)]$ theory.

To find the vacua, we solve 
\begin{equation}
\frac{\tau_2}{\tau_1} \frac{\sigma\eta - \mu_1}{\eta \mu_1- \sigma} \frac{\sigma\eta - \mu_2}{\eta \mu_2- \sigma} = 1\,, 
\label{eq:SL2BAE1r}
\end{equation} a quadratic equation with two solutions for $\sigma$. The solution is not particularly instructive, but it is instructive to rewrite the equation as
\begin{equation}
\frac{\eta \tau_2 - \eta^{-1} \tau_1}{\tau_2 - \tau_1} p_\tau^2+  \frac{\eta \tau_1 - \eta^{-1}  \tau_2}{\tau_1 - \tau_2} p_\tau^1= \mu_1 + \mu_2\,, \qquad p_\tau^1 p_\tau^2 = \mu_1 \mu_2\,, 
\label{eq:wtTSU2}
\end{equation}
where we used $p_\tau^2=\sigma$ and $p_\tau^1 = \frac{\mu_1 \mu_2}{\sigma}$.

We observe a striking fact. If we use the definition of $p_\mu^i$
\begin{equation}
p_\mu^1 = \tau_1 \frac{\eta \mu_1 - \sigma}{\eta \sigma - \mu_1}\,, \qquad p_\mu^2 = \tau_1 \frac{\eta \mu_2 - \sigma}{\eta \sigma - \mu_2}\,,
\label{eq:TSU2ex}
\end{equation}
and plug either of the two roots of the equation for the vacua, we derive a second relation 
\begin{equation}
\frac{-\eta^{-1} \mu_2 + \eta \mu_1}{ \mu_2 - \mu_1} p_\mu^2+  \frac{-\eta^{-1} \mu_1 + \eta \mu_2}{ \mu_1 - \mu_2} p_\mu^1= \tau_1 + \tau_2 \qquad p_\mu^2 p_\mu^1 = \tau_1 \tau_2\,, 
\label{eq:mwtTSU2}
\end{equation}
which shows clearly the self-mirror properties of $T[U(2)]$: mirror symmetry sends $\eta \to - \eta^{-1}$ and $\mu_i \leftrightarrow \tau_i$. 
In a later section, we will show how these equation make the role of $T[U(2)]$ in S-duality manifest. 

We should also look at the definition of $p_\eta$:
\begin{equation}
p_\eta = \frac{(\sigma\eta - \mu_1)(\eta \mu_1- \sigma)(\sigma\eta - \mu_2)(\eta \mu_2- \sigma)}{\sigma^2 \mu_1 \mu_2(1-\eta^2)^2}\,.
\end{equation}
After some algebra, we find two neat relations
\begin{equation}
p_\eta = \frac{(\mu_1 - \mu_2)^2}{(p_\mu^1 - p_\mu^2)^2}\frac{\tau_1 \tau_2}{\mu_1 \mu_2} = \frac{(p_\tau^1 - p_\tau^2)^2}{(\tau_1 - \tau_2)^2}\frac{\tau_1 \tau_2}{\mu_1 \mu_2} \label{eq:TSU2pe}\,,
\end{equation}
which agree with the action of mirror symmetry $p_\eta \to \frac{1}{p_\eta}$. It is useful to observe a related fact:
\begin{equation}
(p_\mu^1 - p_\mu^2)(p_\tau^1 - p_\tau^2)+(\mu_1 - \mu_2)(\tau_1 - \tau_2)=0\,.
\end{equation}
We arrived to an interesting, symmetric description of $\CL$. We know that $\CL$ is a double cover of the space of mass parameters. If we use  \eqref{eq:wtTSU2} to determine the $p_\tau^i$ and  \eqref{eq:mwtTSU2} to determine the $p_\mu^i$, we obtain a degree four cover. Crucially, that locus has two distinct components, and $\CL$ is the one selected by the second equality in \eqref{eq:TSU2pe}.

\subsubsection{The moduli space of ${\cal N} = 2^*$ SQCD}
Our last example before the general formula is ${\cal N} = 2^*$ SQCD,
i.e. a $U(N)$ gauge theory with $M$ flavors. The effective superpotential is 
\begin{align}
\CW &= \sum_{n,a}\ell(s_n - m_a + \epsilon/2)+\ell(-s_n+ m_a+\epsilon/2) + \sum_{n,n'}\ell(s_n - s_{n'} - \epsilon) \notag\\&+ (t_2-t_1) \sum_n s_n+ t_1 \sum_a m_a \,,
\label{eq:EfSQCD}
\end{align}
and the vacuum equations become
\begin{equation}
\frac{\tau_2}{\tau_1} \prod_{n'=1}^N \frac{\sigma_n - \eta^2 \sigma_{n'}}{ \sigma_{n'}- \eta^2 \sigma_n} \prod_{a=1}^M  \frac{ \eta \sigma_n- \mu_a}{\eta \mu_a - \sigma_{n}} = 1\,.
\end{equation}
According to the so-called \textit{Gauge/Bethe correspondence} or \textit{Nekrasov-Shatashvili (NS) duality}, these equations take the form of Bethe equations for a twisted anisotropic XXZ $\alg{su}(2)$ spin chain with $M$ sites with impurity parameters $m_a$ and twist $(-1)^{N+M-1}\frac{\tau_2}{\tau_1}$ around the circle, in a sector with $N$ Bethe roots. Because of the twist, each solution of the Bethe equations, and thus each vacuum of the gauge theory, corresponds to an eigenstate for the spin chain Hamiltonian. 

In particular, a ``good'' theory $M \geq 2N$ corresponds to spin chain states with positive or zero weight $M-2N$. Spin chain states with negative weight are associated to vacua of theories with $2N > M \geq N$. In the spin chain context,  there is an obvious bijection between states of positive weight and states of negative weight in a chain with inverse twist, given by a Weyl reflection which reverses all spins. This induces a bijection between the vacua of a $U(N)$ and a $U(\tilde N)$ gauge theories with $M=N+\tilde N$ flavors. This bijection between massive vacua is not very surprising: after all, for non-zero FI parameter and 
 $\epsilon = m_a=0$, the Higgs branch of both theories coincides with the cotangent 
bundle of the Grassmanian of $N$-planes in $M$ dimensions. Turning on the other mass parameters localizes the theory on the fixed point of the $U(1)_\epsilon \times SU(M)$ isometries of the Higgs branch. 

The observation becomes more interesting if we can match the effective twisted super potentials of the two theories in the corresponding pairs
of vacua, and thus demonstrate a precise Seiberg-like duality between the two $\ssN=4$ theories, analogous to dualities which hold in theories with less supersymmetry
\cite{Aharony:1997gp,Kapustin:2010mh,Bashkirov:2010kz}. We can take inspiration from the action of Weyl symmetry on the spin chain Bethe roots. It is well know that 
two sets of Bethe roots $\sigma_n$ and $\tilde \sigma_n$ related by a Weyl symmetry transformation satisfy the relation
\begin{equation}
 \eta^{N}  \tilde \tau_1 Q(\eta^{-1} \sigma) \tilde Q(\eta \sigma)- (-1)^M \eta^{\tilde N}\tau_1 Q(\eta \sigma) \tilde Q(\eta^{-1} \sigma) = \left(\eta^{\tilde N} \tilde \tau_1-  (-1)^M \eta^N \tau_1\right) M(\sigma)\,.
\end{equation}
We defined $\tilde \tau_1 = (- 1)^{N+1} \tau_2$ and the various $Q$-functions
\begin{equation} Q(\sigma) = \prod_n (\sigma-\sigma_n)\,, \qquad  \tilde Q(\sigma) = \prod_n (\sigma-\tilde \sigma_n)\,, \qquad M(\sigma) = \prod_a (\sigma - \mu_a)\,.
\end{equation}

This simple equation actually implies the Bethe equations for both $\sigma_n$ and $\tilde \sigma_n$: the Bethe equations can be written in the $T-Q$ form 
\begin{equation}
\eta^{N} \tilde \tau_1 Q(\eta^{-2} \sigma) M(\eta \sigma) +  (-1)^M \eta^{\tilde N}\tau_1 Q(\eta^{2} \sigma) M(\eta^{-1} \sigma) = Q(\sigma) T(\sigma)\,,
\end{equation}
and become identically true if we plug in the bilinear form of $M(\sigma)$ and 
\begin{equation}
\eta^{2N} \tilde \tau_1^2 Q(\eta^{-2} \sigma) \tilde Q(\eta^2 \sigma)  - \eta^{2 \tilde N}\tau_1^2 Q(\eta^{2} \sigma) \tilde Q(\eta^{-2} \sigma)= \left(\eta^{\tilde N} \tilde \tau_1-  (-1)^M \eta^N \tau_1\right)T(\sigma)\,.
\end{equation}

The $Q-\tilde Q$ relation gives us almost all information we need about the conjugate momenta which label the vacua in the two theories. 
Indeed, we have
\begin{equation}
p_\mu^a = \tau_1 \prod_n \frac{\mu_a \eta - \sigma_n}{\sigma_n \eta- \mu_a} = (-1)^{N} \frac{\tau_1 Q(\eta \mu_a)}{\eta^N Q(\eta^{-1} \mu_a)}\,.
\label{eq:pmua}
\end{equation}
If we evaluate the $Q-\tilde Q$ relation at $\mu_a$, we learn that the momenta in the dual theory are identical
\begin{equation}
\tilde p_\mu^a = p_\mu^a\,.
\end{equation}
In other words, the effective super-potentials for SQCD with gauge groups $U(N)$ and $U(\tilde N)$ have the same $\mu_a$ dependence.

Next, we can look at $p_\tau^i$:
\begin{equation}
p_\tau^2 = \prod_n \sigma_n = (-1)^N Q(0)\,, \qquad p_\tau^1 = \frac{\prod_a \mu_a}{\prod_n \sigma_n } = (-1)^{M-N}\frac{M(0)}{Q(0)}\,.
\end{equation}
Setting $\sigma=0$ in the $Q-\tilde Q$ relation we get
\begin{equation}
\tilde p_\tau^2 = (-1)^{\tilde N} \tilde Q(0) = \frac{\eta^{\tilde N} \tilde \tau_1-  (-1)^M \eta^N \tau_1}{ \eta^{N}  \tilde \tau_1- (-1)^M \eta^{\tilde N}\tau_1 }p_\tau^1\,,
 \qquad \tilde p_\tau^1 = \frac{ \eta^{N}  \tilde \tau_1- (-1)^M \eta^{\tilde N}\tau_1 } {\eta^{\tilde N} \tilde \tau_1-  (-1)^M \eta^N \tau_1} p_\tau^2\,.
\end{equation}

This relation indicates the presence of extra twisted hypermultiplets in the IR. 
The Coulomb branch of the UV theory $\CN=4$ $U(N)$ SQCD with $2N > M \geq N$ has (hyperk\"ahler) dimension $N$. The Coulomb branch  of the conjectural IR theory $\CN=4$ $U(M-N)$ SQCD has dimension $M-N$. 
The mismatch should correspond to $2N-M$ directions of the UV Coulomb branch 
which will give rise to free twisted hypermultiplets in the IR, possibly belonging to a non-standard representation of the UV R-symmetry group. These free fields should arise from monopole operators hitting a unitarity bound along the RG flow. 

The effective superpotential of the UV theory indeed appears to differ from the one of the IR theory by an amount which contributes  to $p_\tau^2$ as
\begin{align}
& \Delta p_\tau^2 = \frac{\eta^{\tilde N} (- 1)^{N+1} \tau_2-  (-1)^M \eta^N \tau_1} { \eta^{N}  (- 1)^{N+1}  \tau_2- (-1)^M \eta^{\tilde N}\tau_1 }  
= \cr &\frac{\eta^{N-1}  \tau_2 + (-1)^{\tilde N} \eta^{\tilde N+1}\tau_1}{ \eta^{N}  \tau_2 + (-1)^{\tilde N} \eta^{\tilde N}\tau_1 }  
\frac{\eta^{N-2}  \tau_2 + (-1)^{\tilde N} \eta^{\tilde N+2}\tau_1}{\eta^{N-1}  \tau_2 + (-1)^{\tilde N} \eta^{\tilde N+1}\tau_1} \cdots
\frac{\eta^{\tilde N}  \tau_2 + (-1)^{\tilde N} \eta^{N}\tau_1}{\eta^{\tilde N+1}  \tau_2 + (-1)^{\tilde N} \eta^{N-1}\tau_1}\,,
\end{align}
i.e. 
\begin{align}
\Delta \CW &= \sum_{r=1}^{2N-M} \ell\left(t_2 - t_1 + (2N-M-2r)\frac{\epsilon}{2}+ (N-M-1)\frac{i}{2R}\right)\cr
&+\ell\left(-t_2 + t_1 - (2N-M-2r+2 )\frac{\epsilon}{2}- (N-M-1)\frac{i}{2R}\right)\,,
\end{align}
the contribution of $2N-M$ twisted free hypers of appropriately shifted $R$-charges which agree with the quantum numbers of the charge $\pm 1$ monopole operators which are expected to hit the unitarity bound along the RG flow \cite{Borokhov:2002cg, Gaiotto:2008ak}. 

Finally, we can look at $p_\eta$:
\begin{equation}
p_\eta = \frac{\prod_{a,n} (\eta \sigma_n - \mu_a)(\eta/\sigma_n - 1/\mu_a)}{\eta^{N (M-N)} \prod_{n, n'} (\eta^2 \sigma_n - \sigma_{n'})^2}\,.
\end{equation}
It should be possible to rewrite this expression as some sort of discriminant, and compare with the conjectural IR description of the theory. We leave this check for an enthusiastic reader. 

\subsubsection{In full generality}
We can easily write down the effective twisted superpotential for a general $A_{L-1}$ quiver with color and flavor labels $(N_1,M_1),\dots,(N_{L-1},M_{L-1})$:
\begin{align}
\CW &= \sum_{j=1}^{L-2} \sum_{n}^{N_j}\sum_{n'}^{N_{j+1}} \ell\left(s^{(j)}_n - s^{(j+1)}_{n'} + \epsilon/2\right)+\ell\left(-s^{(j)}_n+ s^{(j+1)}_{n'}+\epsilon/2\right) + \notag\\&
\sum_{j=1}^{L-1} \sum_{n}^{N_j}\sum_a^{M_j} \ell\left(s^{(j)}_n - m^{(j)}_a + \epsilon/2\right)+\ell\left(-s^{(j)}_n+ m^{(j)}_a+\epsilon/2\right) + \notag\\&
 \sum_{j=1}^{L-1} \left[\sum_{n,n'}^{N_j} \ell\left(s^{(j)}_n - s^{(j)}_{n'} - \epsilon\right)+  \left(t_{j+1}-t_j + \delta_j \frac{i}{2R}\right) \sum^{N_j}_n s^{(j)}_n+ \sum_{k=1}^j t_k \sum_a^{M_j}   m^{(j)}_a \right] \,,
\label{eq:Efgen}
\end{align}
The corresponding vacuum equations read as follows
\begin{equation}
\frac{\tau_{j+1}}{\tau_{j}}\prod_{n'=1}^{N_{j-1}}\frac{\eta \sigma^{(j)}_n- \sigma^{(j-1)}_{n'}}{\eta \sigma^{(j-1)}_{n'}-\sigma^{(j)}_n} \cdot \prod_{n'\neq n}^{N_{j}}\frac{\eta^{-1} \sigma^{(j)}_n - \eta \sigma^{(j)}_{n'}}{\eta^{-1}\sigma^{(j)}_{n'}-\eta \sigma^{(j)}_n} \cdot \prod_{n'=1}^{N_{j+1}}\frac{\eta \sigma^{(j)}_n - \sigma^{(j+1)}_{n'}}{\eta \sigma^{(j+1)}_{n'}-\sigma^{(j)}_n}\cdot\prod_{a=1}^{M_j}\frac{\eta \sigma^{(j)}_n-\mu^{(j)}_a}{\eta \mu^{(j)}_a-\sigma^{(j)}_n}=(-1)^{\delta_j}\,, 
\label{eq:XXZGen}
\end{equation}
where $j=1,\dots,L-1,\, n=1,\dots, N_j$.
We included the arbitrary sign redefinitions $\delta_j$ of the exponentiated FI parameters. 
Later on, in \secref{sec:higgs} we will fix them in such a way to make sure that the mirror map will simply exchange $\tau$ and $\mu$ parameters, together with $\eta \to - \eta^{-1}$. 

These equations coincide with the equations for an XXZ $SU(L)$ spin chain with $N_i$ Bethe roots at $i$th level of nesting. The spin chain is built out of spins which transform in various antisymmetric powers of the fundamental representation: 
\begin{equation}
\mathcal{R} = \bigoplus_{i=1}^{M_1}\textbf{L} \oplus \bigoplus_{i=1}^{M_2}\Lambda^2\textbf{L} \oplus \dots \oplus \bigoplus_{i=1}^{M_{L-1}}\Lambda^{L-1}\textbf{L}\,,
\label{eq:RepSpinChain}
\end{equation}
i.e. $M_1$ copies of the fundamental representation with impurity parameters $m_a^{(1)}$ , $M_2$ copies of the second antisymmetric power of the fundamental representation
with impurity parameters $m_a^{(2)}$, etcetera.

The weight of a state corresponding to a given set of Bethe roots is such that good quivers correspond to states in the fundamental Weyl chamber. We can act on this state with 
any element of the Weyl group $S_{L}$ of $SU(L)$ and map it to a state for a spin chain with twist parameters appropriately permuted.
Any Weyl group element can be generated by Weyl reflections. Each Weyl reflection acts on the number of Bethe roots as 
$N_i \to M_i + N_{i-1} + N_{i+1} - N_i$. Correspondingly, we can associate states outside the fundamental Weyl chamber to 
the quiver gauge theories which can be reduced to a good quiver by a sequence of 
$N_c \to N_f - N_c$ dualities.  

It is instructive to look at the example of a two-node quiver. Starting from a good quiver with ranks $(N_1,N_2)$, we can consider the RG flows 
\begin{align}
(N_1,N_2) \leftarrow &(N_1, M_2 + N_1 - N_2) \leftarrow (M_1 + M_2 -N_2, M_2 + N_1 - N_2) \leftarrow\cr
&\leftarrow (M_1 + M_2 -N_2, M_1 + M_2 -N_1)\,.
\end{align}
Acting with the dualities in reverse order, we get an alternative path 
\begin{align}
(M_1 + M_2 -N_2, M_1 + M_2 -N_1)& \to \cr
\to (M_1 + N_2 -N_1, M_1 + M_2 -N_1)  \to (M_1 + & N_2 -N_1, N_2) \to (N_1,N_2)\,.
\end{align}
We obtain a Weyl orbit of six theories which flow to the same IR fixed point, up to 
appropriate sets of free twisted hypermultiplets. 

The most familiar spin chain setting involves fundamental spins only and corresponds to a {\it triangular} quiver, with fundamental matter at the first node only. 
We will see later in this section that a generic quiver may be obtained from a triangular quiver by an appropriate Higgsing procedure. 
On the spin chain side, this will correspond to the statement that the space of states for a spin chain with generic anti-symmetric representations can be embedded 
in the space of states of a spin chain with fundamental representations only, by a certain fusion procedure on the Lax matrices. 

This is a very concrete operation on the 
Bethe roots: a spin in the $i$-th antisymmetric representation, with impurity $m$, is represented by a string of auxiliary fundamental spins, with impurities 
\begin{equation}
(m + (i-1) \epsilon/2, m+(i-3)\epsilon/2, \cdots, m-(i-1) \epsilon/2)
\end{equation}
accompanied by a string of auxiliary first level Bethe roots 
\begin{equation} 
(m + (i-2) \epsilon/2, m+(i-3)\epsilon/2, \cdots, m-(i-2) \epsilon/2)
\label{eq:aux}
\end{equation}
etcetera, all the way to a single auxiliary $(i-1)$-th level Bethe root of value $m$.
The contributions of all the auxiliary fugacities telescopes out from the Bethe equations, leaving the correct contribution in the $i$-th equation. We can see immediately that this locus in parameter space is the origin of a Higgs branch in flat space: 
the alignment of mass parameters and Coulomb branch vevs gives zero mass to a collection of chiral multiplets 
inside the fundamental and bi-fundamental hypermultiplets. With help from the brane construction, we will see that these chiral multiplets can get a large vev, which Higgses part of the gauge group and reduces the matter content of the quiver to match 
the final spin chain.  

In the triangular case, we can use some standard integrability lore to deepen our understanding of the moduli space of vacua. 
In particular, there is a beautiful analogue of the $Q-\tilde Q$ equation, which we will briefly sketch here. 
Let's denote the $Q$-function for the $i$-th node of the quiver as $Q_i(\sigma)$.
We can apply generic Weyl transformations on the Bethe roots, and 
consider the $Q$-functions for the last node of quiver in all the possible Weyl group images 
of the original Bethe roots. It is easy to argue that these $Q$-functions are polynomials of degree $N_{L-1}$, 
$N_{i}- N_{i+1}$ and $M_1 - N_1$, and we can denote them respectively $q_L(\sigma)$, $q_{i+1}(\sigma)$, $q_1(\sigma)$.

All other $Q$-functions and the generating function $M(\sigma)$ for the impurity parameters can be written 
as determinants of the basic $q_i$ functions with shifted parameters, and the Weyl group permutes the $q_i$ functions. 
The Bethe equations are equivalent to the constraint that  $M(\sigma)$ is the appropriate determinant of the shifted $q_i$ functions.  
This construction can be extended to the general, non-triangular case simply by imposing the presence of appropriate strings in the $Q$-functions. The ranks of the $q_i$ polynomials become 
\begin{equation}
\mathrm{rank}\, q_i = N_{i-1}- N_{i} + \sum_{j=i}^{L-1} M_j\,.
\end{equation}

\subsubsection{Conjugate momenta}
For completeness, we will write down the explicit form of the conjugate momenta.
The momenta conjugate to FI parameters are
\begin{equation}
p_\tau^j = \frac{\displaystyle\prod\limits_{n=1}^{N_{j-1}} \sigma^{(j-1)}_n}{\displaystyle\prod\limits_{n=1}^{N_j} \sigma^{(j)}_n} \prod_{k \geq j}^{L^\vee-1} \prod_{a=1}^{M_k} \mu_a^{(k)}\,.
\label{eq:ptaugen}
\end{equation}
The momenta conjugate to masses are 
\begin{equation}
p_\mu^{(j),a}= \prod_{k=1}^j \tau_k \cdot  \prod\limits_{n=1}^{N_j} \frac{\eta \mu_a^{(j)} - \sigma_n^{(j)}}{\eta \sigma_n^{(j)} - \mu_a^{(j)}}\,.
\label{eq:pmugen}
\end{equation}
Mirror duals to the above momenta are the following 
\begin{equation}
\tilde p_\tau^j = \frac{\displaystyle\prod\limits_{n=1}^{N^\vee_{j-1}} \tilde\sigma^{(j-1)}_n}{\displaystyle\prod\limits_{n=1}^{N^\vee_j} \tilde \sigma^{(j)}_n} \prod_{k \geq j}^{L-1} \prod_{a=1}^{M^\vee_k} \tau_a^{(k)}\,,
\label{eq:ptaugenMir}
\end{equation}
for dual FI parameters and 
\begin{equation}
\tilde p_\mu^{(j),a}= \prod_{k=1}^j \mu_k \cdot  \prod\limits_{n=1}^{N^\vee_j} \frac{\eta^{-1} \tau_a^{(j)} + \tilde \sigma_n^{(j)}}{\eta^{-1} \tilde\sigma_n^{(j)} + \tau_a^{(j)}}\,,
\label{eq:pmugenMir}
\end{equation}
for dual masses.

\subsection{3d Mirror Symmetry and Bispectral Duality}\label{sec:3dMirrBispec}
Mirror symmetry acts on a three-dimensional SCFT by exchanging the $SU(2)_H$ and $SU(2)_C$ R-symmetry groups, 
and thus the Higgs and Coulomb branches of vacua. If we are given a UV gauge theory description of the SCFT, 
which computes the Higgs branch exactly, we can look for a mirror UV gauge theory description, which computes the Coulomb branch exactly. Clearly, mirror symmetry exchanges the $G_C$ and $G_H$ flavor groups, and the corresponding mass parameters, 
i.e. the masses and FI parameters of the two mirror UV descriptions. The sign of the R-symmetry twisted 
mass parameter $\epsilon$ needs to be flipped as well. 

In what follows we shall discuss XXZ chains which are NS dual to the mirror quivers and then compare them with those originating from the original quivers. Given an $A_{L^\vee-1}$ quiver theory with some color and flavor labels $(N_1, M_1)\,\dots, (N_{L^\vee-1}, M_{L^\vee-1})$ the mirror dual is an $A_L$ quiver theory with some other labels $(N^\vee_1, M^\vee_1),\dots, (N^\vee_{L-1}, M^\vee_{L-1})$ for some integer $L$. An explicit prescription of getting the dual labels $(N^\vee_a, M^\vee_a)$ from the original ones $(N_i, M_i)$ exists and we shall review it later in this section when the brane constructions for 3d quiver theories will be discussed. As of now let us assume that a dual quiver with some labels is already constructed and study how the moduli space of vacua of the two mirror quivers are related to each other.

As the $T[U(2)]$ example illustrates, the vacua of the two mirror theories are in bijection, as long as we map $\epsilon \to - \epsilon$ and exchange masses and FI parameters. More precisely, and perhaps surprisingly, we need to map $\eta \to -\eta^{-1}$ to match the mirror theories.\footnote{We could trade such extra sign for extra signs scattered through the $\mu$, $\sigma$, $\tau$ variables. 
More precisely, we can change the sign of $\mu_a$ for even $a$, $\sigma_n^{(a)}$ for odd $a$, and for appropriate $\tau_i$} If we include appropriate background terms in the definition of the twisted superpotential, the super potentials in corresponding vacua coincide. In particular, we can relate the conjugate momenta $p_\eta$, $p_\mu$, $p_\tau$ for the moduli space of vacua of the two theories. 

The mirror symmetry statement clearly implies a bijection between the solutions of Bethe equations for two distinct 
XXZ spin chains, with appropriate representation content, under the exchange of twist and impurity parameters, 
and $\eta \to -\eta^{-1}$. Furthermore, as the gauge theory twisted superpotential coincides with the Yang-Yang functional 
of the spin chain, we learn that the bijection relates solutions of Bethe equations with the same value of the Yang-Yang functional, and of whatever spin chain observables we can relate to  $p_\eta$, $p_\mu$, $p_\tau$.

In the literature on integrable systems a statement which relates solutions of one integrable system with the solutions of the other such that all states are in one-to-one correspondence is often called a \textit{bispectral duality}. Using the NS limit applied to our construction we can conclude that
\begin{center}
\framebox{Bispectral duality in XXZ spin chains = Mirror symmetry in 3d gauge theories}
\end{center}
The general mirror symmetry relation for $\ssN=4$ linear quivers of unitary gauge groups enlarges and enriches the known bispectral duality statements for the XXZ spin chains. In particular, gauge theory provides a complete answer to the question: what is the bispectral dual system to an $SU(L^\vee)$  XXZ spin chain with spins in the representation $\mathcal{R}$ \footnote{Note that we have changed $L$ in (\ref{eq:RepSpinChain}) to $L^\vee$ here to adopt the notations for the mirror symmetry.} \eqref{eq:RepSpinChain}, in a sector of specific weight $w$? As the recent literature on integrable system suggests (see e.g. \cite{MR2409414}) the question appears to be quite nontrivial for high rank. 

Since bispectrally dual spin chains are associated to mirror dual quivers, the duality maps a sector of a $SU(L^\vee)$ XXZ chain on $L$ sites given by the quiver labels $(N_i,M_i)$ to a $SU(L)$ XXZ chain on $L^\vee$ sites with dual labels $(N^\vee_j, M^\vee_j)$. 
In order to compute the dual quiver labels, it is useful to express the data of $\mathcal{R}$ and $w$ as two sets of {\it linking numbers}, which are simply exchanged by mirror symmetry/bispectral duality. The first set of linking numbers $r_a$ encodes $\mathcal{R}$: $r_a=L^\vee-i$ if the $a$-th spin belongs to the $i$-th antisymmetric power of the fundamental representation. We order the $r_a$ so that they are non-decreasing. 

The second set of linking numbers $r^\vee_i$ encode the weight $w$. In terms of the number of Bethe roots, we have 
\begin{equation}
r_1^\vee = N_1\,, \qquad r_{i+1}^\vee - r_{i}^\vee = M_i + N_{i-1} + N_{i+1} - 2 N_i=\Delta_i\,.
\label{eq:linkingnumbd}
\end{equation}
where $\Delta_i$ was introduced in \eqref{eq:deltaqeight}. Thus these linking numbers are positive and non-decreasing for a good quiver, i.e. for a weight in the fundamental Weyl chamber. 

In the next subsection we shall review the universal construction of a mirror quiver using a type IIB brane description. Before plunging into the details of the construction let us make a possibly useful observation about the bispectral duality.

We have expressed bispectral duality as a relation between sectors of a given weight in a spin chain with a given 
representation content. There is an alternative perspective which we find very suggestive. 
Remember that the Fock space of $L^\vee$ fermionic operators is the direct sum of all antisymmetric powers of the 
$SU(L^\vee)$ fundamental representation:
\begin{equation}
{\cal F}_{L^\vee} = \oplus_n \Lambda^n\textbf{L}^\vee\,.
\end{equation}
We can thus define an $SU(L^\vee)$ XXZ spin chain with $L$ spins valued in the Fock space ${\cal F}_{L^\vee}$, 
simply by using the appropriate spin chain Hamiltonian for each irreducible $SU(L^\vee)$ representation in the Fock space.
Clearly, the space of states of this ${\cal F}^{\otimes L}_{L^\vee}$ spin chain includes once all the possible 
choice of $M_i$ labels for a given total number of sites $L$. In other words, the state of spaces includes 
all the possible compatible pairs of linking numbers $r_a$ and $r_i^\vee$. 
Each of the ${\cal F}_{L^\vee}$ Fock spaces carries an additional $U(1)$ fermion number action. The linking numbers $r_a$ 
are the charges under the corresponding $U(1)^L$ symmetries of ${\cal F}^{\otimes L}_{L^\vee}$.

This makes bispectral duality into a bijection between the eigenstates of the ${\cal F}^{\otimes L}_{L^\vee}$ $SU(L^\vee)$ spin chain and the 
${\cal F}^{\otimes L^\vee}_{L}$ $SU(L)$ spin chain. These Fock spaces are identical: they are the Fock space ${\cal F}_{L L^\vee}$ of $L L^\vee$ fermionic operators. 
The constraint on linking numbers tell us that bispectral duality relates sectors in the two spin chains with the same weights under the Cartan
generators of the $SU(L^\vee) \times SU(L)$ symmetries of ${\cal F}_{L L^\vee}$.

It is natural to speculate that bispectral duality could be given an economical proof by realizing the two spin chain Hamiltonians as commuting operators on the same ${\cal F}_{L L^\vee}$ Fock space. It would be interesting to pursue this point further. 

\subsection{Brane Description of Quiver Theories and Mirror Symmetry}\label{sec:BraneMirror}
\begin{figure}
\begin{center}
\includegraphics[height=5cm, width=7.5cm]{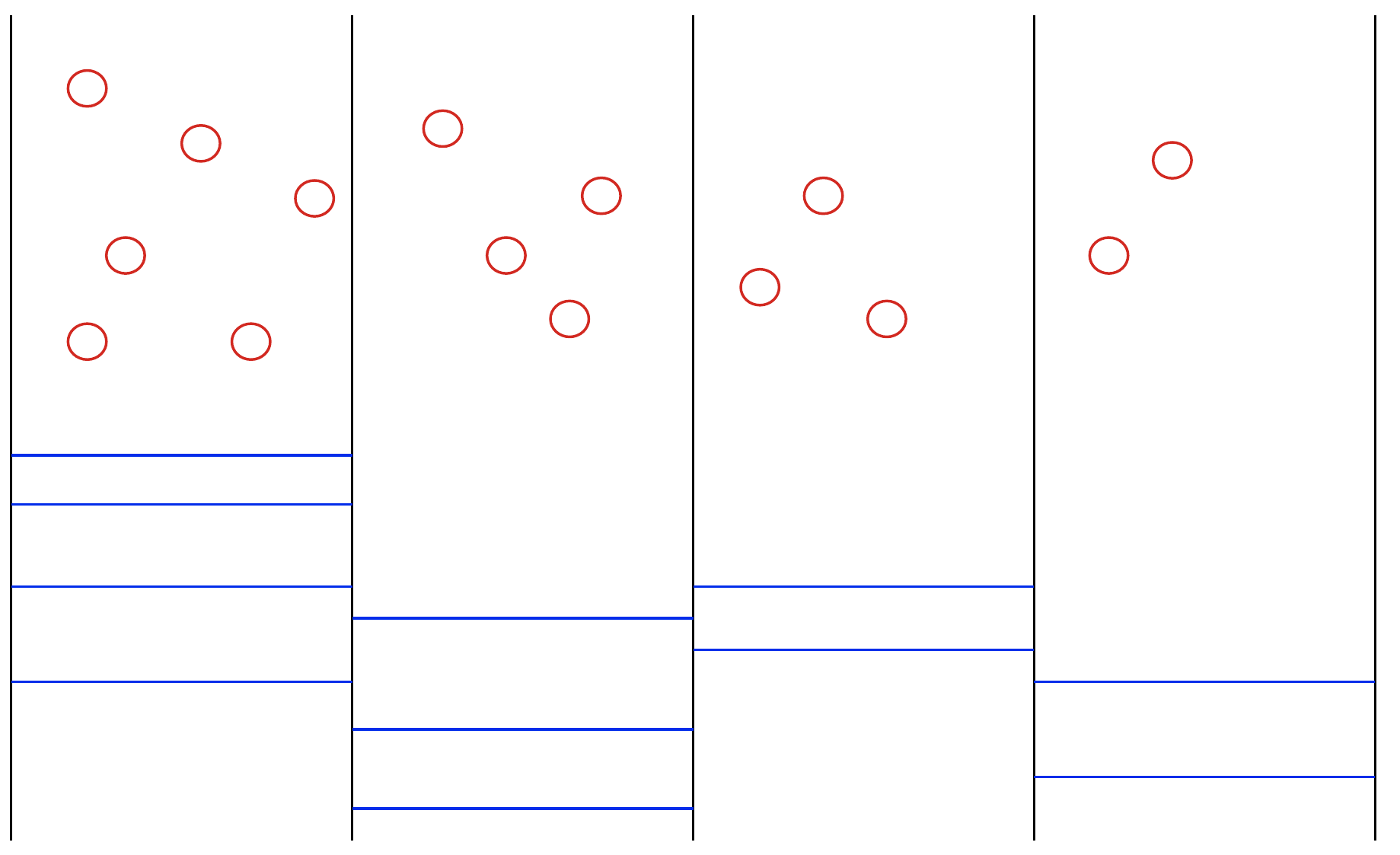}
\caption{Brane realization of $A_4$ quiver with $(N_i,M_i)$ labels $(4,6),(3,4),(2,3),(2,2)$. Here and further on in the paper red ovals denote D5 branes, horizontal blue lines show D3 branes and vertical black lines designate NS5 branes.}
\label{fig:branequiv}
\end{center}
\end{figure}

Many three dimensional $\ssN=4$ supersymmetric quiver gauge theories can be described by type IIB brane constructions of Hanany-Witten type \cite{Hanany:1996ie}. The construction involves D3 branes which are stretched between NS5 and D5 branes as shown in an example in \figref{fig:branequiv}, 
and the mutual orientation of the branes is shown in the table below. 
\begin{center}
\begin{tabular}{|c|c|c|c|c|c|c|c|c|c|c|}
\hline
         & 0 & 1 &  2 & 3 & 4 & 5 & 6 & 7 & 8 & 9 \\
\hline \hline
NS5 & x &  x &  x &   &   &  &    &  x     &  x & x  \\
\hline
D5    & x &  x & x  &   & x  & x  & x &    &     &    \\
\hline
D3    & x &  x & x & x &    &     &    &   &      &    \\
\hline
\end{tabular}\label{tab:HWbranes}
\end{center}
A specific quiver gauge theory is engineered in a simple way \cite{Gaiotto:2008ak}. We start with $L^\vee$ NS5 branes, and stretch $N_i$ D3 branes in the interval between the $i$-th and $(i+1)$-th NS5 brane. We also add groups of $M_i$ D5 branes in each interval. In a field theory limit, which squeezes the picture in the $456789$ direction, the world-volume theory on the D3 branes in a $U(N_i)$ $\ssN=4$ four-dimensional SYM gauge theory. The fivebranes give rise to codimension one defects in the four-dimensional gauge theory. The NS5 branes become an interface between the $U(N_{i-1})$ and $U(N_i)$ gauge theories, where both gauge groups have Neumann boundary conditions and are coupled to a set of 3d bifundamental hypermultiplets. The D5 branes become a simple 3d defect: a set of 3d fundamental hypermultiplets coupled to the 4d gauge group. At energy scales below the scale set by the interval lengths, the four-dimensional gauge theories 
reduce further to three-dimensional $U(N_i)$ $\ssN=4$ SYM gauge theories, coupled to the same three-dimensional hypermultiplet content. 
We thus recover the desired quiver. 

The precise positions of the branes along the $x^3$ direction does not affect the infrared physics, but the relative order of the fivebranes is important. A D5 brane and a NS5 brane can be transported across each other 
only if the number of D3 branes in the system is changed appropriately, in order to keep constant the {\it linking number} of each fivebrane. The linking number $r_i^\vee$ for an NS5 brane can be defined as 
the number of D3 branes ending from the left, plus the number of D5 branes on the right of it, minus the number of D3 branes emerging from the right
\begin{equation}
r_i^\vee=\# D3(L)-\# D3(R)+\# D5(R)\,.
\label{eq:linkNS5}
\end{equation}
The linking number $r_i$ for a D5 branes is defined in the same way, but it is useful to switch left and right in the definition
\begin{equation}
r_i=\# D3(R)-\# D3(L)+\# NS5(L)\,.
\end{equation}
With this definition of linking number, the sum over all NS5 linking numbers equals the sum over all D5 linking numbers. 
The linking number of a D5 brane in the $i$-th interval in a quiver gauge theory setup is $i$. 
The linking numbers of the NS5 branes are given by \eqref{eq:linkingnumbd}
\begin{equation}
r_{i}^\vee-r_{i+1}^\vee = M_i + N_{i-1} + N_{i+1} - 2 N_i=\Delta_i\,, 
\label{eq:weights}
\end{equation}
and are thus positive and increasing to the left from $r_{L^\vee+1} =N_{L^\vee}$ if the quiver is good. The boundary conditions for linear quivers assume $N_0=M_0=N_{L^\vee+1}=M_{L^\vee+1}=0$. 

Notice that if we start from a quiver setup and we move the D5 branes around, we will generically arrive to configurations where a net number of D3 branes end on a D5 brane. These configurations to not have an immediate UV complete three-dimensional field-theoretic interpretation. They have a four-dimensional gauge theory interpretation, involving certain interfaces which we will review in a later section. These four-dimensional configurations flow in the infrared to the same 3d SCFT as the original quiver, but will allow us 
to give a neat geometric interpretation of the moduli space of vacua. 

Mirror symmetry for the quiver gauge theory involves an S-duality transformation on the brane system \cite{Gaiotto:2008ak}, which 
transforms D5 branes into NS5 branes and vice versa. It is also useful to do a reflection on the $x^3$ coordinate 
in order to preserve the definition of linking number \footnote{In the context of the bispectral duality brane setup was first effectively used in \cite{Gorsky:1997jq, Gorsky:1997mw}. The authors employed a 90 degrees turn of the construction to visualize the duality.}. After an S-duality transformation, we can attempt to reorder the fivebranes to arrive to a mirror quiver configuration. This is always possible if the original quiver was good, and gives us the mirror data $N^\vee_i$, $M^\vee_i$. Thus for the mirror quiver one has
\begin{equation}
r_{i} - r_{i+1} = M^\vee_i + N^\vee_{i-1} + N^\vee_{i+1} - 2 N^\vee_i=\Delta_i^\vee\,,
\label{eq:dualweights}
\end{equation}
which can be used to find color and flavor labels of the dual theory. 

\subsection{Examples of the Bispectral Duality}\label{sec:BispecExamples}
At this point it is instructive to give several examples.

\begin{enumerate}
\item
Let us start with a rather well known example due to Intriligator and Seiberg \cite{Intriligator:1996ex}. The original theory is the SQED with $M$ flavors, i.e. $A_1$ quiver with labels $(1,M)$. The mirror dual is $A_{M-1}$ quiver gauge theory with the following labels $(1,1),(1,0),\dots,(1,0),(1,1)$, so there are two flavors, one at the first node, one on the last node. 
The D5 and NS5 linking numbers for the SQED are $r_1=\dots=r_M=1$ and $r^\vee_2=1,r^\vee_1=M-1$, whereas for the $A_{M-1}$ quiver they are the opposite, so the two theories are indeed mirror to each other. In the spin chain language the SQED gives a sector of an $\alg{su}(2)$ spin chain on $M$ sites with a single Bethe root, whereas the dual quiver represents a sector of an $\alg{su}(M)$ chain with one fundamental and one anti fundamental spins. 

\item
Then let us look at $T[U(2)]$ theory. The linking numbers are $r_1 = r_2 = 1$ and $r^\vee_1 = 1$, $r^\vee_2 =1$. Hence the theory is self-mirror. It is useful to consider a more general self-mirror theory: $T[U(L)]$, with $r_a = r_i^\vee =1$ and $L^\vee = L$. The corresponding quiver has all fundamental flavors at the first node. The $L-1$ nodes of the quiver have rank $N_i = L-i$. Thus, the $T[U(L)]$ theory is an example of self-mirror theory; it will play an important role in the next sections. 

\item
We now address an $A_1$ quiver with generic color and flavor labels $(N,M)$.  We have $L=M\,,L^\vee =2$ thus all $r_i=1$. Then we get $r_1^\vee=M-N$ and $r_2^\vee=N$. Dually, we find the color labels of the dual $A_{M-1}$ quiver \begin{equation}
N^\vee_i= 
\begin{cases} 
i\quad & 0 <i<N+1\,, \\
N\quad & N<i<M-N-1\,, \\
M-i \quad & M-N-1<i<M\,.
\end{cases}
\end{equation}

\item
For a generic good quiver one may iteratively solve \eqref{eq:weights, eq:dualweights} in order to extract $(N^\vee_j,M^\vee_j)$ out of $(N_i,M_j)$. We refrain from giving the complete set of formulae here, instead we provide a dual quiver to the one given by the brane construction in \figref{fig:branequiv}. The dual labels of this $A_{14}$ quiver are the following
\begin{equation}
(1,0)\,(2,2)\,(2,0)\,(2,0)\,(2,0)\,(3,1)\,(3,0)\,(3,0)\,(3,0)\,(4,1)\,(4,1)\,(3,0)\,(2,0)\,(1,0)\,.
\end{equation}
\end{enumerate}

\subsection{Higgsing a Quiver}
\label{sec:higgs}
There is an intricate network of RG flows relating different quiver gauge theories. These RG flows are initiated by giving a vev to some Higgs branch or Coulomb branch operators. 
The Higgsing procedure has a simple pictorial interpretation in the brane system. The field theoretic description 
is necessary to understand fully Higgsing in the $\ssN=2^*$ context, as the $\ssN=2^*$ mass deformation is not readily 
turned on in the brane setup itself. Mirror symmetry also helps understand the precise adjustment of parameters needed for Higgsing in the Coulomb branch. 

In order to do a Higgsing in the brane setup, we simply adjust the parameters so that two NS5 branes are at the same position in the $456$ directions, or two D5 branes are at the same position in the $789$ directions. The transverse positions of the fivebranes correspond to the FI parameters $t_i$ and the masses $m_a$ respectively. Once the fivebranes are aligned, we can pick a D3 brane segment stretched between them, and bring it to infinity. This changes the linking numbers of the two branes by $1$ unit, increasing the largest and decreasing the smallest. 

If we align NS5 branes and move a single D3 segment, we are turning on vevs for adjoint Coulomb branch scalar fields 
and freezing some hypermultiplets. The brane picture shows that the rank of the gauge groups in between the NS5 branes goes down by $1$. The matter content adjusts accordingly, leaving the $M_i$ parameters invariant.  

\begin{figure}
\begin{center}
\includegraphics[height=2.5cm, width=6cm]{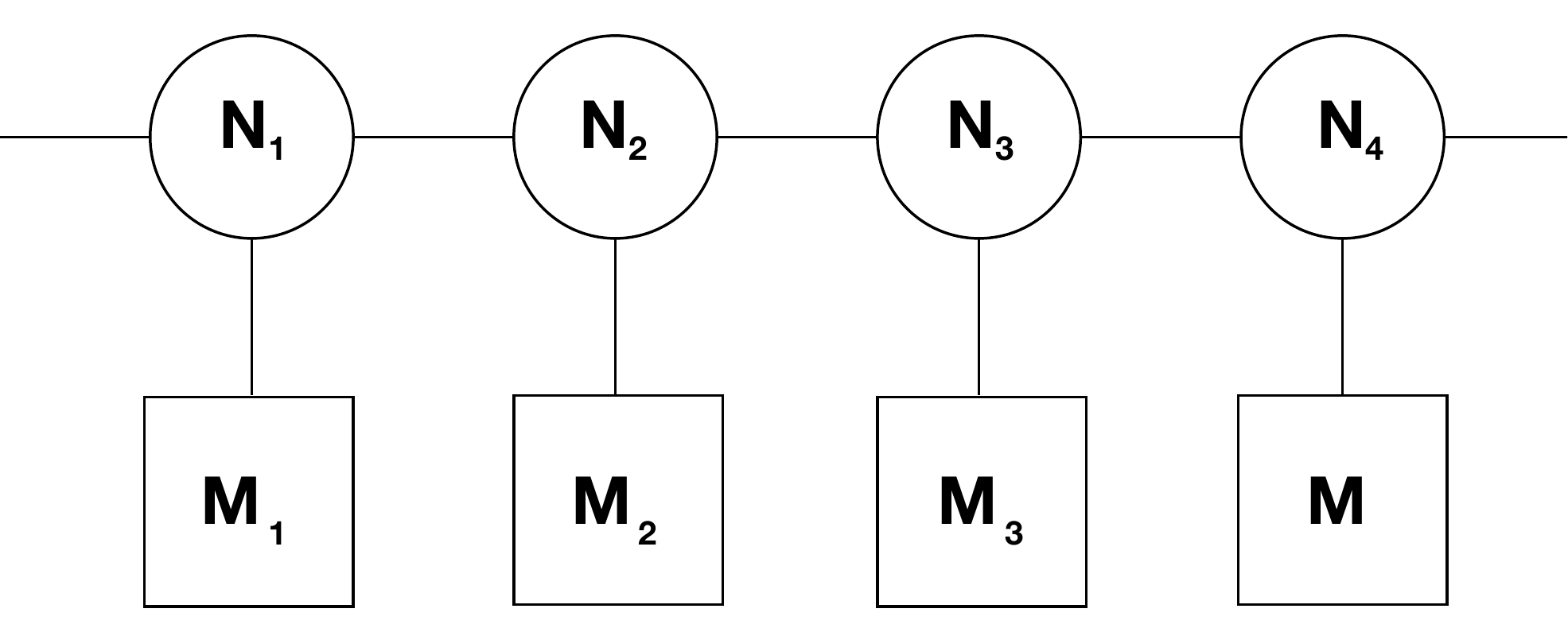}\qquad\includegraphics[height=2.4cm, width=6.2cm]{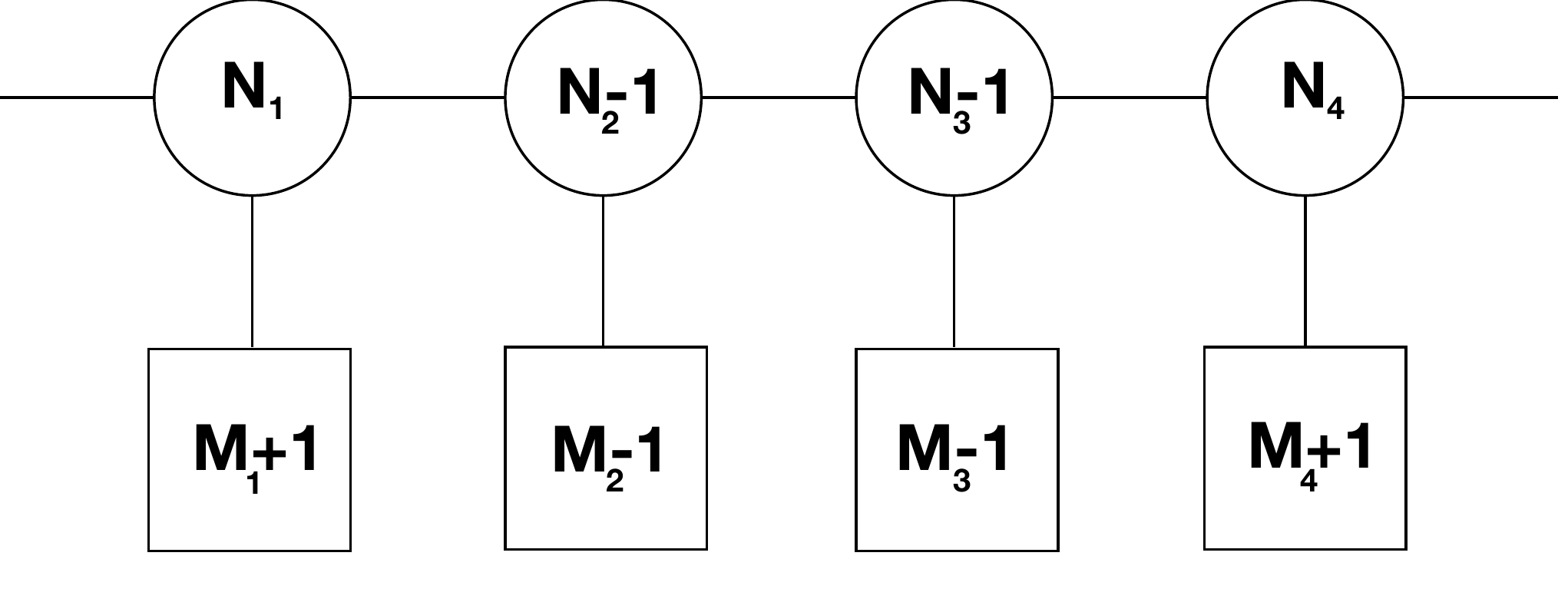}
\caption{The transformation of a portion of a linear quiver induced by Higgsing along the D5 branes at positions $2$ and $3$.}
\label{fig:quiverfrag}
\end{center}
\end{figure}

\begin{figure}
\begin{center}
\includegraphics[height=4cm, width=6cm]{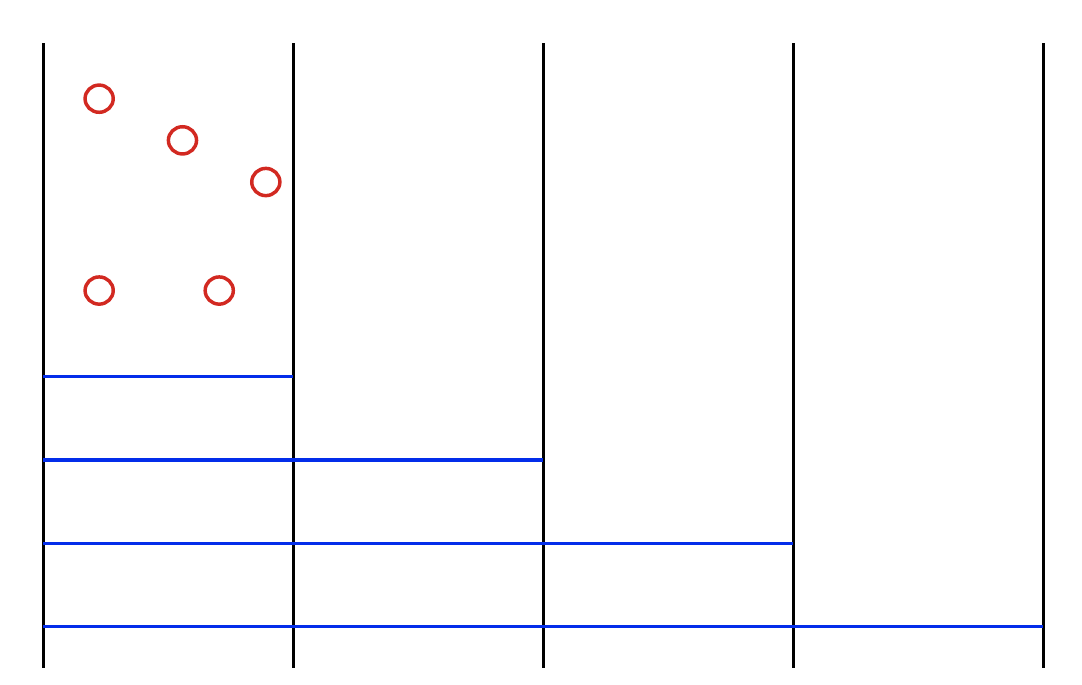}\qquad\includegraphics[height=4cm, width=6cm]{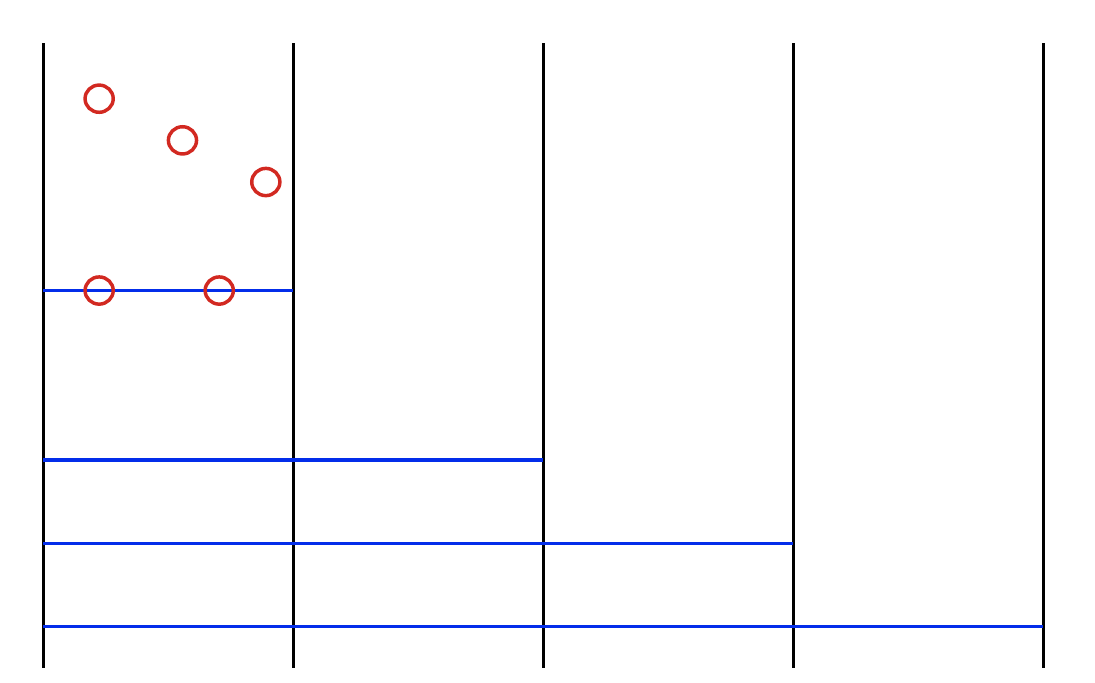}\\
\includegraphics[height=4cm, width=6cm]{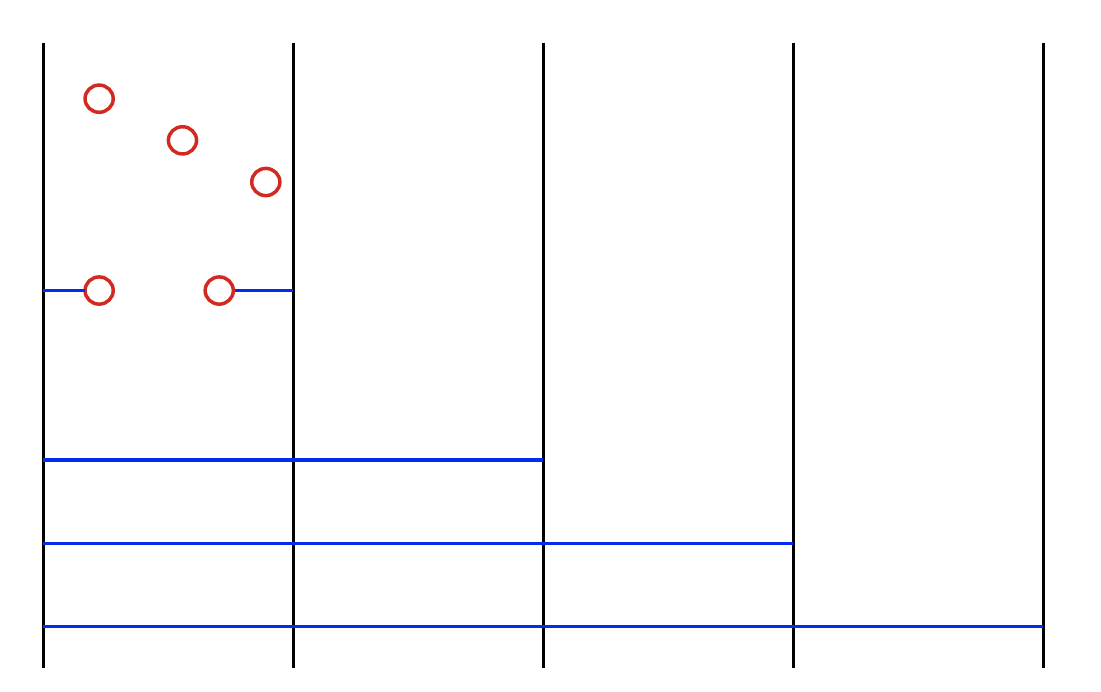}\qquad\includegraphics[height=4cm, width=6cm]{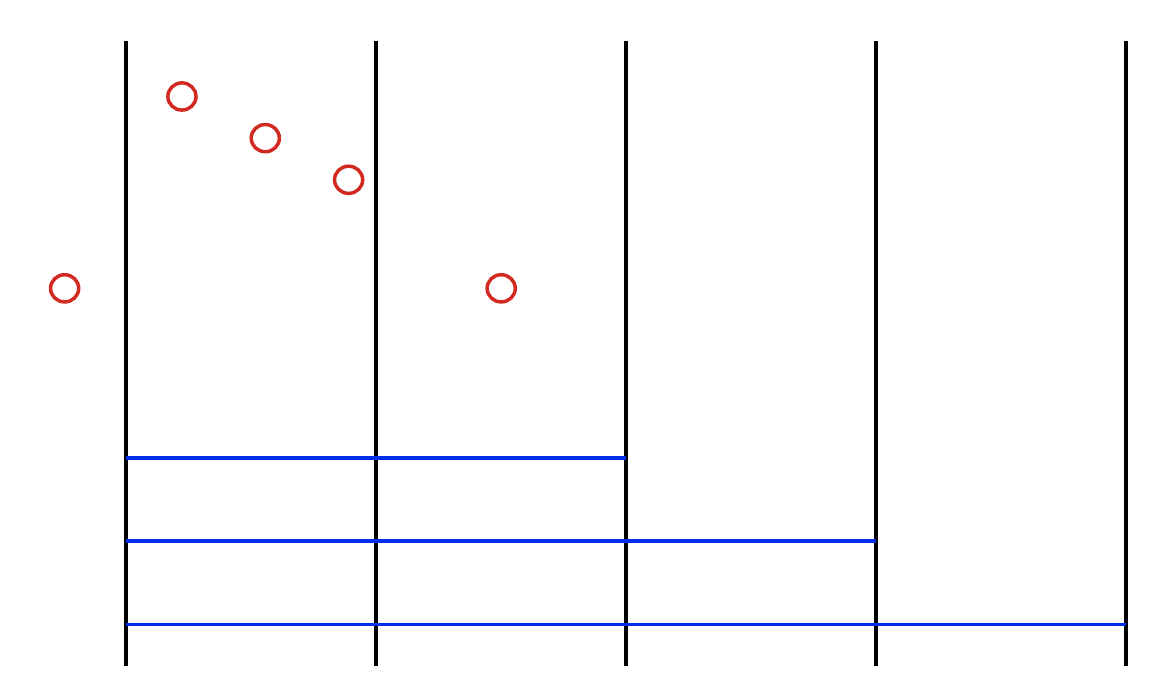}
\caption{The sequence of Hanany-Witten moves on $T[U(5)]$, which Higgses the quiver gauge theory as in figure \ref{fig:TSUNHWmoves1}.}
\label{fig:TSUNHWmoves}
\end{center}
\end{figure}

If we align D5 branes and move a single D3 segment, we are turning on a vev for a Higgs branch operator, 
which is a gauge-invariant monomial in fundamental and bi-fundamental fields which follows a path along the quiver, starting and ending with the (anti)fundamental flavors associated to the two D5 branes. The vev will break some gauge symmetry, and Higgs the gauge group to a subgroup. 
A few Hanany-Witten moves allow us to read off the final gauge group. Removing the $D3$ segment
leaves us with the rest of the D3 brane hanging on the $D5$ brane. We need to move the D5s outwards across one NS5 brane each in order to go back to a quiver setup. Thus the process moves two flavors, say at the $i$ and $j$-th nodes with $i\leq j$, outwards to the $i-1$ and $j+1$ nodes, and lowers by $1$ the rank of the gauge groups between the $i$ and $j$ nodes (included). \figref{fig:quiverfrag} exhibits this phenomenon for $j=i+1$.
\figref{fig:TSUNHWmoves} illustrates the idea of the Higgsing procedure we described above for the brane construction of $T[U(5)]$ theory.
\begin{figure}
\begin{center}
\includegraphics[height=2.5cm, width=5.7cm]{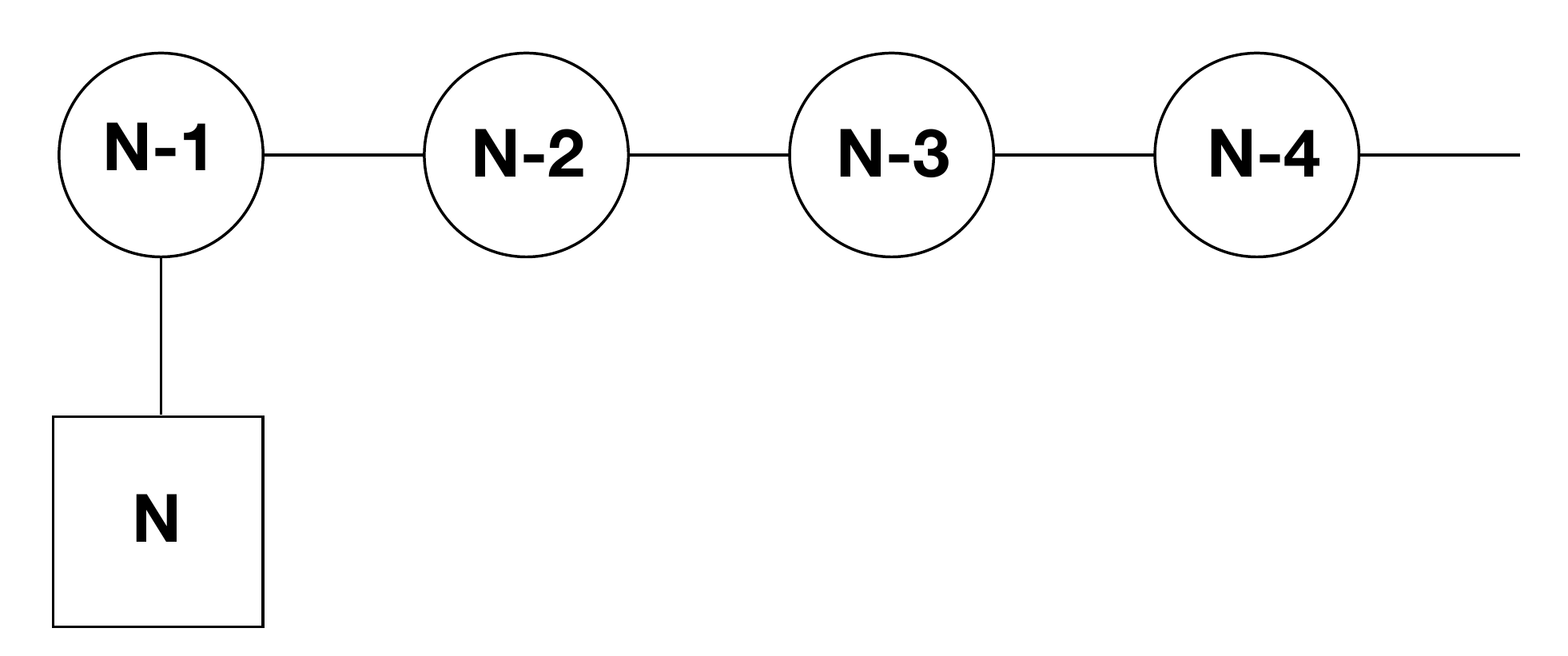}\qquad\includegraphics[height=2.5cm, width=5.7cm]{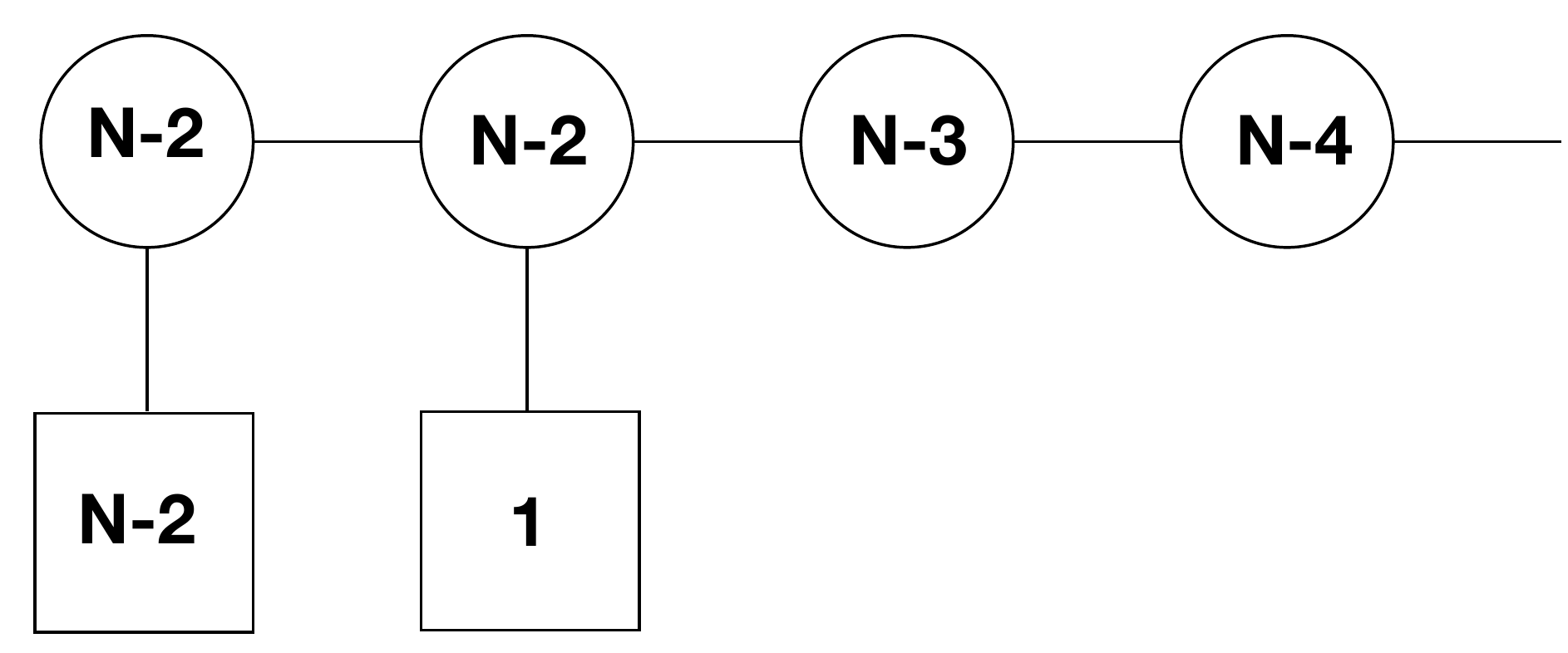}
\caption{Transformation of the $T[U(N)]$ quiver corresponding to the brane manipulations in Fig. \ref{fig:TSUNHWmoves}.}
\label{fig:TSUNHWmoves1}
\end{center}
\end{figure}

The corresponding transformation of the quiver data is shown in \figref{fig:TSUNHWmoves1}. The resulting theory after Higgsings and Hanany-Witten moves will have non-generic FI parameters or masses, as the fivebranes are still aligned. The only exception are cases where one of the aligned fivebranes ends up disconnected from the rest of the theory. For example, if we apply the transformation to a D5 brane at the first node of the quiver, it will end up to the left of the whole system, and we will have traded a flavor at the $i$-th node and a flavor at the first node for a flavor at the $(i+1)$-th node only, as in \figref{fig:quiverfrag}. If we apply the transformation repeatedly on a group of $n+1$ D5 branes at the first node, we can end up with a single flavor at the $n$-th node. 

In the $\ssN=2^*$ context, in order to move along the Higgs branch, we need to make all the chiral multiplets involved in the process massless. In order to cancel the $\epsilon/2$ contribution to the hypermultiplet mass, the sequence 
\begin{equation}
\left( m^{(i)}_a, \sigma^{(i)}_1,  \sigma^{(i+1)}_1, \cdots,  \sigma^{(j)}_1, m^{(j)}_{a'} \right)
\end{equation}
of masses and gauge-multiplet scalar vevs must align itself with precise spacing $\epsilon/2$ as we have already mentioned in \eqref{eq:aux}. In particular, the Higgs branch 
may open up if and only if the masses for the two flavors involved in the transformation differ by $\frac{j-i+2}{2} \epsilon$. 
This alignment leads to exact cancellations in the Bethe equations, which precisely follow the expected rearrangement of the quiver. 

Let us demonstrate Higgsing for the $T[U(N)]$ theory using Bethe equations. One starts with
\begin{equation}
\frac{\tau_{j+1}}{\tau_{j}}\prod_{i=1}^{N-j+1}\frac{\eta \sigma^{(j)}_s- \sigma^{(j-1)}_i}{\eta \sigma^{(j-1)}_i-\sigma^{(j)}_s} \cdot \prod_{k\neq s}^{N-j}\frac{\eta^{-1} \sigma^{(j)}_s - \eta \sigma^{(j)}_k}{\eta^{-1}\sigma^{(j)}_k-\eta \sigma^{(j)}_s} \cdot \prod_{k=1}^{N-j-1}\frac{\eta \sigma^{(j)}_s - \sigma^{(j+1)}_k}{\eta \sigma^{(j+1)}_k-\sigma^{(j)}_s} = (-1)^{\delta_j}\,, 
\label{eq:XXZtauGen}
\end{equation}
where $j=1,\dots, N-1$. We denoted $\sigma^{(0)}_i=\mu_i$. The move described in \figref{fig:TSUNHWmoves} corresponds to
\begin{equation}
\sigma^{(1)}_1=\eta^{-1}\mu_1=\eta\mu_2=\tilde{\mu}\,,
\label{eq:Higgsrho1}
\end{equation}
which opens up a Higgs branch in the theory and enables the Hanany-Witten phenomena to occur. As \figref{fig:TSUNHWmoves} shows, one D5 brane moves to the right of the second NS5 brane, therefore the corresponding hypermultiplet is now coupled to the next gauge group. This transformation of the theory corresponds to the change of the representation $\CR$ \eqref{eq:RepSpinChain} on the XXZ side: we traded two fundamental spins for a single spin transforming in a two-index antisymmetric representation. Indeed, for the $T[U(N)]$ theory the first set of Bethe equations \eqref{eq:XXZtauGen} can be written as 
\begin{equation}
\frac{\tau_{2}}{\tau_{1}}\prod_{k\neq s}^{N-1}\frac{\eta^{-1} \sigma^{(1)}_s - \eta \sigma^{(1)}_k}{\eta^{-1}\sigma^{(1)}_k-\eta \sigma^{(1)}_s} \prod_{k=1}^{N-2}\frac{\eta \sigma^{(1)}_i - \sigma^{(2)}_k}{\eta \sigma^{(2)}_k-\sigma^{(1)}_i} = (-1)^{\delta_1+1} \frac{ \eta^{-1}\sigma^{(1)}_s-\eta \sigma^{(1)}_1}{\eta^{-1}\sigma^{(1)}_1-\eta\sigma^{(1)}_s}\prod_{a=3}^{N}\frac{ \sigma^{(1)}_s-\eta \mu_a}{\mu_a-\eta \sigma^{(1)}_s}\,, 
\label{eq:XXZtauGen1}
\end{equation}
and the other $N-2$ equations which include higher nested Bethe roots have the form \eqref{eq:XXZtauGen}. We see from \eqref{eq:Higgsrho1} that the first ratio in the r.h.s. of the above equation cancels with one of the rations in the first product in the l.h.s. (for $k=1$). Also, the equation for second nesting level will look like \eqref{eq:XXZGen}, with $M_2=1$:
\begin{equation}
\frac{\tau_{3}}{\tau_{2}}\prod_{k=2}^{N-1}\frac{\eta \sigma^{(2)}_i - \sigma^{(1)}_k}{\eta \sigma^{(1)}_k-\sigma^{(2)}_i}
\prod_{k\neq s}^{N-2}\frac{\eta^{-1} \sigma^{(2)}_s - \eta \sigma^{(2)}_k}{\eta^{-1}\sigma^{(2)}_k-\eta \sigma^{(2)}_s} \prod_{k=1}^{N-3}\frac{\eta \sigma^{(2)}_i - \sigma^{(3)}_k}{\eta \sigma^{(3)}_k-\sigma^{(2)}_i} = (-1)^{\delta_2}\frac{ \sigma^{(2)}_s-\eta \tilde{\mu}}{\tilde{\mu}-\eta \sigma^{(2)}_s}\,. 
\label{eq:XXZtauGen2}
\end{equation}
Thus we reproduce the Bethe equations expected from the new quiver, as long as our sign redefinition $\delta_1$ changes by one unit when $N_1$ decreases by $1$. We will see momentarily that this is indeed the case with our chosen definition for $\delta_j$. 

\begin{figure}
\begin{center}
\includegraphics[height=6.5cm, width=16.5cm]{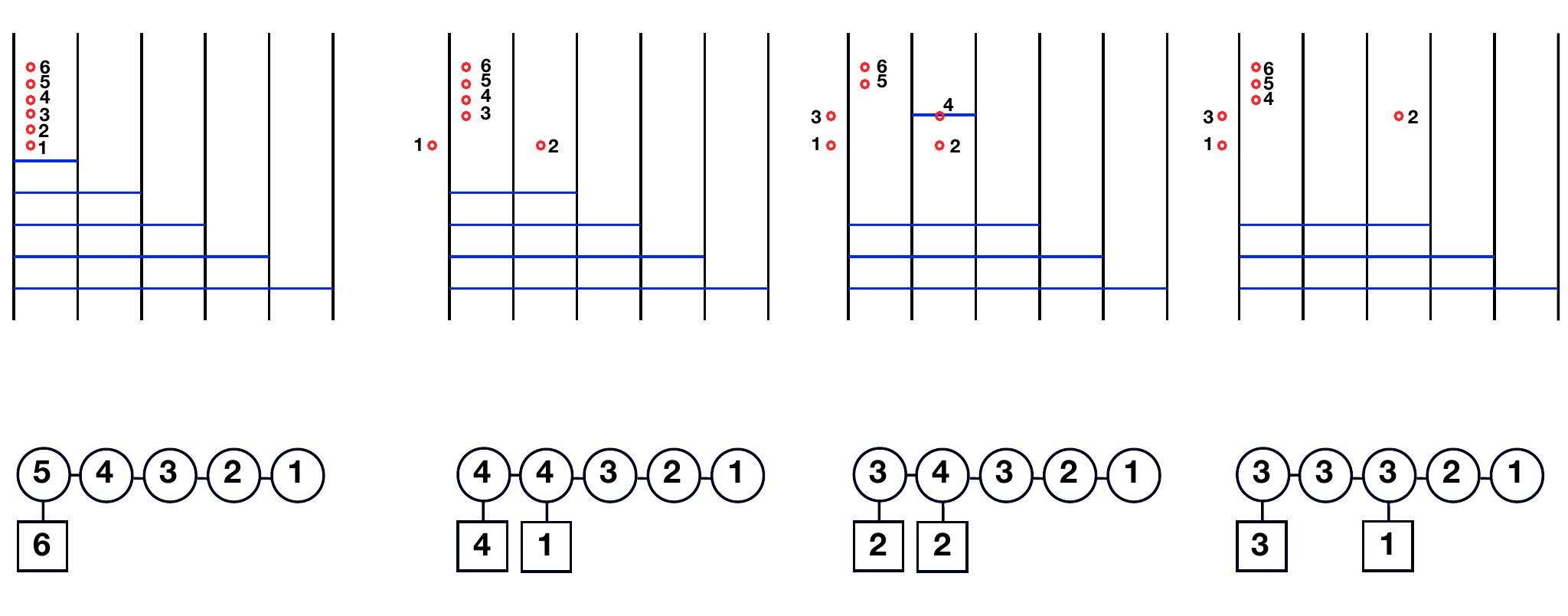}
\caption{Several quiver theories originating from $T[U(6)]$ (left). The quiver transformations in this figure are accomplished by moves similar to the one described in Fig.\ref{fig:TSUNHWmoves}. According to the Nekrasov-Shatashvili duality, quiver diagrams in the third row of this figure provide us with the data for the Bethe ansatz equations of the corresponding XXZ spin chains.}
\label{fig:HWmovesYoung}
\end{center}
\end{figure}

One can analyze other possible deformations of $T[U(N)]$ theories by iteratively applying Higgsings. Some examples of such quivers together with the corresponding representations $\CR$ are shown in \figref{fig:HWmovesYoung}. 

Interestingly it appears that the Higgsing transformation in the quiver gauge theory corresponds to the fusion of R-matrices in the XXZ spin chain: if the impurity parameters of two spins in representations $R_1$ and $R_2$ are aligned appropriately, the Hilbert space of the spin chain and the spin chain Hamiltonian can often be projected to the Hilbert space and Hamiltonian for a sub-representation in the product 
$R_1 \otimes R_2$. The canonical example is the product of two fundamental spins with impurities $m\pm \epsilon/2$, which can be projected to the second antisymmetric power with impurity $m$. From the point of view of the Bethe equations, the projection 
focuses on solutions of the Bethe equations which include the appropriate string of Bethe roots we see in the gauge theory 
description. 

Notice that appropriate patterns of shifts in the impurity parameters and projection operations allow one to build more general 
representations in the spin chain, such as symmetric powers of the fundamental. Spin chains with such general representations have been proposed as NS duals of $\ssN=2$ gauge theories with a similar matter content as the $\ssN=4$ quivers, but more general superpotential couplings. In some simple examples we have looked at, the alignment of mass parameters allows one to add 
extra superpotential terms to the $\ssN=4$ Lagrangian, such as complex mass terms linear in the mesons. Integrating away 
such terms may reproduce the conjectural $\ssN=2$ duals of XXZ spin chains with more general representation content. 
It would be interesting to pursue this direction of inquiry, but for the remainder of this paper we will restrict ourselves to $\ssN=2^*$ theories with no additional super-potentials. 

\subsubsection{Fixing the signs}
In the remainder of this section, we will use Higgsing considerations to fix the 
sign ambiguities in the mirror transformation. We know that Higgsing is possible in the Higgs branch if 
two mass parameters $\mu_a$ and $\mu_b$ at nodes $i$ and $i+k$ differ by an $\eta^{2 + k}$ factor, 
say $\mu_b = \eta^{2 + k} \mu_a$, 
and happens if we tune the Coulomb branch parameters as
\begin{equation}
\sigma^{(i)}_1 = \eta \mu_a \qquad \cdots \qquad \sigma^{(i+k)}_1 = \eta^{k+1} \mu_a
\end{equation}
We get telescopic cancellations in the Bethe equations as we approach these values, 
which give the expected Bethe equations for the new quiver up for an extra factor of $(-1)$ 
at all nodes from $i$ to $i+k$. This indicates that our sign redefinitions $\delta_j$ should shift accordingly. 
Notice also that the two new flavor parameters in the new quiver will be $\eta \mu_a$ and $\eta^{-1} \mu_b$.

On the other hand, we can study the Higgsing in the Coulomb branch by sending, say, $\sigma^{(j)}_1$ to $\infty$. \footnote{We could also send it to $0$. This would be mirror to $\mu_b = \eta^{-2 - k} \mu_a$ in the previous example}
This limit is sensible only if $\tau_{j+1}/\tau_j$ has a very specific value. Indeed, Bethe equations \eqref{eq:XXZGen} at the $j$-th node for $n=1$ give 
\begin{equation}
\frac{\tau_{j+1}}{\tau_{j}}(-\eta)^{2+\Delta_{j}}(-1)^{N_j+1} =(-1)^{\delta_j}\,, 
\end{equation}
or, using \eqref{eq:linkingnumbd} we arrive at
\begin{equation}
\tau_{j+1} = (-\eta^{-1})^{2 + r^\vee_{j+1}-r^\vee_{j}} (-1)^{\delta_j -N_j-1} \tau_j.
\end{equation}
We conclude that in order for the mirror map to take the simplest form $\tau \leftrightarrow\mu$ and 
$\eta \leftrightarrow -\eta^{-1}$ and be compatible with Higgs branch and Coulomb branch Higgsing transformations, 
we need to choose our sign conventions as follows:
\begin{equation}
\delta_j = N_j + 1\,,\qquad \delta^\vee_j = N^\vee_j + 1\,.
\label{eq:deltas}
\end{equation}

As a final check, we can look at the other Bethe equations at the $j$-th node
\begin{equation}
\eta^2 \frac{\tau_{j+1}}{\tau_{j}}\prod_{n'=1}^{N_{j-1}}\frac{\eta \sigma^{(j)}_n- \sigma^{(j-1)}_{n'}}{\eta \sigma^{(j-1)}_{n'}-\sigma^{(j)}_n} \cdot \prod_{n'=2}^{N_{j}}\frac{\eta^{-1} \sigma^{(j)}_n - \eta \sigma^{(j)}_{n'}}{\eta^{-1}\sigma^{(j)}_{n'}-\eta \sigma^{(j)}_n} \cdot \prod_{n'=1}^{N_{j+1}}\frac{\eta \sigma^{(j)}_n - \sigma^{(j+1)}_{n'}}{\eta \sigma^{(j+1)}_{n'}-\sigma^{(j)}_n}\cdot\prod_{a=1}^{M_j}\frac{\eta \sigma^{(j)}_n-\mu^{(j)}_a}{\eta \mu^{(j)}_a-\sigma^{(j)}_n}=(-1)^{\delta_j+1}\,, 
\end{equation}
and at the nearby nodes (say at $(j+1)$st)
\begin{align}
&(-\eta^{-1})\frac{\tau_{j+2}}{\tau_{j+1}}\prod_{n'=2}^{N_{j}}\frac{\eta \sigma^{(j+1)}_n- \sigma^{(j)}_{n'}}{\eta \sigma^{(j)}_{n'}-\sigma^{(j+1)}_n} \cdot \prod_{n'}^{N_{j+1}}\frac{\eta^{-1} \sigma^{(j+1)}_n - \eta \sigma^{(j+1)}_{n'}}{\eta^{-1}\sigma^{(j+1)}_{n'}-\eta \sigma^{(j+1)}_n}\cr
& \cdot \prod_{n'=1}^{N_{j+2}}\frac{\eta \sigma^{(j+1)}_n - \sigma^{(j+2)}_{n'}}{\eta \sigma^{(j+2)}_{n'}-\sigma^{(j+1)}_n}\cdot\prod_{a=1}^{M_{j+1}}\frac{\eta \sigma^{(j+1)}_n-\mu^{(j+1)}_a}{\eta \mu^{(j+1)}_a-\sigma^{(j+1)}_n}=(-1)^{\delta_{j+1}}\,.
\end{align}
The prefactors agree with the idea that the new quiver after Higgsing has FI parameters $(-\eta^{-1})\tau_j$,  $(-\eta^{-1})^{-1}\tau_{j+1}$ and $N_j$, and thus $\delta_j$, reduced by one unit. 
Everything is consistent with the mirror map.

\subsection{From Quiver Gauge theories to Four-Dimensional Gauge Theory on a Segment}
As we mentioned before, brane configurations where some D3 branes end on a D5 brane do not have a direct three-dimensional interpretation. Rather, they can be interpreted in the context of a four-dimensional setup, where the fivebranes map to half-BPS domain walls for four-dimensional $\ssN=4$ SYM gauge theory \cite{Gaiotto:2008sa}. The six adjoint scalar fields on $\ssN=4$ SYM decompose into two groups of three, transforming into triplets of $SU(2)_C$ and $SU(2)_R$ respectively. Roughly, Neumann boundary conditions on the gauge fields require by supersymmetry Dirichlet b.c. on the $SU(2)_H$ fields and Neumann on the $SU(2)_C$ fields. The converse is true for Dirichlet boundary conditions. The fivebranes engineer domain walls which generalize Neumann and Dirichlet boundary conditions in interesting ways. 

The NS5 branes engineer a simple domain wall: the four-dimensional gauge theories on the two sides of the wall have 
Neumann boundary conditions, and are coupled at the interface to a set of 3d bifundamental hypermultiplets. 
If no D3 branes are present on one side, we recover the standard Neumann boundary conditions. 

The D5 branes are more subtle. If the number of D3 branes on the two sides of the fivebrane is equal, the domain wall does not affect directly the four-dimensional fields, but simply couples to them a three-dimensional fundamental hypermultiplet. If the number of D3 branes $N_+$ on one side is bigger than the number $N_-$ on the other side, fields have a generalized, partial Dirichlet boundary condition: 
the $U(N_-)$ gauge fields are glued at the interface to a $N_- \times N_-$ block of the $U(N_+)$ gauge fields. The remaining $U(N_+)$ gauge fields have Dirichlet boundary conditions, deformed by 
a special boundary condition on the three scalar fields $X^{1,2,3}_H$ which are rotated by $SU(2)_H$: they should  have a Nahm pole in the complementary $(N_+ - N_-) \times (N_+ - N_-)$ block
\begin{equation}
X_H^{i} \sim \frac{t_\rho^i}{x^3}\,,
\end{equation}
where the $t_\rho^i$ are the generators of an irreducible $(N_+ - N_-)$-dimensional representation of $\mathfrak{su}(2)$.
In particular, a single D5 brane with no D3 branes on one side will {\it not} engineer Dirichlet boundary conditions, but rather the deformation involving a Nahm pole of maximal size. 
Conversely, Dirichlet boundary condition for a $U(N)$ gauge theory can only be engineered by using $N$ D5 branes in sequence, and letting 
a single D3 brane end on each, so that the rank of the gauge group is progressively reduced one step at the time all the way to nothing, without Nahm poles. 

This illustrates the intricacies of the action of S-duality on half-BPS boundary conditions \cite{Gaiotto:2008ak}. A Neumann boundary condition is S-dual to a maximal Nahm pole. 
The S-dual of a Dirichlet boundary condition involves a sequence of NS5 branes, and thus a sequence of segments with gauge group $U(i)$ in the $(N-i)$-th
segment, with appropriate 3d bifundamentals. In the IR, we find a specific linear quiver gauge theory at the boundary, with $N_i = N-i$, coupled to the $U(N)$ four-dimensional gauge theory 
by gauging the flavor symmetry of $N$ fundamentals at the first node. This theory is denoted conventionally as $T[U(N)]$.\footnote{There is a very small difference between the theory denoted as $T[U(N)]$ and the theory denoted as $T[SU(N)]$. They have the same matter content, and only $SU(N)_H \times SU(N)_C$ flavor symmetry acting on the Higgs and Coulomb branch operators. 
The theory $T[U(N)]$ is equipped with a specific promotion of the flavor symmetry to $U(N)_H \times U(N)_C$, which is the same we introduced for convenience in earlier sections.}

This statement admits a simple generalization: a set of D5 branes with linking numbers $r_i$ engineer a (left) Nahm pole boundary condition where 
the $t_\rho^i$ are the generators of an $\mathfrak{su}(2)$ embedding $\rho$ in $\mathfrak{u}(N)$, defined by the sum of irreducible representations of dimension $r_i$. 
If we S-dualize the brane setup, we arrive to a (right) boundary condition, where NS5 branes with linking numbers $r_i$ engineer a triangular quiver 
gauge theory with $N$ flavors at the first node only, and appropriate $U(N_i)$ gauge groups. 

\begin{figure}
\begin{center}
\includegraphics[height=5cm, width=8.5cm]{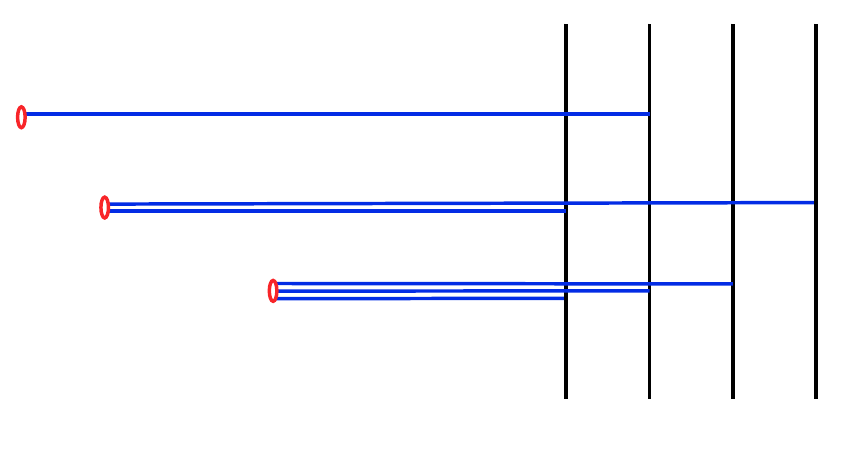}
\caption{Brane construction for $T[U(6)]_\rho^{\rho^\vee}$ theory with $\rho=(3,2,1)$ and $\rho^\vee=(2,2,1,1)$. It also gives rise to $A_3$ quiver with labels $(1,1),(1,1),(1,1)$.}
\label{fig:TSUNrrdbrane}
\end{center}
\end{figure}

After this preparation, we can do a useful transformation on the brane system which engineers a linear quiver gauge theory: 
we can bring all NS5 branes to the far right of the system and all D5 branes to the far left as shown in \figref{fig:TSUNrrdbrane}. 
By construction, the linking numbers tell us exactly how many D3 branes end on each fivebrane. Define 
\begin{equation}
Q = \sum_i r_i^\vee= \sum_{j=1}^{L^\vee-1} j M_j\,.
\label{eq:Qdef}
\end{equation}
The field theory limit of this configuration is an $U(Q)$ four-dimensional $\ssN=4$ SYM gauge theory with two boundary conditions, labelled by the D5 linking numbers $\rho = (r_a)$ on the left side, and the NS5 linking numbers $\rho^\vee = (r^\vee_i)$ on the right side (see \figref{fig:TSUNrrdbrane}). 
We organized the linking numbers into two Young Tableaux $\rho$ and $\rho^\vee$. 
In a sense, this setup decouples the two sets of linking numbers: we can first study the two individual boundary conditions for $\ssN=4$ SYM, and then combine their effect into a new description of the 3d SCFTs associated originally to the quiver gauge theories. This presentation of the SCFTs is denoted as $T[U(Q)]_\rho^{\rho^\vee}$.

Let us understand the connection between the partitions $\rho$ and $\rho^\vee$ of $Q$ \eqref{eq:Qdef} with the data of the 3d quiver $(N_i,M_j),\,j=1,\dots,L^\vee-1$ in more detail using the brane language. We start from the configuration which directly gives us the gauge and matter content of the theory, like in \figref{fig:branequiv}, in other words when no D3 brane ends on a D5 brane. So once we know $(N_i,M_j)$ one can easily get the partition $\rho$ (actually its transposition) by moving all D5 branes to the left of NS5s and counting the number of D3 branes which are picked up on the way due to Hanany-Witten phenomena. Let us first find the linking number for the leftmost NS fivebrane. Every D5 brane will acquire a D3 brane attached to it. In order to get the actual linking number one has to subtract the number of D3 branes for the first gauge group, as they are attached to the first NS5 brane from the right side. Thus we get $r_1^\vee = L-N_1$. Next, for the second from the left NS5 brane all but first $M_1$ D5 branes, which belong to the first gauge group, namely $\sum_{k=2}^L M_k$ D5 branes. By construction this number is $r^T_2$ -- the second from the bottom row of transposed partition $\rho^T$. Thus we can find the linking number for the second NS5 brane: $r_2^\vee=r_2^T+N_1-N_2$ is the total sum of D3 branes and antibranes which end on this NS5 brane. So generically for linear $A_{L-1}$ quivers
\begin{equation}
r_1^\vee = L-N_1\,,\quad r_j^\vee=r_j^T+N_{j-1}-N_j\,, \quad j=2,\dots, L^\vee\,,
\label{eq:linkingnum}
\end{equation}
where $N_L=0$. By construction the transposed partition $\rho^T$ gives flavor labels $M_i$ one has the following formula for its rows
\begin{equation}
r_i^T=\sum\limits_{k=i}^{L^\vee-1} M_k\,.
\label{eq:transposedpho}
\end{equation}
One can see that \eqref{eq:linkingnum} combined with the above formula reproduces the definition of the NS5 brane linking numbers  \eqref{eq:linkNS5}. Also \eqref{eq:linkingnum,eq:transposedpho} we can extract the weights of representation $\CR$ in order to get \eqref{eq:weights}. Indeed,
\begin{equation}
r_j^\vee-r_{j+1}^\vee = r_j^T-r_{j+1}^T+N_{j-1}-N_{j}-N_j+N_{j+1}=M_j+N_{j-1}+N_{j+1}-2N_j=\Delta_j\,.
\end{equation}
The total number of D3 branes in the configuration when D5s and NS5s are pulled apart, as in \figref{fig:TSUNrrdbrane}, is equal to the size of the partition \eqref{eq:Qdef}.

For example, $A_3$ quiver with $N_1=N_2=N_3=M_1=M_2=M_3=1$ after moving branes to the sides gives rise to the configuration in \figref{fig:TSUNrrdbrane} with $\rho=(3,2,1)$ and $\rho^\vee=(2,2,1,1)$.

\begin{figure}
\begin{center}
\includegraphics[height=6.5cm, width=16.5cm]{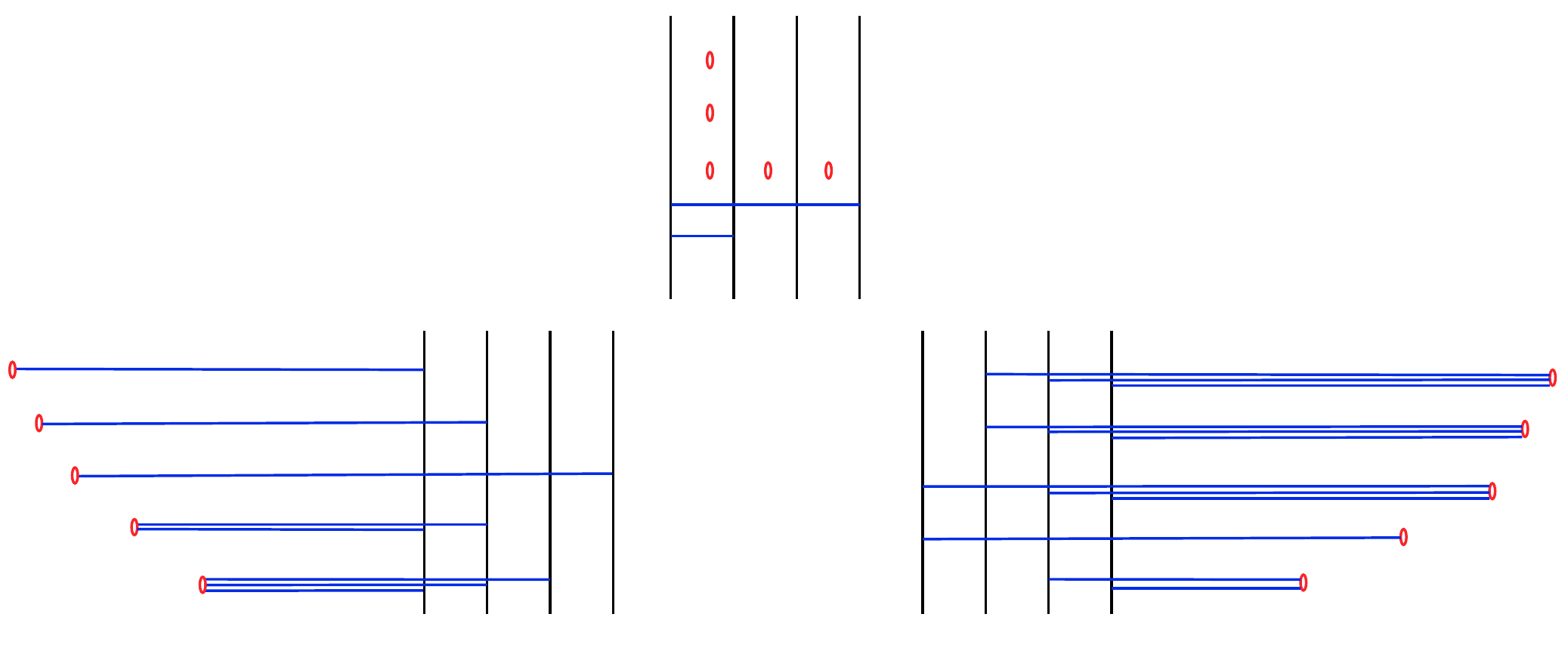}
\caption{Different boundary conditions obtained from the $A_3$ theory with labels $(2,3),(1,1),(1,1)$ whose brane diagram is depicted in the center. The diagram in the lower left corner represents $T[U(8)]_{\rho_1}^{\rho^\vee_1}$ theory with $\rho_1=(3,2,1,1,1)$ and $\rho^\vee_1=(3,3,1,1)$. In the lower right corner we have $T[U(12)]_{\rho_2}^{\rho^\vee_2}$ theory with $\rho_2=(3,3,3,2,1)$ and $\rho^\vee_2=(4,4,2,2)$.}
\label{fig:LRrrd}
\end{center}
\end{figure}

Notably there are two different theories $T[U(Q_1)]_{\rho_1}^{\rho^\vee_1}$ and $T[U(Q_2)]_{\rho_2}^{\rho^\vee_2}$, which can be obtained from the same $A_{L^\vee}$ quiver with labels $(N_i, M_i)$. Above we just described what happens when all NS5 branes are moved to the right and all D5 branes are moved to the left. Let us see what will happen when we do the opposite. An example of such action is shown in \figref{fig:LRrrd}.
Thus we can see that the same quiver theory can be derived from two completely different sets of boundary conditions.

\subsubsection{Boundary conditions and the space of vacua}
In the next section we will look in detail at the moduli space of vacua of these four-dimensional configurations upon circle compactification and mass deformation to $\ssN=2^*$. 
Here, we would like to review a few facts about the moduli space of vacua in flat space, for the original $\ssN=4$ setup on a half-line or segment. 

The Higgs branch of the system is still derived from classical configuration. It is parameterized by $x^3$-dependent configurations of  the three  scalar fields rotated by $SU(2)_H$ and the $A_3$ component of the gauge field. These fields satisfy a set of first-order differential 
equations called Nahm equations, with appropriate boundary conditions at domain walls. If we only care about the complex structure of the Higgs branch, the equations reduce to the statement that the complex field ${\cal X} = X_H^1 + i X_H^2$ is covariantly constant. 

The Nahm boundary conditions force ${\cal X}$ to live in the so-called Slodowy slice for $\rho$, i.e. 
\begin{equation}
{\cal X} = \mathfrak{t}^+_\rho + x^*
\end{equation} 
where $\mathfrak{t}^+_\rho$ is the raising generator of the $\mathfrak{su}(2)$ embedding and the matrix $x^*$ should be a lowest weight for the $\mathfrak{su}(2)$ action. 

The boundary condition which is given by coupling to the triangular quiver $T[U(Q)]^{\rho^\vee}$ forces ${\cal X}$ to coincide with 
the moment map for the $U(Q)_H$ flavour symmetry of $T[U(Q)]^{\rho^\vee}$. The moment map parameterizes faithfully the Higgs branch 
of $T[U(Q)]^{\rho^\vee}$, and lies in a specific nilpotent orbit of $GL(Q)$, labelled by the transposed partition $\rho_T^\vee$ to $\rho^\vee$. Overall, the Higgs moduli space of vacua of a general linear quiver $T[U(Q)]^{\rho^\vee}_\rho$ is the intersection of the 
Slodowy slice for $\rho$ and the nilpotent orbit for $\rho_T^\vee$. By S-duality, the Coulomb branch has the opposite characterization.

\section{BPS Boundary Conditions and S-duality}\label{sec:bcS}
In this section we investigate moduli spaces of $\ssN=2^*$ 3d gauge theories from a different perspective, namely we start with the $\ssN=4$ super Yang-Mills theory in four dimensions and study the moduli space upon compactification on a circle and $\ssN=2^*$ mass deformation. We then introduce BPS boundary conditions and domain walls and study their moduli space of vacua. 

In particular, the $T[U(Q)]^{\rho^\vee}_\rho$ construction of the 3d SCFTs associated to quiver gauge theories will elucidate the geometric structure of the problem, and a surprising connection between the XXZ spin chain Bethe equations and the \textit{trigonometric Ruijsenaars-Schneider (tRS)} \cite{MR1329481} integrable model which describes the 
vacuum geometry of the four-dimensional theory \cite{Nekrasov:2011bc}. In what follows we shall heavily rely on the analysis of \cite{Gaiotto:2008sa,Gaiotto:2008ak}, albeit adopted to $\ssN=2^*$ theories. 

\subsection{Generalities and Localization}
For a general $\CN=2$ four-dimensional gauge theory, the space of vacua $\CM$ upon compactification on a circle is parameterized by the four-dimensional scalar fields and by the electric and magnetic Wilson lines for the gauge fields. It is an hyperk\"ahler 
manifold, and has a $\mathbb{CP}^1$ worth of possible complex structures. In this section we focus on half-BPS boundary conditions, and the choice of supercharges preserved by the boundary condition selects a specific complex structure. Roughly speaking, the real and imaginary parts of the four-dimensional scalar fields are promoted to complex variables, whose imaginary parts are the electric and magnetic Wilson lines. 

A half-BPS boundary condition constrains the set of supersymmetric vacua of the system to lie on a complex Lagrangian submanifold $\CL \subset \CM$. Similarly, domain walls between two theories with moduli spaces $\CM_{L,R}$ are associated to complex Lagrangian submanifolds $\CL \subset \CM_L \times \CM_R$, with complex symplectic form $\Omega_L - \Omega_R$ on the product manifold. For example, a trivial domain wall will correspond to a diagonal embedding $\CM \subset \CM \times \CM$. Thus if we consider the four-dimensional theory on a segment, with boundary conditions $\CL_L$ and $\CL_R$, we obtain a system with a discrete set of vacua, the intersection of $\CL_L \cap \CL_R$. If the setup flows to the infrared to some 3d SCFT, we expect the vacua of the 3d SCFT on the circle to match the four-dimensional result. 

The geometry of the manifold $\CM$ does not follow in a simple way from the UV description of the 4d theory. Classically, 
every $\CN=2$ 4d Coulomb branch scalar field in the Cartan sub-algebra of the UV gauge group 
gives rise to an $(\mathbb{C}^*)^2$ factor, modulo the Weyl group symmetry. Integrating away the 
massive matter fields modifies this naive geometry. 
The quantum-corrected complex geometry in the appropriate complex structure may be probed by the expectation value of BPS line defects 
wrapping the circle. These expectation values provide a natural set of holomorphic functions on moduli space. 
The expectation values can be computed for a Lagrangian theory through localization. Although localization can be done directly in flat space \cite{Ito:2011ea}, most work has been done on spheres \cite{Pestun:2007rz, Gomis:2011pf}. Luckily, the partition function and correlation functions on the ellipsoid $S^4_b$ \cite{Hama:2012bg} have a useful $b \to 0$ limit which can be used to reproduce flat space results. 

The general structure of an $S^4_b$ correlation function is that of a matrix model-like integral over half of the 
vectormultiplet scalar fields $a$ in the UV Cartan subalgebra, i.e. the half which enters the definition of a BPS Wilson loop. The integrand has a factorized form, 
\begin{align}
\CZ_{S^4_b} &= \int d\nu(a) |\CZ_{\mathrm{inst}}(a, b, 1/b)|^2\,, \cr
\langle L \rangle &= \CZ_{S^4_b}^{-1} \int d\nu(a) \bar \CZ_{\mathrm{inst}}(a,  b, 1/b) \hat L\, \CZ_{\mathrm{inst}}(a,  b, 1/b)\,.
\end{align} 
We will discuss the integration measure momentarily. Here $\CZ_{\mathrm{inst}}$ is Nekrasov's instanton partition function \cite{Nekrasov:2002qd} with equivariant parameters $\epsilon_{1,2} =  b^{\pm 1}$. We will not need any information about it. The interesting part for us is the operator $\hat L$, which implements the insertion of a generic BPS line defect $L$ is inserted on a specific circle in the $S^4_b$ geometry. The operator $\hat L$ for a Wilson loop in representation ${\cal R}$ acts by multiplication by a simple function
\begin{equation}
\hat W_{\cal R} = \mathrm{Tr}_{\cal R} e^{2 \pi b a}\,,
\end{equation}
while 't Hooft-Wilson loops involve a sum of shift operators $e^{i n b\partial_a}$ with coefficients which are rational functions of $e^{\pi b a}$.

For example, the fundamental Wilson loop for an $U(N)$ gauge theory would be 
\begin{equation}
\hat W_{0,1} = \sum_n e^{2\pi b a_n}\,,
\end{equation}
while the naive classical 't Hooft loop of minimal charge (which is correct for an $\ssN=4$ theory in the absence of mass deformation)
would be \begin{equation}
\hat W_{1,0} = \sum_n e^{i b \partial_{a_n}}\,,
\end{equation}
and a 't Hooft-Wilson loop of minimal magnetic charge
\begin{equation}
\hat W_{1,q} = \sum_n e^{2\pi b q a_n}  e^{i b \partial_{a_n}}\,.
\end{equation}
In a specific $\ssN=2$ theory, the coefficients of 't Hooft loops with small enough charge will be modified by one-loop corrections from the matter field. Loops of higher charge also contain intricate contributions from monopole bubbling phenomena \cite{Gomis:2011pf}.
 
We can also insert domain walls on an ellipsoid parallel to the equator,
or replace one or both hemispheres of $S^4_b$ by a boundary condition \cite{Drukker:2010jp}. 
For example, Dirichlet boundary conditions simply force the variables $a$ to a fixed value. 
For a generalized Neumann boundary condition which involves a coupling to 3d degrees of freedom, one should replace the corresponding instanton partition function with the partition function of the 3d theory on the 3d ellipsoid $\CZ_{S^3_b}(a)$ \cite{Hama:2011ea,Imamura:2011wg}, with mass parameters replaced by 4d scalar vevs. 

To make this statement meaningful, we need to make two choices. On one hand, we need to specify the measure. 
On the other hand, we need to specify the boundary condition for bulk hypermultiplets. For a through discussion of this 
problem, we refer to \cite{Drukker:2010jp} and \cite{Dimofte:2012pd}. Here, we will use a simple shortcut to arrive to the correct answer for the $\ssN=2^*$ theory. All we need to know about the choice of boundary condition is which 3d degrees of freedom arise from the reduction of the 4d theory on a segment 
when the segment is very short. Indeed, we know that the $S^4_b$ partition function with both hemispheres replaced by boundary conditions
\begin{equation}
\CZ_{S^4_b,L,R} = \int d\nu(a) \CZ_{S^3_b,L}(a) \CZ_{S^3_b,R}(a)
\end{equation} 
should reproduce to the $S^3_b$ partition function for the 3d theory which arises from the reduction on a segment. 
This shows that $d\nu(a)$ should contain the contribution to the $S^3_b$ partition function of the 3d degrees of freedom which arise from the reduction of 4d fields on the segment. In the case of interest for us, we know that 
whenever we reduce the ${\cal N}=4$ theory on a segment with generalized Neumann boundary conditions,
the 4d fields give rise to a 3d $\ssN=4$ gauge multiplet. 

In the $b \to 0$ limit, several simplifications occur. First of all, the 3d partition function is related to the twisted effective superpotential $\CW$ with $R = b$
\begin{equation}
\CZ_{S^3_b} \sim e^{-2 \pi i {\cal W}}\,.
\label{eq:WeffZ}
\end{equation}
For example, the partition function for a single chiral field of mass $m$ involves the special function
\begin{equation}
\CZ_{S^3_b}(m) =  s_b \left(\frac{i}{2} b + \frac{i}{2} b^{-1} - m \right)\,,
\end{equation}
which satisfied the recursion relation \cite{Kharchev:2001rs, Hama:2011ea}
\begin{equation}
\CZ_{S^3_b}(m + i b) = 2 \sinh(\pi b m) \CZ_{S^3_b}(m)\,,
\end{equation}
and thus in the $b \to 0$ limit
\begin{equation}
i b\, \partial_m \log \CZ_{S^3_b}(m) = \log 2 \sinh \pi b m
\end{equation}
to be compared with the function which appears in the effective twisted superpotential: 
\begin{equation}
2 \pi R \partial_x \ell(x) = \log 2\sinh \pi R x\,.
\end{equation}

In the $b \to 0$ limit, the Coulomb branch integration variables $a$ are identified with 
half of the holomorphic coordinates on $\CM$, which are essentially {\it defined} 
by the Wilson loop vevs in some representation $\CR$
\begin{equation}
W_{\cal R} = \sum_{w \in \cal R} e^{2 \pi R a \cdot w} \equiv \sum_{w \in \cal R} \alpha^w\,.
\end{equation}
The shift operators in the line defect vevs $e^{i b \partial_a}$ are identified in the $b \to 0$ limit
with a dual set of coordinates $p_\alpha$ on $\CM$.

Inside the partition function, the line defect operator acts on the instanton partition function, or the 
contribution of the right boundary condition in 
\begin{align}
\langle L \rangle_{L,R} &= \int d\nu(a) \CZ_{S^3_b,L}(a) \hat L\, \CZ_{S^3_b,R}(a)\,.
\end{align} 
In either case (at least for generalized Neumann boundary conditions) in the limit $b \to 0$, the right boundary condition  fixes the value of $p_a$ to lie on the Lagrangian submanifold $\CL_R$ 
\begin{equation}
p_a = \exp 2 \pi R \frac{\partial{\cal W}_R}{\partial a}\,.
\end{equation}
where ${\cal W}_R$ comes from the $b \to 0$ limit of $\CZ_{S^3_b,R}(a)$ \eqref{eq:WeffZ}. This is our familiar Lagrangian moduli space of vacua of the boundary 3d theory, embedded in $\CM$ by the coordinates $(a, p_a)$. 

On the other hand, we can bring the operator to act on the left boundary condition, by shifting the integration variable. Including the effect of the shift in the integration measure, which survives in the $b \to 0$ limit, 
the boundary condition becomes the Lagrangian submanifold $\CL_L$
\begin{equation}
p_\alpha = \exp 2 \pi R \left[- \frac{\partial{\cal W}_L}{\partial a} - \frac{\partial{\cal W}_{\mathrm{4d}}}{\partial a}\right]\,.
\label{eq:palphaL}
\end{equation}
where ${\cal W}_{\mathrm{4d}}$ indicates the contribution from the 3d fields which arise from the reduction of 4d fields on the segment and ${\cal W}_L$ comes from the $b \to 0$ limit of $\CZ_{S^3_b,L}(a)$.
We can also introduce a new variable $\tilde p_a$ such that the left boundary condition is defined as 
\begin{equation}
\tilde p_\alpha = \exp 2 \pi R \frac{\partial{\cal W}_L}{\partial a}\,.
\label{eq:palphatilde}
\end{equation}
For the $\ssN=2^*$ theory, after combining \eqref{eq:palphaL} and \eqref{eq:palphatilde} we thus have 
\begin{equation}
p_\alpha^i \tilde p_\alpha^i = \frac{\prod_{k}(\alpha_k \eta^{-1}-\alpha_i \eta)}{\prod_{k}( \alpha_i \eta^{-1}-\alpha_k \eta)}\,.
\end{equation}

\subsection{Boundary Conditions for $\ssN=2^*$ Four-dimensional SYM on $S^1$}\label{Sec:BCN2}
The $\CN=2^*$ mass deformation of four-dimensional $\CN=4$ SYM is completely analogous to the $\CN=2^*$ mass deformation of the $\CN=4$ three-dimensional SCFTs: it is the mass deformation associated to an $SU(2)_f$ subgroup 
of the $\CN=4$ R-symmetry, which commutes with the $\CN=2$ R-symmetry subgroup. Indeed, we claim that in the presence of half-BPS domain walls or boundary conditions, the $SU(2)_f$ subgroup is broken exactly to the $U(1)_f$ 
subgroup which we used to mass-deform three-dimensional $\CN=4$ SCFTs. 

Whenever we have a composite system 
which includes four-dimensional $\CN=4$ SYM and half-BPS domain walls or boundary conditions, we can use the overall 
$U(1)_f$ for the system to introduce a three-dimensional real mass deformation. This will induce a (real) $\CN=2^*$ mass deformation in the bulk theory, and a canonical $\CN=2^*$ mass deformation of the domain walls and boundary conditions. 
We will denote the mass deformation parameter as $\epsilon$, as in the previous sections. It is important to observe that 
all three-dimensional theories coupled to the same four-dimensional theory will have the same deformation parameter $\epsilon$, which will coincide with the bulk mass deformation parameter as well.  

For completeness, we can describe a few more group-theoretic details of the mass deformation. 
The bulk theory $\CN=4$ SYM is an $\CN=2$ gauge theory coupled to an adjoint hypermultiplet, and the $\CN=2^*$ mass deformation parameter is associated canonically to the $SU(2)_f$ flavor symmetry of that hypermultiplet. The $\CN=2$ R-symmetries $U(1)_r \times SU(2)_R$ combine with $SU(2)_f$ to an $SO(2) \times SO(4)$ subgroup of the overall $SO(6)_R$ R-symmetry of the theory. Boundary conditions and domain walls break $SO(6)_R$ to the familiar $SO(3)_C \times SO(3)_H$ subgroup. The two subgroups of $SO(6)$ are both block diagonal, and intersect along an $U(1)_R \times U(1)_f$ subgroup.
 
The moduli space $\CM$ for an $U(N)$ $\CN=2^*$ SYM theory is a rather interesting manifold: the space of flat $GL(N)$ flat connections on a torus with one simple puncture (i.e. a puncture with minimal simple monodromy) \cite{Donagi:1995cf}
with $N-1$ monodromy eigenvalues $\eta^{-2} = e^{-2 \pi R\epsilon}$ and one $\eta^{2N-2}$, where $R$ is the compactification radius (cf. \eqref{eq:nus}). The space is $2N$ complex dimensional, and can be written as 
\begin{equation}
M T M^{-1} T^{-1} = E\,
\label{eq:FlatnessStand}
\end{equation}
modulo conjugation by gauge transformations, with $E$ being the monodromy at the simple puncture, and $M$, $T$ the monodromies around the $A$ and $B$ cycles of the torus. The moduli space is locally independent of the gauge coupling (modular parameter of the torus). 

The natural functions on this space are the traces of holonomies on the torus, in appropriate representations. 
The Wilson loop vevs in the gauge theory map to traces of $M$ in the appropriate representation. 
Thus the eigenvalues of $M$ coincide with the $a_i$ variables we encountered from localization. 
We will denote as before $\alpha_i = \exp 2 \pi R a_i$. 
The 't Hooft loop vevs in the gauge theory should coincide with traces of $T$ in appropriate representations. 
The localization calculations for the 't Hooft loop operators suggest a simple interpretation for the $(\alpha_i, p^i_\alpha)$ 
Darboux variables on $\CM$: they are Fenchel-Nielsen (FN) coordinates. 

The flatness condition \eqref{eq:FlatnessStand} can be conveniently rewritten in a slightly different, useful form
\begin{equation}
\eta M T - \eta^{-1} T M = u v^T\,,
\label{eq:FlatConNew}
\end{equation}
where $u$, $v$ are vectors, defined up to a $\mathbb{C}^*$ rescaling in opposite directions. 
If we gauge-fix by making $M$ diagonal with eigenvalues $\alpha_i$ 
we get 
\begin{equation} \label{eq:lax1}
T_{ij} = \frac{u_i v_j}{\eta \alpha_i - \eta^{-1} \alpha_j}\,.
\end{equation}
The conjugate momenta are defined through
\begin{equation}\label{eq:lax2}
u_i v_i \prod_{k \neq i} (\alpha_i - \alpha_k) = - p_\alpha^i \prod_{k}( \alpha_i \eta^{-1}-\alpha_k \eta) = - (\tilde p_\alpha^i)^{-1} \prod_{k}( \alpha_k \eta^{-1}-\alpha_i \eta)\,.
\end{equation}
This defines the FN momenta $p_\alpha^i$. The matrix $T$ coincides with the Lax matrix of the tRS model \cite{MR1329481} (in an appropriate gauge, see \cite{Gorsky:1993dq,Fock:1999ae} for more details, where a slightly different gauge is used). The determinant of the Lax matrix is $\det T = \prod_i p_\alpha^i$. 
The alternative set of momenta $\tilde p_\alpha^i$ appear naturally in $T^{-1}$. 

Notice that trace of $T$ 
\begin{equation}
\mathrm{Tr}\, T= \sum_i p_\alpha^i  \prod_{k \neq i} \frac{\alpha_i \eta^{-1}-\alpha_k \eta}{\alpha_i - \alpha_k}
\label{eq:trLaxtRS}
\end{equation}
coincides with the localization expression for the fundamental 't Hooft loop. 
Let us review an example of such localization computation \cite{Ito:2011ea}. The authors computed vevs of Wilson and 't Hooft loops of $U(2)$ $\CN=2^*$ SYM theory on $\mathbb{R}^3\times S^1$. For the Wilson loop in the spin $\half$ representation one gets\footnote{We absorbed the factors of $i$ which appear in \cite{Ito:2011ea} into $m_i$s}
\begin{equation}
\mathrm{Tr}\, M = \mathrm{e}^{2\pi R m_1}+\mathrm{e}^{2\pi R m_2}=:\alpha_1+\alpha_2\,,
\end{equation}
whereas for the 't-Hooft loop one gets
\begin{equation}
\mathrm{Tr}\, T = (\mathrm{e}^{2\pi R \nu_1}+\mathrm{e}^{2\pi R \nu_2})\frac{\sqrt{(\eta\alpha_1-\eta^{-1}\alpha_2)(\eta^{-1}\alpha_1-\eta\alpha_2)}}{\alpha_1-\alpha_2}\,.
\label{eq:TrTStandard}
\end{equation}
It was noticed in \cite{Nekrasov:2011bc} that the latter expression gives the Hamiltonian function of the tRS model and $m_i$ and $\nu_i$ should be treated as the Darboux coordinates on the Lagrangian submanifold given by fixing the eigenvalues of $M$. Following \cite{Dimofte:2011jd} one may define the FN twisted coordinates as follows
\begin{equation}
\mathrm{e}^{2\pi R \nu_1}=\frac{\eta\alpha_1-\eta^{-1}\alpha_2}{\eta^{-1}\alpha_1-\eta\alpha_2}p_\alpha^1\,,\quad
\mathrm{e}^{2\pi R \nu_2}=\frac{\eta\alpha_1-\eta^{-1}\alpha_2}{\eta^{-1}\alpha_1-\eta\alpha_2}p_\alpha^2\,,
\end{equation}
in order to reproduce \eqref{eq:trLaxtRS}.

Finally, it is useful to look carefully at how S-duality is realized in this context. The flatness equation 
can be rewritten as 
\begin{equation}
\eta T M^{-1} - \eta^{-1} M^{-1} T = u_\vee v^T_\vee\,,
\label{eq:SFlatCon}
\end{equation}
and thus the standard S-duality transformation is $T \to M$, $M^{-1} \to T$, preserving the symplectic form. 
On the other hand, in the context of engineering 3d SCFTs and mirror symmetry, it is natural to accompany 
S-duality with a permutation of scalar fields and reflection in the $x^3$ coordinates (which reflects the symplectic form). 
This corresponds to the S-duality transformation 
\begin{equation}
T \to M\,, \qquad M \to T\,, \qquad \eta \to - \eta^{-1}\,,
\end{equation}
which also leaves \eqref{eq:FlatConNew} invariant. 

After this somewhat lengthy preparation, we are ready to look at interesting boundary conditions and their duality properties. Some examples are in order.

\subsection{Boundary Conditions and Domain Walls in the $U(1)$ Theory}
It is useful to start with some Abelian examples. Neumann and Dirichlet boundary conditions are rather obvious. 
For Dirichlet boundary conditions we set the the scalar vev $\alpha$ to some constant $\mu$. 
For Neumann b.c. we set the momentum $p_\alpha$ to a constant $\tau$, the FI parameter at the boundary. 
The basic NS5 and D5 domain walls are more interesting. 

\begin{figure}
\begin{center}
\includegraphics[height=5.5cm, width=11.5cm]{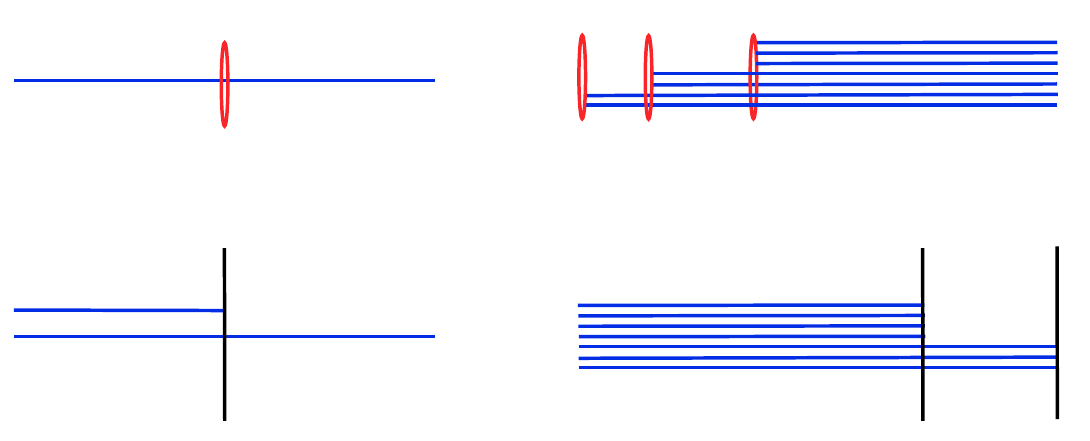}
\caption{Brane constructions for various boundary conditions and domain walls inside $\CN=2^*$ 4d theory. The top left figure shows a D5 domain wall between two $U(1)$ theories; the bottom left figure depicts a NS5 domain wall between a $U(2)$ and a $U(1)$ theory; the top right figure shows seven D3 branes ending one different D5 branes, thereby describing a generic Nahm pole given by $\rho=(3,2,2)$; the bottom right figure does the same of the boundary conditions S-dual to Nahm b.c. with $\rho^\vee=(4,3)$.}
\label{fig:bcall}
\end{center}
\end{figure}

For the D5 domain wall between two $U(1)$ gauge theories (see \figref{fig:bcall})
we simply add a 3d hypermultiplet of unit charge to a 4d $U(1)$ gauge theory. The scalar vev is continuous across the interface, while the conjugate momentum jumps across the interface: 
the momentum to the left $p_\alpha^1$ is related to the momentum to the right $p_\alpha^2$ as
\begin{equation}
p_\alpha^1 = \frac{\eta \alpha - \mu}{\eta \mu - \alpha} p_\alpha^2\,.
\end{equation}
We can invert this relationship 
\begin{equation}
\alpha = \frac{\eta p_\alpha^1+ p_\alpha^2}{\eta p_\alpha^2+ p_\alpha^1} \mu\,.
\end{equation}

Let us apply our S-duality prescription, renaming $\mu$ as $\tau$ for convenience. We exchange $\alpha_i$ and $p_\alpha^i$, $\eta \to - \eta^{-1}$. 
We get 
\begin{equation}
p_\alpha^1 = p_\alpha^2 = \frac{\eta \alpha_2- \alpha_1}{\eta \alpha_1- \alpha_2} \tau\,.
\end{equation}
Now the momentum to the left $p_\alpha^2$ corresponds to coupling to a bifundamental hypermultiplet
and FI parameter $\tau$. The momentum to the right, 
\begin{equation}
\tilde p_\alpha^1 = \frac{\eta \alpha_1- \alpha_2}{\eta \alpha_2- \alpha_1} \tau^{-1}
\end{equation}
also corresponds to coupling to a bifundamental hypermultiplet, with opposite FI parameter $\tau^{-1}$. 

The coupling to a bifundamental, and the opposite FI parameters on the two sides of the NS5 domain wall 
will recur at higher rank.  

\subsection{Boundary Conditions and Domain Walls in the $U(2)$ Theory}
The first, basic example are Neumann b.c. We can add a boundary FI parameter for the $U(1)$ factor, by the condition 
\begin{equation}
p_\alpha^i = \tau\,.
\end{equation}
The Lax matrix $T$ \eqref{eq:trLaxtRS}  turns out to have has eigenvalues $\tau \eta$ and $\tau \eta^{-1}$, and very simple eigenvectors. In particular, we can diagonalize 
$T$ by a rational gauge transformation, and look at $M$ in that gauge. We find from \eqref{eq:FlatConNew} that $M_{11}=0$, and we can set $M_{21}=1$. 
For left boundary conditions, we find similar eigenvalues $-\tau^{-1} \eta$ and $-\tau^{-1} \eta^{-1}$ and $M_{22}=0$ in the diagonal gauge for $T$.

Now we can apply S-duality on Neumann b.c. and derive the description of Nahm boundary conditions. 
If we set the eigenvalues of $M$ to $\alpha_1 = \mu \eta$ and $\alpha_2 = \mu \eta^{-1}$ for some $\mu$, we are clearly in a somewhat special situation, as some denominators in $T_{ij}$ go to zero. If we go back to \eqref{eq:FlatConNew} we see that $T_{21}$ is undetermined, and can be gauge-fixed to $1$. The other elements are given by the usual ansatz, but we need to set $u_2 v_1=0$.  If we choose $v_1 =0$, then $T_{11}=0$. This is a left Nahm boundary condition, the S-dual to the right Neumann b.c. If we choose $u_2 = 0$, then $T_{22}$ is zero. This is a right Nahm boundary condition, the S-dual to the left Neumann b.c. 
  
We can understand part of these boundary conditions classically. The Nahm pole forces one to combine gauge and R-symmetry rotations to preserve the boundary conditions. Hence the $SU(2)$ part of the Coulomb branch parameter should align with the R-symmetry mass parameter $\epsilon$. This is exactly the $\alpha_1 = \mu \eta$ and $\alpha_2 = \mu \eta^{-1}$ condition. The extra $T_{11}=0$ constraint is less obvious to understand physically. It closely resembles the Slodowy slice condition on ${\cal X}$ we reviewed in the previous section, but 
we do not understand why that should be the case: the field ${\cal X}$ is massive in $\ssN=2^*$ and is not obviously related to the 't Hooft loop generator $T$.
At most, we can point out that monopole operators can be brought successfully to a Nahm boundary condition \cite{Witten:2011zz} but they may differ from bulk operators in the spectrum of Abelian magnetic charges which they may carry on the Coulomb branch 
of the theory. This fact may ultimately explain the $T_{11}=0$ condition. It would be interesting to develop this point further. 

What about the S-dual of Dirichlet boundary conditions? The Dirichlet boundary conditions fix the Coulomb branch parameters $\alpha_i$ up to permutations and leave the momenta unconstrained. Dually, we need to impose a condition on the momenta such that the eigenvalues of $T$ are fixed. We already know how to do that: couple a $T[U(2)]$ theory to the gauge theory at the boundary, and thus set
\begin{equation}
p_\alpha^i = (-1)^{\delta_0} \tau_1 \frac{\eta \alpha_i - \sigma}{\eta \sigma - \alpha_i}\,.
\label{eq:pu2bif}
\end{equation}
We included the possibility of a sign redefinition for the $\tau_i$, analogous to the ones we introduced for 
3d gauge theories. In analogy to that, we will pick $\delta_0=1$.
This parameterization alone insures that one eigenvector of $T$ is $\tau_1$. 
The other eigenvalue
\begin{equation}
\tau_2 = \tau_1 \frac{(\eta \alpha_1 - \sigma)(\eta \alpha_2 - \sigma)}{(\eta \sigma - \alpha_1)(\eta \sigma - \alpha_2)}
\label{eq:2ndEigen}
\end{equation}
is fixed if we impose the Bethe equations for the XXZ chain on two sites with one Bethe root \eqref{eq:SL2BAE1r}. Independently of the Bethe equations, it is also automatically true that in a gauge where $T$ is diagonal, $M$ takes the form 
\begin{equation}
M_{ij} = \frac{\tilde u_i \tilde v_j}{\eta^{-1} \tau_i - \eta \tau_j}\,,
\end{equation}
where
\begin{equation}
p_\tau^1 = \tilde u_1 \tilde v_1 = \frac{\alpha_1 \alpha_2}{\sigma}\,,\quad p_\tau^2 = \tilde u_2 \tilde v_2 = \sigma\,. 
\end{equation}

We can summarize the geometric interpretation of these results. 
Inside the moduli space $\CM$ of vacua of the four-dimensional $U(2)$ gauge theory, i.e. the space of flat $GL(2)$ connections on the one-punctured torus, we can identify two natural families of Lagrangian submanifolds. The family $\CL_\mu$ fixes the A-cycle monodromy eigenvalues to specific $\mu_i$, and is associated to either left or right Dirichlet b.c. At the special values of the coordinates specified by\footnote{cf. \eqref{eq:Higgsrho1}} 
\begin{equation}
\alpha_1 = \eta^2 \alpha_2\,,
\label{eq:alignmenta}
\end{equation} 
the Lagrangian $\CL_\mu$ decomposes into two submanifolds, which correspond to a Nahm pole b.c. on the left and on the right respectively. The family $\CL_\tau$ fixes the B-cycle monodromy eigenvalues to specific $\tau_i$ and is associated to either left or right generalized Neumann b.c., which involve coupling to $T[U(2)]$. At special values of the parameters $\tau_1 = \eta^2 \tau_2$ 
the Lagrangian $\CL_\tau$ decomposes into two sub-manifolds, which correspond to left or right Neumann b.c. 

It is also useful to discuss the properties of the elementary domain walls associated to single fivebranes. 
We consider an NS5 domain wall between a $U(2)$ gauge theory and a $U(1)$ gauge theory \figref{fig:bcall}, i.e. bifundamental fields and opposite FI parameters on the two sides. Thus we parameterize the momenta on the $U(2)$ side as in \eqref{eq:pu2bif} but identify $\sigma$ with the Coulomb branch parameter of the bulk $U(1)$ theory, and 
\begin{equation} 
\tilde p_\sigma  = \tau_1^{-1} \prod_i \frac{\eta \sigma - \alpha_i}{\eta \alpha_i - \sigma}\,. \end{equation} 
These conditions fix one eigenvalue of $T_{U(2)}$ to $\tau_1$, while the other becomes $p_\sigma$, i.e. $T_{U(1)}$. The constraint from this domain wall is clearly S-dual to the constraint from the corresponding D5 domain wall, which fixes $M_{U(2)}$ to have a block-diagonal form, 
with one eigenvalue fixed, and the other identified with $M_{U(1)}$. 

We should also consider an NS5 domain wall between two $U(2)$ theories. If we plug the bifundamental contribution
\begin{equation}
p_\alpha^i = \tau \frac{\eta \alpha_i - \sigma_1}{\eta \sigma_1 - \alpha_i}\frac{\eta \alpha_i - \sigma_2}{\eta \sigma_2 - \alpha_i}\,.
\end{equation}
and corresponding contribution for $\tilde p_\sigma^n$, we obtain two $U(2)$ $T$ matrices, say $T_L$ and $T_R$
which have the same eigenvalues. With more work, it should be possible to show that this is dual to the appropriate D5 domain wall, i.e. 
the S-dual momenta in $M$ jump by the contribution of a fundamental hyper. It would be interesting to understand better the geometric meaning of this 
domain wall. We will not pursue the matter further, for $U(2)$ or at higher rank. 

\subsection{Boundary Conditions in the $U(N)$ Theory}\label{sec:bcUN}
It is straightforward to study the Neumann  b.c. in the general $U(N)$ gauge theory. 
If we set $p_\alpha^i=1$ in $T$, and pick an appropriate gauge choice for the $u_i$, $v_i$,
we can find explicit eigenvectors of the form $(\alpha_i^k)$ and identify the eigenvalues 
as $\tau \eta^{s}$, for $s$ taking values $(N-1, N-3, \cdots, 1-N)$. Because of the form of the eigenvalues, 
in the gauge where $T$ is diagonal, $M$ has a very simple form, with most elements equal to zero, 
except for the elements just below the diagonal, which can be set to $1$, and the last column, where we find the 
coefficients of the characteristic polynomial of $M$.   

By S-duality, we learn the conditions imposed by a full Nahm pole. The mixing of R-symmetry and gauge symmetry imposed by the pole restricts the Coulomb parameters to the form $\mu \eta^{N-1-2i}$. Then we must have $u_{i+1} v_i =0$ in $T$. Out of all the components of this locus, the right Nahm pole appears to choose $v_i=0$ for $i<N$. The left Nahm pole chooses $u_i=0$ for $i>0$. $T_{i+1,i}$ is undetermined, and can be gauge-fixed to any desired 
non-zero value. The only non-trivial elements are the last column for the right boundary condition or the first row for the 
left boundary condition.  

We have now enough information to make an educated guess for the boundary condition imposed by a general Nahm pole
associated to a general $\mathfrak{su}(2)$ embedding $\rho$. The embedding instructs us of how to constrain the Coulomb branch parameters: inside the $a$-th irreducible block of $\rho$, corresponding to a group of $r_a$ D3 branes ending on the $a$-th D5 brane in the brane setup \figref{fig:bcall}, the Coulomb branch parameters, and thus the eigenvalues of $M$, take the form $\mu_a \eta^{r_a-1-2i}$. The equations for $T$ then enforce $u_{i+1} v_i =0$ inside each block, and we pick the component with all necessary $v_i=0$ for right boundary conditions, $u_i=0$ for left boundary conditions. 

We can write these constraints in a familiar form: $T = \mathfrak{t}^+_\rho + T^*$, where $\mathfrak{t}^+_\rho$ is the raising generator of the $\mathfrak{su}(2)$ embedding and $T^*$ is built as usual from the $u_i$ and $v_j$. The Nahm pole requires $T^*$ to be a lowest weight vector for the $\mathfrak{su}(2)$
action. This is again analogous to the Slodowy slice condition from \cite{Gaiotto:2008ak}, although it takes place in the group manifold rather than the Lie algebra as in flat space, and involves a different set of variables. 

We expect the S-dual boundary condition to correspond to the coupling of the gauge theory to 
$T[U(N)]_\rho$, i.e. a 3d quiver gauge theory with $N$ flavors at the first node only, and gauge groups $N_i$, such that 
the linking numbers $(N - N_1, N_1 - N_2, \cdots)$ match the dimensions of irreducible blocks 
in $\rho$. In particular, this must mean that the parameterization of the momenta $p^\mu_a$ 
from the 3d gauge theory, inserted into $T$, together with the Bethe equations must enforce 
that the eigenvalues of $T$ will take the form $\tau_i \eta^{N-1-2i}$.
It must also enforce that in a gauge where $T$ is diagonal, only the appropriate elements of $M$ are non-zero. 

It is useful to think at the $T[U(N)]_\rho$ boundary condition as a sequence of domain walls, separating 
$U(N_i)$ four-dimensional gauge theories on segments, coupled at the interfaces to 3d bifundamental hypermultiplets. 
This presentation, corresponding to the field theory limit of the NS5 brane system, makes the recursive nature of the problem clear. In other words, we only need to understand the basic NS5 brane domain wall between 
consecutive segments and the S-duality relation to the basic D5 domain wall. 

In the following analysis, we will not keep track carefully of the signs associated to the $\delta_i$ shifts. 
We can start by coupling $N_1$ fundamental hypers at a Neumann boundary for the 4d theory, with FI $t_1$. The momenta set the boundary conditions to 
\begin{equation}
p_\alpha^i = \tau_1 \prod_{n=1}^{N_1} \frac{\eta \alpha_i - \sigma_n}{\eta \sigma_n - \alpha_i}\,.
\end{equation}
We expect the following two facts to be true.
First, $N-N_1$ eigenvalues of $T$ take values $\tau_1 \eta^{N-N_1-1-2i}$. We can put $T$ in a block-diagonal form 
\begin{equation}
g^{-1} T g=  \begin{pmatrix}\tau_1 \eta^{N-N_1-1-2i}  &0\cr 0 & \tau_1 T_\sigma \end{pmatrix}\,,
\end{equation}
with $T_\sigma$ coinciding with the Lax matrix for $U(N_1)$, with Coulomb branch parameters $\sigma_n$, and 
\begin{equation}
\tilde p_\sigma^n = \prod_{i=1}^{N}  \frac{\eta \sigma_n - \alpha_i}{\eta \alpha_i - \sigma_n}\,.
\end{equation}
Second, in such a gauge we should find 
\begin{equation}
g^{-1} M g = \begin{pmatrix} \mathfrak{t}^+ + a& b \cr c & M_\sigma \end{pmatrix}\,, 
\end{equation}
with $a$ of lowest weight under $\mathfrak{t}^+$ and $M_\sigma$ diagonal with $\sigma$ eigenvalues. 
 
We will sketch here the derivation of these facts for the case $N-N_1=1$. The generalization is straight forward. 
It is useful to proceed backwards. 
We seek a matrix $g$ which conjugates $T$ to a $(1,N-1)$ block-diagonal form 
\begin{equation}
g^{-1} M g = \begin{pmatrix} p^\xi & a^T \cr b & M' \end{pmatrix}\,, \qquad g^{-1} T g=  \begin{pmatrix} \xi &0\cr 0 & T' \end{pmatrix}\,.
\end{equation}
The flatness condition on $M$ and $T$ implies the corresponding rank $N-1$ flatness constraint on $M'$ and $T'$, together with the equations for the components $(u_1, u')$ of $u$ and $(v_1, v')$ of $v$
\begin{align}
(\eta^{-1} - \eta)\xi p^\xi = u_1 v_1 \cr
\eta^{-1} \xi a^T - \eta a^T T' = u_1 (v')^T \cr
\eta^{-1} T' b - \eta \xi b =u' v_1\,. 
\end{align}
These equations determine $a$ and $b$ and $u_1 v_1$ in terms of $\xi$, $p^\xi$ and the solution to the flatness constraint. 
The gauge transformation left are gauge symmetry $GL(1) \times GL(N-1)$, and the residual $GL(1)$ gauge symmetry eliminates the ratio between $u_1$ and $v_1$. Thus if we can find the gauge transformation $g$ which makes $T$ block-diagonal, we would have reduced the rank $N$ problem to the rank $N-1$ problem. We parameterize the solution of the rank $N-1$ problem in terms of the eigenvalues $\sigma_i$ of $M'$ and the momenta $p_\sigma^i$ which enter the ansatz for $T'$. 

We have the linear equations for $g$, decomposed as $g = (g_1, g')$ ($g_1$ is the first column of the matrix and $g'$ is formed out of the $N-1$ others)
\begin{equation}
(M g_1, Mg') = ( p^\xi g_1 + g' b , a^T g_1 +g' M' )\,, \qquad (T g_1,T g')=  (\xi g_1 ,g' T' )\,.
\label{eq:MTdecomp}
\end{equation}
In particular from the second equation we discover that $g_1$ is an eigenvector of $T$, with eigenvalue $\xi$ and 
\begin{equation}
\sum\limits_{b=1}^N T_{ab} g'_{bi} = \sum\limits_{j=1}^{N-1}g'_{aj}T'_{ji}\,,
\label{eq:RedT}
\end{equation}
where we can parameterize
\begin{equation}
g'_{ai} = \frac{g_{a1}a_i }{\alpha_a - \sigma_i}\,, \quad a=1,\dots,N\,,\quad i=1,\dots, N-1\,.
\end{equation}
Using a simple gauge for $T$ and for $T'$,  with $\alpha_a$s replaced by $\sigma_i$s and $p_\alpha^a$s replaced by $p_\sigma^i$, we can rewrite \eqref{eq:RedT} as follows
\begin{align}
&\sum\limits_{b=1}^N p_\alpha^a \alpha_a \frac{\eta^{-1} - \eta}{\alpha_a \eta^{-1} - \alpha_b \eta}\prod_{c \neq a}^N \frac{\alpha_a \eta^{-1}-\alpha_c \eta}{\alpha_a - \alpha_c} \frac{g_{b1}a_i}{\alpha_b - \sigma_i} \notag \\
&= \sum\limits_{j=1}^{N-1} \frac{g_{a1} a_j}{\alpha_a - \sigma_j} p_\sigma^j \sigma_j \frac{\eta^{-1} - \eta}{\sigma_i \eta^{-1} - \sigma_j \eta}\prod\limits_{k \neq j}^{N-1} \frac{\sigma_j \eta^{-1}-\sigma_k \eta}{\sigma_j - \sigma_k}\,.
\label{eq:rankreduction}
\end{align}
One can verify that the following parameterization of $p_\alpha^a$ and $p_\sigma^i$ 
\begin{equation}
p_\alpha^a = \tau_1\prod_{k=1}^{N-1}\frac{\alpha_a - \eta^{-1} \sigma_k}{\eta^{-1} \alpha_a - \sigma_k}\,, \qquad p_\sigma^i = \tau_1\prod_{b=1}^N\frac{\eta^{-1} \sigma_i- \alpha_b}{ \sigma_i-\eta^{-1} \alpha_b}\prod_{k\neq i}^{N-1}\frac{\eta \sigma_i - \eta^{-1} \sigma_k}{\eta^{-1} \sigma_i - \eta\sigma_k}\,,
\label{eq:pmupsigmasub}
\end{equation}
allows us to solve \eqref{eq:rankreduction} for appropriate $a_i$ and $g_{a1}$. In this case from \eqref{eq:pmupsigmasub} we conclude that
\begin{equation}
g_{a1} =\sum\limits_{k=1}^{N-1}g_k \frac{\alpha_a(\alpha_a-\sigma_k)}{\eta^{-1}\alpha_a-\sigma_k}\,,
\end{equation}
where $g_k$ are some constants. At the next step of the recursion, the next domain wall/boundary set $p_\sigma^n$ to appropriate values 
\begin{equation}
p_\sigma^n = \tau_2 \prod_{n'=1}^{N_2} \frac{\eta \sigma_n - \sigma^{(2)}_{n'}}{\eta \sigma^{(2)}_{n'} - \sigma_n}\,,
\end{equation}
and we get the level one Bethe equations for the $\sigma_n$ roots. 
Recursive application of this reasoning show that the $T[U(N)]_\rho$ boundary condition is S-dual to a Nahm pole $\rho$,
and in particular $T[U(N)]$ is dual to Dirichlet b.c., and implements S-duality. 

It is interesting to verify the expected properties of $T$ by looking at the 't Hooft loop $\mathrm{Tr} T$ and 
plugging in the appropriate expressions for the momenta. The desired form of the answer follows from some neat 
rational function identities which we review in \appref{sec:GLnFlatApp}. \footnote{These identities were first derived 
in the course of a project on line defects for the 2d Toda CFT, in collaboration with  Bruno Le Floch and Jaume Gomis}

\subsection{Putting the Pieces Back Together: the Vacua of $T[U(Q)]_\rho^{\rho^\vee}$ and the XXZ/tRS duality}
At this point, we are ready to formulate a geometric description of the vacua of the linear quiver gauge theories 
through the $T[U(Q)]_\rho^{\rho^\vee}$ realization, and thus of the solutions of the Bethe equations for the generic 
XXZ spin chain with antisymmetric spins. Remember that $Q = \sum_i M_i$.

\begin{figure}
\begin{center}
\includegraphics[height=3.5cm, width=9.5cm]{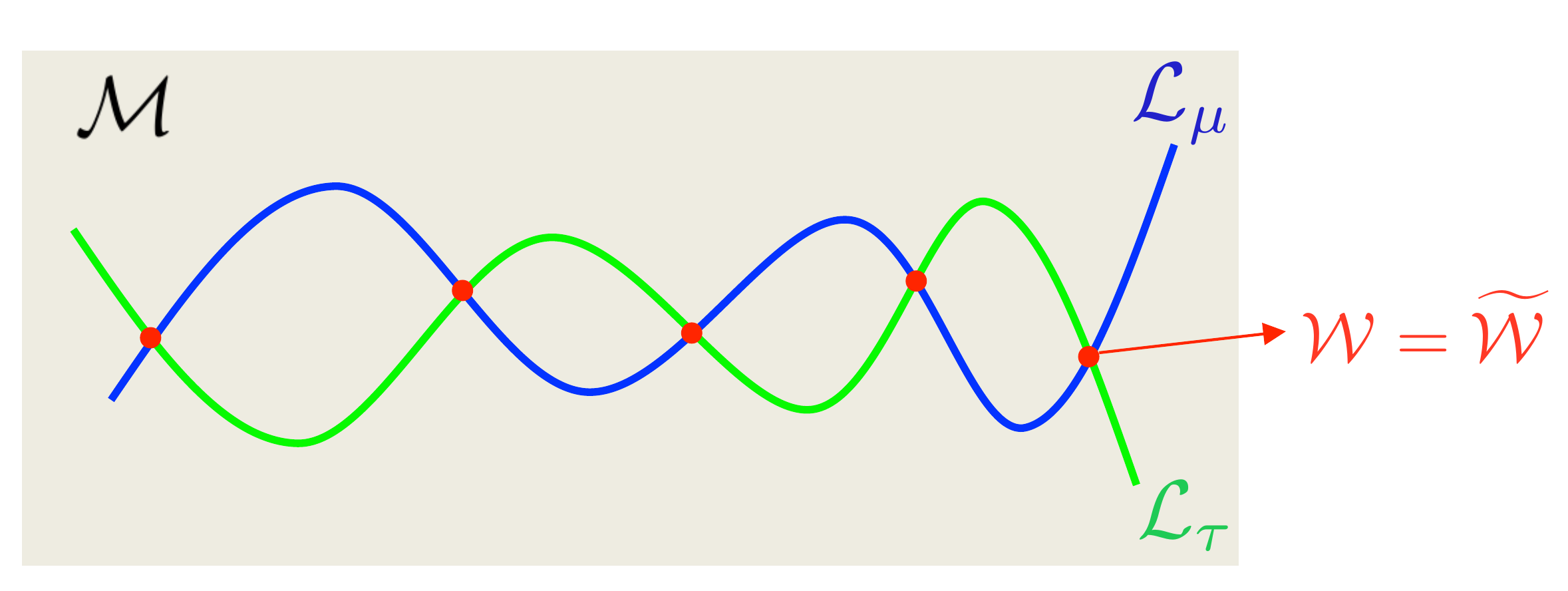}
\caption{Two Lagrangian submanifolds $\CL_{L,\mu}^{\rho}$ and $\CL_{R,\tau}^{\rho^\vee}$ intersect at loci which coincide with the moduli space of vacua for the corresponding $T[U(Q)]_\rho^{\rho^\vee}$ theory. The effective twisted superpotential $\CW$ for the XXZ chain and its mirror dual $\CW^\vee$ coincide at those loci.}
\label{fig:lagrint}
\end{center}
\end{figure}

We have the four-dimensional $U(Q)$ theory on a segment, with $\rho$ Nahm pole on the left and $\rho^\vee$ generalized Neumann b.c. on the right \figref{fig:TSUNrrdbrane}. The left b.c. give a Lagrangian manifold $\CL_{L,\mu}^\rho$ in $\CM$, which fixes the eigenvalues of $M$ in terms of the masses and $\rho$ and imposes a Slodowy structure on $T$. 
The right b.c. gives Lagrangian manifold $\CL_{R,\tau}^{\rho^\vee}$ which fixes the eigenvalues of $T$ 
in terms of the FI parameters and $\rho^\vee$ and imposes a Slodowy structure on $M$.
The vacua of the $T[U(Q)]_\rho^{\rho^\vee}$ theory are the intersection points of the two Lagrangian submanifolds \figref{fig:lagrint}.
\begin{equation}
\CL = \CL_{L,\mu}^{\rho}\cap\CL_{R,\tau}^{\rho^\vee}\,.
\label{eq:LeqInt}
\end{equation}
It is instructive to summarize this correspondence in the following table \tabref{tab:quiverN2ast}.
\begin{table}[!h]
\begin{center}
\begin{tabular}{|c|c|}
\hline
3d $\CN=2^*$ $A_L$ quiver  & 4d $U(Q)$ $\CN=2^*$ SYM \\
gauge theory & on segment with $\half$ BPS b.c. \\
\hline \hline
Moduli space of vacua & Intersection of Lagrangians \\
of a quiver theory $\CL$ & $\CL_{L,\mu}^{\rho}\cap\CL_{R,\tau}^{\rho^\vee}$   \\
\hline
Twisted masses $\mu_i$  &  Eigenvalues of $M$ \\
\hline 
Complexified FI parameters $\tau_a$  &  Eigenvalues of $T$ \\
\hline
Twisted mass for $U(1)_\epsilon$ R-symmetry & Eigenvalue of $E$  \\
\hline
Color and flavor labels & Embeddings $\alg{su}(2)\hookrightarrow\alg{u}(Q)$   \\
$(N_i,M_i)$  &  $\rho$ and $\rho^\vee$ \\
\hline
\end{tabular}\label{tab:quiverN2ast}
\caption{The duality table between quiver gauge theories and segment compactifications of SYM theories.}
\end{center}
\end{table}
We shall promptly discuss the consequences of relation \eqref{eq:LeqInt} in the language of integrable systems.

\section{Applications to Integrable Systems}\label{Sec:IntAppl}
\begin{figure}[!h]
\begin{center}
\includegraphics[height=5cm, width=11.5cm]{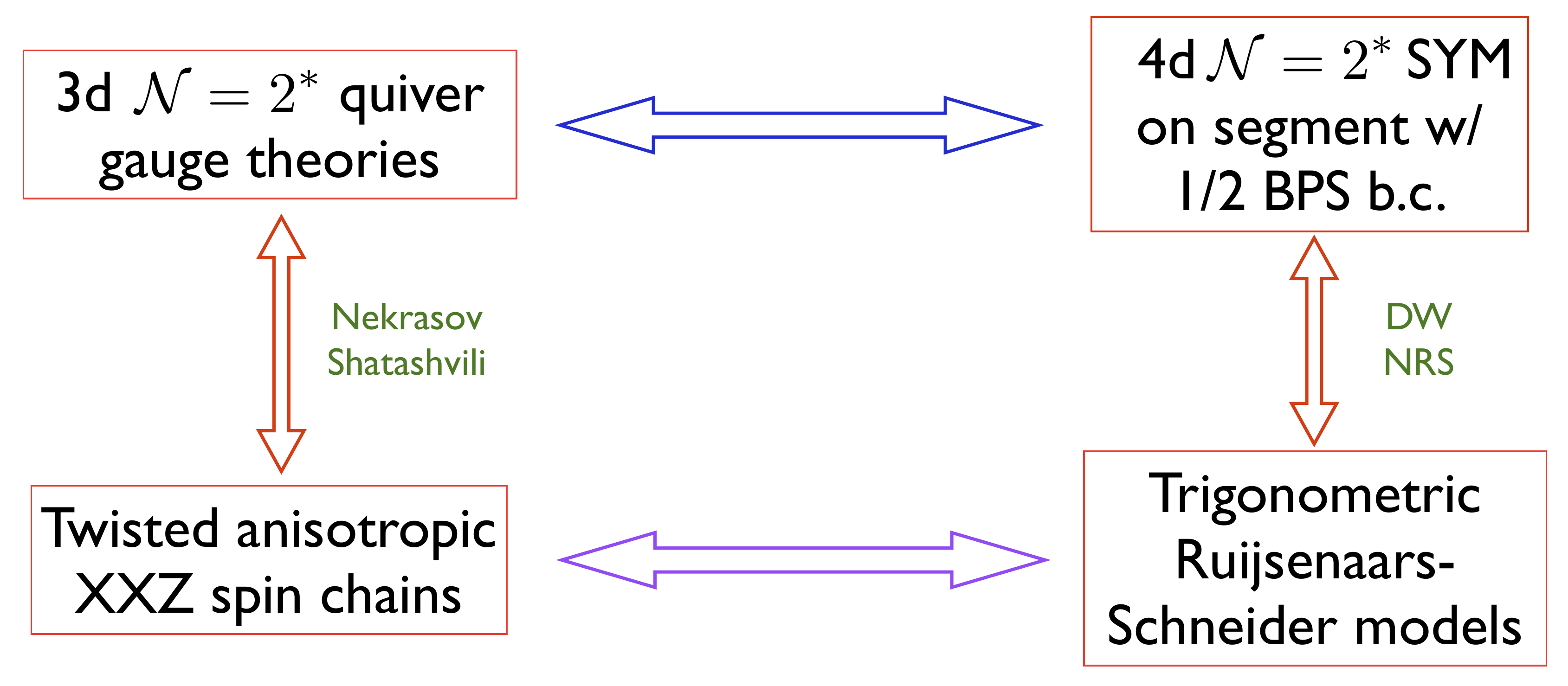}
\caption{Dualities between tRS and XXZ via equivalences of vacua moduli spaces of gauge theories. Abbreviations on the vertical right arrow stand for the contributions of Donagi-Witten \cite{Donagi:1995cf} and Nekrasov-Rosly-Shatashvili \cite{Nekrasov:2011bc}.}
\label{fig:tRSXXZfrom3d}
\end{center}
\end{figure}

In the last couple of decades dualities between various integrable systems have been discussed extensively \cite{1996alg.geom.12001G, 2009arXiv0906, 2012arXiv1201.3990M,MR2729945, MR851627, MR887995,2009JPhA...42r5202F,2010JMP....51j3511F,2011CMaPh.301...55F,2012NuPhB.860..464F}. The network of dualities between various integrable systems we are about to present widely generalizes results from the literature. 
In the main text we have connected XXZ spin chains and tRS models in a rich circle of dualities. See figure \figref{fig:tRSXXZfrom3d} for a sketch of the gauge theory origin of these dualities.
We can summarize it as follows. A reasonable starting point is the Lax matrix description of the tRS model: the Hamiltonians of the tRS
model are built from the positions $\alpha_i$ and the momenta $p_\alpha^i$ by taking traces of powers of the Lax matrix $T$ described by \eqref{eq:lax1} and \eqref{eq:lax2}.
The Lax matrix and the diagonal matrix $M$ built out of the $\alpha_i$ satisfy the flatness condition \eqref{eq:SFlatCon}, which treats $M$ and $T$ in a symmetric fashion (up to $\eta \to -\eta^{-1}$). 

This suggests a natural question: how do we map into each other the phase spaces of the original tRS model, and of the S-dual tRS$^\vee$ model which is defined by a gauge transformation to a basis where $T$ is diagonal? Our analysis gives a surprising answer to this question: this $\CL_S$ Lagrangian submanifold in the product of the two phase spaces $\CM \times \CM^\vee$ 
coincides with the moduli space of the $T[U(Q)]$ theory (for $Q$ particles in the tRS model), i.e. with the solution of Bethe equations for an XXZ $SU(Q)$ spin chain with $Q$ fundamental 
spins, in a weight zero sector. More precisely, the Lagrangian submanifold admits two distinct (S-dual) descriptions in terms of a pair of identical XXZ spin chains. 
In one description the positions $\alpha_i$ are impurities in the spin chain, and the S-dual positions are twists, in the other description they play an opposite role. 
In either case, the Yang-Yang functional $\CW$ \eqref{eq:Efgen} for the spin chain plays the role of generating function for the Lagrangian submanifold, and thus encodes the conjugate momenta as functions of the positions. 

This question admits a generalization. We could define a {\it restricted} tRS model, labelled by a general $\mathfrak{su}(2)$ embedding $\rho$.
In the restricted model, the positions are subdivided into as many blocks as irreducible blocks in $\rho$, and constrained to take the form $\mu_a \eta^{r_a - 1 - 2 i}$ inside 
a block of size $r_a$, with $i = 0 \dots r_a-1$. This subset of the phase space consists of several components, and we pick a specific one by restricting the Lax matrix $T$
 to take the Slodowy form, i.e. to differ from the raising operator of $\rho$ by a lowest weight vector for the 
$\mathfrak{su}(2)$ action. The resulting Lax matrix still depends on $Q$ extra degrees of freedom, which include the conjugate momenta to the $\mu_a$ 
variables. In other words, the restricted tRS model corresponds to a specific Lagrangian submanifold $\CL_\rho$ in the product of the original tRS phase space, and the 
phase space $\CM_\rho$ defined by the $\mu_a$ and their conjugate momenta. 

If we consider restrictions $\rho$ of the tRS model and and $\rho^\vee$ of the S-dual tRS$^\vee$ model, with parameters $\mu_a$ and $\tau_i$, 
and we combine (intersect) the Lagrangian correspondences $\CL_\rho$, $\CL_S$ and $\CL_{\rho^\vee}$, we arrive to a Lagrangian manifold 
$\CL_\rho^{\rho^\vee}$, which describes pairs of matrices $M$, $T$ which satisfy both restrictions. Our claim is that this manifold 
coincides with the moduli space of the $T[U(Q)]_\rho^{\rho^\vee}$ theory, i.e. with the solutions of Bethe equations for an XXZ spin chain of impurities $\mu_a$ 
and twists $\tau_i$, with representation content encoded by $\rho$, in a sector of weight encoded by $\rho^\vee$. 
It will of course also coincide with the space of solutions of the bispectral dual spin chain. See \tabref{tab:tRSXXZ} for a summary of the relations between parameters of the models. 


\begin{table}[!h]
\begin{center}
\begin{tabular}{|c|c|}
\hline
$SU(L+1)$ XXZ spin chain & $GL(Q)$ tRS model \\
\hline \hline
Impurities $\mu_i$  &  Eigenvalues of $M$ \\
\hline
Twists $\tau_a$  &  Eigenvalues of $T$ \\
\hline
Anisotropy parameter &  \\
(quantum deformation) &  Eigenvalue of $E$ \\
\hline
Sector of the Hilbert space of the spin chain &  Patterns of degeneracy of   \\
and representation $\CR$ & the eigenvalues of $M$ and $T$  \\
\hline
Parameter space of solutions & Intersection of Lagrangian  \\
of the Bethe equations &  submanifolds \\
\hline
\end{tabular}\label{tab:tRSXXZ}
\end{center}
\caption{Dualities between XXZ and tRS models.}
\end{table}

\subsection{Interesting limits of the XXZ bispectral duality}
\begin{figure}[!h]
\begin{center}
\includegraphics[height=6cm, width=14cm]{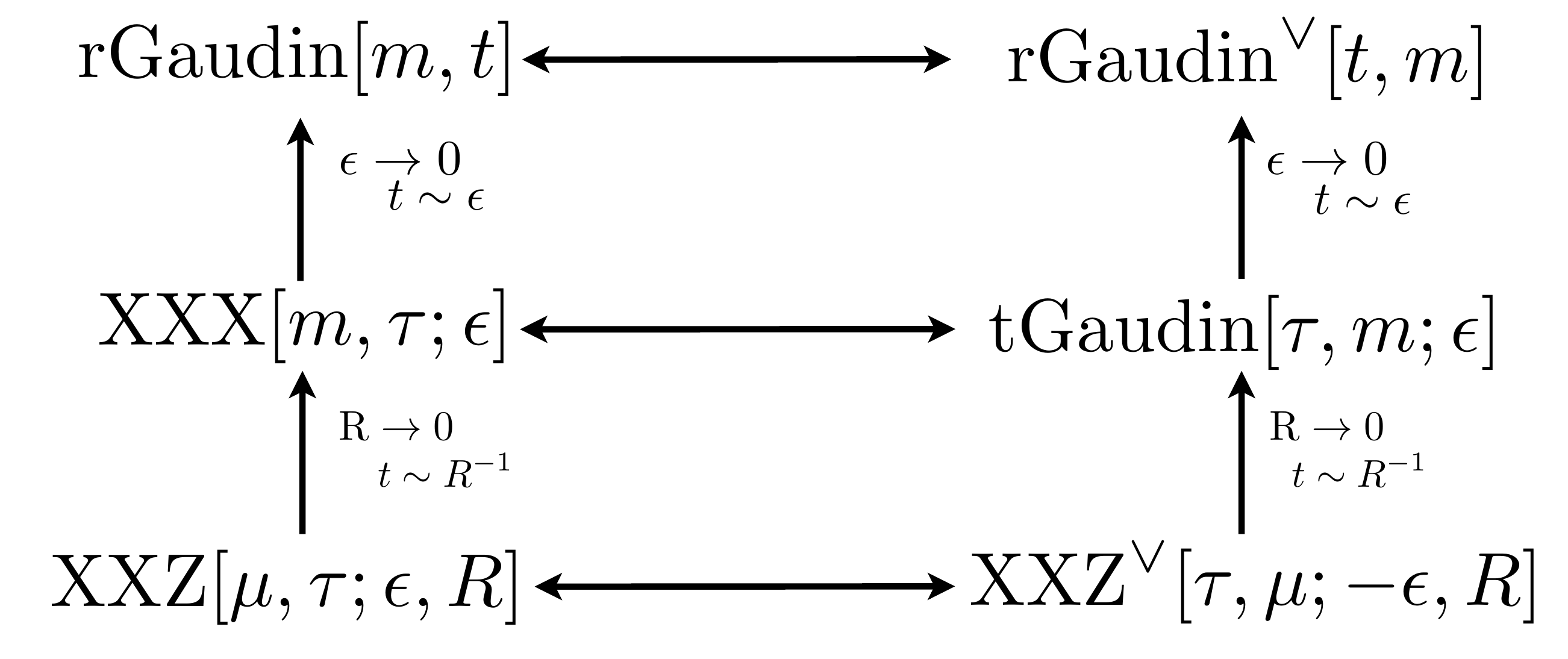}
\caption{Network of dualities between quantum models emerging from XXZ chains and their mirror duals. These models are the XXX chain and trigonometric and rational Gaudin models (tGaudin and rGaudin). The sets of parameters each model depends on are shown in brackets. }
\label{fig:XXZduals}
\end{center}
\end{figure}

In this section, we will take careful $R \to 0$ and $\epsilon \to 0$ limits of the Bethe equations 
for the bispectral dual pairs of XXZ spin chains \cite{Drinfeld:1986in, Jimbo:1985kc}. At the first step, we will send $R$ to zero, keeping the
scaling the other parameters in such a way that the $\tau$ parameters remain fixed, but $\mu_i \sim \exp 2 \pi R m_i$ and $\eta \sim \exp \pi R \epsilon$.
On one side of the duality, the XXZ spin chain Bethe equations will reduce to XXX spin chain Bethe equations \cite{Heisenberg:1928aa, Bethe:1931hc}. 
This corresponds to a standard 2d limit of the corresponding gauge theories. 
On the other side of the duality, the Bethe equations for the XXZ spin chain will reduce to the Bethe equations for a trigonometric Gaudin system \cite{refId0,Feigin:1994in}.
Thus we will discover a general bispectral duality statement between XXX spin chains and tGaudin systems. 
From there we take another limit $\epsilon\to 0$, with $m_i$ fixed and $\tau_a \sim \exp \epsilon t_a$. The result of this procedure is a pair of rational Gaudin systems,
which should still be bispectral dual. 
The conjugate momenta and thus the Yang-Yang functional will have nice limits as well, and will still 
coincide under the proposed bispectral dualities. 
In \eqref{fig:XXZduals} we show the network of proposed dualities.

It may be useful to review briefly the Gaudin models and their Bethe equations.
The rational Gaudin model with symmetry group $G$ with $L$ spins is a simple example of the Hitchin system on $S^2$ with $L$ marked points $z_1,\dots, z_L$ \cite{Nekrasov:1995nq}. The trigonometric Gaudin model includes two more punctures at zero and at infinity. At each puncture we fix representations $V(\nu_1),\dots,V(\nu_L)$ of $\alg{g}$ with weights $\nu_a,\, a =1,\dots, L$. Now the task is to diagonalize Gaudin Hamiltonians which read
\begin{equation}
\mathcal{H}_a=\sum\limits_{b\neq a}\sum\limits_{\alpha=1}^{\text{dim}(\alg{g})}\frac{\mathfrak{J}_\alpha^{(b)}\mathfrak{J}^{\alpha\,(b)}}{z_a-z_b}\,,
\end{equation}
where $\mathfrak{J}_\alpha^{(b)}$ of the acts with $\mathfrak{J}_\alpha\in\alg{g}$ on the $b$-th spin and with identity on the others. 
Let us for simplicity take $\alg{g}=\alg{su}(2)$. Then diagonalization of Gaudin Hamiltonians leads to the system Bethe ansatz equations for the sector with some $\kappa$ Bethe roots $\zeta_i$ 
\begin{equation}
\sum_{b=1}^L \frac{\nu_b\epsilon}{\zeta_i-z_b} -\sum_{\substack{j=1\\ j\neq i}}^{\kappa} \frac{2\epsilon}{\zeta_i-\zeta_j}=0, \,\quad i=1,\dots,\kappa\,.
\label{eq:RationalGaudGen}
\end{equation}
The Bethe equations for tGaudin include an extra term corresponding to the extra puncture at $z_0=0$ (see \cite{2006JSMTE..08..002M} and references therein for more details). 

\subsubsection{XXX/tGaudin duality}\label{sec:XXXtGaud}
The Bethe equations for the XXX chain can be obtained as the $R\to 0$ limit of \eqref{eq:XXZGen} 
\begin{align}
&\frac{\tau_{j+1}}{\tau_{j}}\cdot\prod_{a=1}^{M_j}\frac{s^{(j)}_n-m^{(j)}_a+\sfrac{\epsilon}{2}}{-s^{(j)}_n+m^{(j)}_a+\sfrac{\epsilon}{2}}\cr
&\cdot\prod_{n'=1}^{N_{j-1}}\frac{s^{(j)}_n- s^{(j-1)}_{n'}+\sfrac{\epsilon}{2}}{-s^{(j)}_n+ s^{(j-1)}_{n'}+\sfrac{\epsilon}{2}} \cdot \prod_{n'}^{N_{j}}\frac{s^{(j)}_n - s^{(j)}_{n'}-\epsilon}{-s^{(j)}_n + s^{(j)}_{n'}-\epsilon} \cdot \prod_{n'=1}^{N_{j+1}}\frac{s^{(j)}_n - s^{(j+1)}_{n'}+\sfrac{\epsilon}{2}}{-s^{(j)}_n + s^{(j+1)}_{n'}+\sfrac{\epsilon}{2}}=(-1)^{\delta_j}\,, 
\label{eq:XXXGen}
\end{align}
where $j=1,\dots,L^\vee-1$\footnote{Notice the slight change of the notation compared to (\ref{eq:XXZGen}): $L$ is replaced with $L^\vee$. For the purposed of this section the latter choice is more symmetric.} runs through the elements of the Cartan subalgebra of $A_{L^\vee-1}$. 

In order to get tGaudin Bethe equations we start with the set which is mirror dual to \eqref{eq:XXZGen}. According to the mirror symmetry prescription parameters $(N_i, M_i)$ of the original spin chain are replaced by the corresponding set $(N^\vee_i, M^\vee_i)$ (see \eqref{eq:dualweights} for the direction to derive the data of the dual model, for explicit examples of dual sets $(N_i, M_i)$ and $(N^\vee_i, M^\vee_i)$  see \secref{sec:BispecExamples}). With the usual mirror map, the mirror XXZ$^\vee$ equations will have the following form
\begin{align}
&\frac{\mu_{j+1}}{\mu_{j}}\prod_{n'=1}^{N^\vee_{j-1}}\frac{\tilde\sigma^{(j)}_n+ \eta^{-1}\tilde\sigma^{(j-1)}_{n'}}{\tilde\sigma^{(j-1)}_{n'}+\eta^{-1}\tilde\sigma^{(j)}_n} \cdot \prod_{n'}^{N^\vee_{j}}\frac{-\eta^{-1} \tilde\sigma^{(j)}_n + \eta \tilde\sigma^{(j)}_{n'}}{-\eta^{-1}\tilde\sigma^{(j)}_{n'}+\eta \tilde\sigma^{(j)}_n} \cr
&\cdot \prod_{n'=1}^{N^\vee_{j+1}}\frac{\tilde\sigma^{(j)}_n + \eta^{-1}\tilde\sigma^{(j+1)}_{n'}}{\tilde\sigma^{(j+1)}_{n'}+\eta^{-1}\tilde\sigma^{(j)}_n}\cdot\prod_{k=1}^{M^\vee_j}\frac{\tilde\sigma^{(j)}_n+\eta^{-1}\tau^{(j)}_k}{\tau^{(j)}_k+\eta^{-1} \tilde\sigma^{(j)}_n}=(-1)^{\delta^\vee_j}\,, 
\label{eq:XXZGenMirr}
\end{align}
where $j=1,\dots L-1,\, n=1,\dots,N^\vee_j$. 

Let us look carefully at the $R\to 0$ limit of \eqref{eq:XXZGenMirr} with the proper scaling of twists $Rt_a=\hat{t}_a$. The l.h.s. of \eqref{eq:XXZGenMirr} in the limit we are discussing is a product of some rational functions of $\tilde\sigma, m_i,\epsilon,\tau_j$. By doing the Taylor expansion at small $R$ it is easy to realize that \eqref{eq:XXZGenMirr} will have the following schematic form
\begin{equation}
(-1)^{N^\vee_j+1}+R(\dots)+O(R^2) = (-1)^{\delta^\vee_j}\,,
\end{equation}
Thanks to our choice $\delta^\vee_j = N^\vee_j+1$ \eqref{eq:deltas}, the order $1$ term cancels out, and we are left with the order $R$ coefficients,  
\begin{align}
&\frac{m_{j+1}-m_j- \frac{1}{2}(\Delta^\vee_{j}+2)\epsilon}{\tilde\sigma^{(j)}_n}+\sum_{n'=1}^{N^\vee_{j-1}}\frac{\epsilon}{\tilde\sigma^{(j)}_n+ \tilde\sigma^{(j-1)}_{n'}}\cr
&- \sum_{n'\neq n}^{N^\vee_{j}}\frac{2 \epsilon}{\tilde\sigma^{(j)}_n - \tilde\sigma^{(j)}_{n'}} +\sum_{n'=1}^{N^\vee_{j+1}}\frac{\epsilon}{\tilde\sigma^{(j)}_n + \tilde\sigma^{(j+1)}_{n'}}+\sum_{k=1}^{M^\vee_j}\frac{\epsilon}{\tilde\sigma^{(j)}_n+\tau^{(j)}_k}=0\,, 
\label{eq:BAEGaudinslNGen}
\end{align}
which will give us to the corresponding Gaudin Bethe equations, up to a sign re-definition 
\begin{equation}
\tau^{(j)}_k \to (-1)^j \tau^{(j)}_k\,, \qquad \tilde\sigma^{(j)}_n \to (-1)^{j+1} \tilde\sigma^{(j)}_n\,.
\end{equation}
Thus the bispectral duality states that the solutions of 
\eqref{eq:XXXGen} and \eqref{eq:BAEGaudinslNGen} are in one to one correspondence to each other. 
The correspondence will still relate the Yang-Yang functions evaluated on the solutions, as long as we take appropriate limits of the $\ell(x)$ functions \eqref{eq:Wchiralell}, which are main building blocks of the Yang-Yang functions, or, better to say, of the expressions for the conjugate momenta to masses and twists. Thus momenta conjugate to FI parameters \eqref{eq:ptaugen} on the XXX side behave as $p_\tau^j = \exp 2 \pi R P_\tau^j$ with 
\begin{equation}
P_\tau^j = \sum_n s^{(j-1)}_n- \sum_n s^{(j)}_n+ \sum_{k \geq j} \sum_a m_a^{(k)}\,.
\label{eq:PtauXXX}
\end{equation}
Momenta conjugate to masses \eqref{eq:pmugen} have a finite limit 
\begin{equation}
p_\mu^{(j),a}= \prod_{k=1}^j \tau_k \cdot  \prod_{n=1}^{N_j} \frac{m_a^{(j)} - s_n^{(j)}+ \sfrac{\epsilon}{2}}{s_n^{(j)} - m_a^{(j)}+ \sfrac{\epsilon}{2}}\,.
\label{eq:pmuXXX}
\end{equation}
On the mirror Gaudin side we have momenta conjugate to masses \eqref{eq:ptaugenMir} which have a finite limit\footnote{Here and in the following formula we omit tildes for the mirror momenta.}
\begin{equation}
p_\mu^j = \frac{\displaystyle\prod\limits_{n=1}^{N^\vee_{j-1}} \tilde\sigma^{(j-1)}_n}{\displaystyle\prod\limits_{n=1}^{N^\vee_j} \tilde\sigma^{(j)}_n} \prod_{k \geq j}^{L-1} \prod_{a=1}^{M^\vee_k} \tau_a^{(k)}\,.
\label{eq:pmutG}
\end{equation}
The momenta conjugate to FI parameters \eqref{eq:pmugenMir} behave as $p_\tau^{(j),a} = \exp 2 \pi R P_\tau^{(j),a}$ with 
\begin{equation}
P_\tau^{(j),a}= \sum_{k=1}^j m_k +\frac{\epsilon}{2}\sum_{n=1}^{N^\vee_j} \frac{\tilde\sigma_n^{(j)}-\tau_a^{(j)}}{\tilde\sigma_n^{(j)}+\tau_a^{(j)}}\,.
\label{eq:PtautGaudin}
\end{equation}

\subsubsection{rGaudin/rGaudin duality}
At the next step of our limiting procedure (see \figref{fig:XXZduals}), in order to derive the bispectral duality for the rational models, 
we need to take an $\epsilon \to 0$ limit. We will need to redefine appropriately the 
signs of the $\tau_i$ parameters before we take the limit. We will write
\begin{equation}
\tau_i = (-1)^{r^\vee_i} e^{\epsilon t_i}
\end{equation}
for some new renormalized $t_i$ parameters which are kept fixed in the limit. 
 On the XXX side the sign choice guarantees the cancellation of the order $1$ terms and leaves us with the following set of Bethe equations for the rational Gaudin model
 \begin{align}
&t_{j+1}- t_j + \sum_{a=1}^{M_j}\frac{1}{s^{(j)}_n-m^{(j)}_a}\cr
&+\sum_{n'=1}^{N_{j-1}}\frac{1}{s^{(j)}_n- s^{(j-1)}_{n'}}- \sum_{n'\neq n}^{N_{j}}\frac{2 }{s^{(j)}_n - s^{(j)}_{n'}}+ \sum_{n'=1}^{N_{j+1}}\frac{1}{s^{(j)}_n - s^{(j+1)}_{n'}}=0\,, 
\label{eq:XXXlim}
\end{align}
On the mirror side, the $\tau_i$ parameterization arises naturally as
\begin{equation}
\tau_n^{(i)} = (-1)^{i} e^{\epsilon t^{(i)}_n}\,,
\end{equation}
and we take the limit of \eqref{eq:BAEGaudinslNGen} making sure that 
 \begin{equation}
\tilde\sigma_n^{(i)} = (-1)^{i+1} e^{\epsilon \tilde s_n^{(i)}}.
\end{equation}
We get
\begin{equation}
m_{j+1}-m_j+\sum_{n'=1}^{N^\vee_{j-1}}\frac{1}{\tilde s^{(j)}_n-\tilde s^{(j-1)}_{n'}}- \sum_{n'\neq n}^{N^\vee_{j}}\frac{2}{\tilde s^{(j)}_n - \tilde s^{(j)}_{n'}} +\sum_{n'=1}^{N^\vee_{j+1}}\frac{1}{\tilde s^{(j)}_n -\tilde s^{(j+1)}_{n'}}+\sum_{k=1}^{M^\vee_j}\frac{1}{\tilde s^{(j)}_n-t^{(j)}_k}=0\,.
\label{eq:BAEGaudinslNGenlim}
\end{equation}
At this point we have reached a statement about bispectral duality of the rational Gaudin model. 
The correspondence will relate the Yang-Yang functionals evaluated on the solution, as long as we 
take appropriate limits of the superpotentials or, better to say, of the conjugate momenta. 
On the XXX side the FI momenta remain unchanged 
\begin{equation}
P_\tau^j = \sum_n s^{(j-1)}_n- \sum_n s^{(j)}_n+ \sum_{k \geq j} \sum_a m_a^{(k)}\,,
\end{equation}
whereas momenta conjugate to masses \eqref{eq:pmuXXX} behave in the limit as $p_\mu^{(j),a} \sim (-1)^{\lambda_j}\exp \epsilon P_\mu^{(j),a}$, with 
\begin{equation}
P_\mu^{(j),a}= \sum_{k=1}^j t_k+ \sum_n \frac{1}{s_n^{(j)}-m_a^{(j)}}\,,
\label{eq:PmurGaudin}
\end{equation}
where 
\begin{equation}
\lambda_j=N_j + \sum_{k=1}^j r_k^\vee.
\label{eq:lambdaj}
\end{equation}

On the mirror Gaudin side we have momenta conjugate to masses \eqref{eq:pmutG} to behave in the limit as $p_\mu^{(l),a} \sim (-1)^{\tilde \lambda^{(l),a}}\exp \epsilon P_\mu^{(l),a}$, with
\begin{equation}
P_\mu^{(l)} = \sum_{n=1}^{N^\vee_{l-1}} \tilde s^{(l-1)}_n-\sum_{n=1}^{N^\vee_l} \tilde s^{(l)}_n+\sum_{k \geq l} \sum_a t_a^{(k)}\,,
\end{equation}
where the sign factor is given by
\begin{equation}
\tilde \lambda^{(l),a}=l N^\vee_{l-1} - (l+1)N^\vee_l+ \sum\limits_{k \geq l}k M^\vee_{k}\,,
\end{equation}
and coincides with the appropriate $\lambda_j$ \eqref{eq:lambdaj} with help of the S-duality formulae we used in the main text. 
The momenta conjugate to FI parameters \eqref{eq:PtautGaudin} have a finite limit
\begin{equation}
P_\tau^{(l),a}= \sum_{k=1}^l m_k +\sum_{n=1}^{N^\vee_l} \frac{1}{\tilde s_n^{(l)}-t_a^{(l)}}\,.
\end{equation}
At this point the mirror symmetry between the rGaudin systems becomes completely obvious.

The Bethe equations for rGaudin appeared in \cite{Gaiotto:2011nm} in the study of irregular conformal blocks of the Virasoro algebra (see also \cite{2006math.....12798F}). 

\subsection{Classical tRS Model and its Limits}
\begin{figure}[!h]
\begin{center}
\includegraphics[height=6cm, width=14cm]{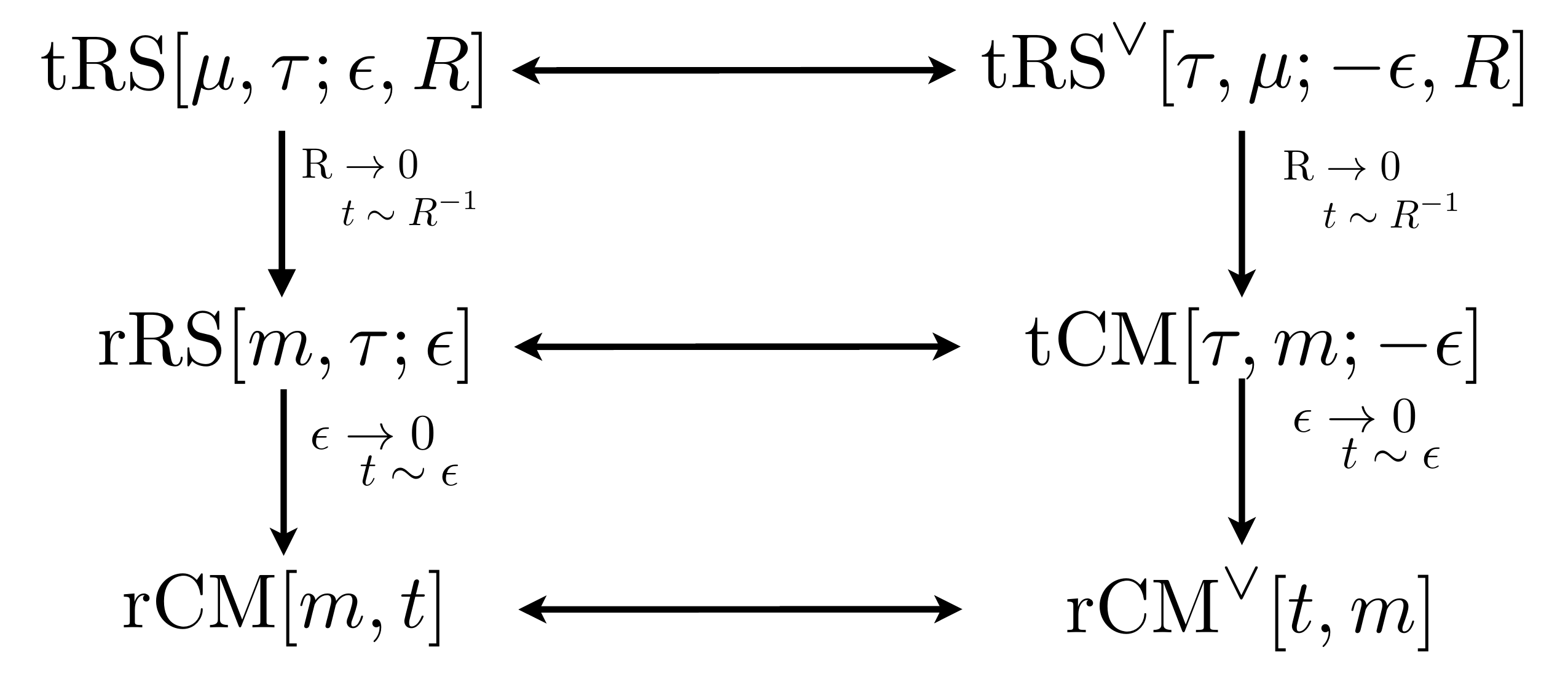}
\caption{Network of dualities between classical models emerging from tRS models. They are rational RS as well as trigonometric and rational Calogero-Moser (t(r)CM) models. Horizontal arrows denote bispectral dualities. In the brackets we put the collections of parameters which each model depends on.}
\label{fig:tRSduals}
\end{center}
\end{figure}

We can take in a similar way the $R \to 0$ and then $\epsilon \to 0$ limits of the 
tRS model and the S-dual tRS model. The limit produces simpler pairs of S-dual 
models: the rational Ruijsenaars-Schneider model (rRS) \cite{MR929148, MR1322943,MR1329481} and trigonometric Calogero-Moser 
 (tCM) first \cite{MR0280103, MR0375869, Sutherland:1971ks}, and then rational Calogero-Moser (rCM). This is sketched in \figref{fig:tRSduals},
 and was studied in great details by Fock et al in \cite{Fock:1999ae}. 
\footnote{In their construction the parameter $\epsilon$ is replaced by the inverse of the speed of light $c^{-1}$, thus the $\epsilon\to 0$ limit corresponds to the nonrelativistic limit of the corresponding system.}

Our results then show that the diagonalization of the Lax matrices involved in the S-duality relations is controlled by the Bethe equations for the XXX spin chain, tGaudin and rGaudin respectively. Of course, it is also natural to consider restricted models, with $\mathfrak{su}(2)$ embeddings $\rho$ and $\rho^\vee$. 
Various special cases of our proposal have already been suggested in the literature, such as \cite{2009arXiv0906, 2012arXiv1201.3990M}, where various dualities between XXX, Gaudin systems on one side and and trigonometric CM and rational RS on the other have been proposed (see also \cite{Bulycheva:2012ct}). Our analysis confirms previous results and provides us with the complete network of dualities.

\subsubsection{rRS/tCM duality}
It is useful to focus our attention on the flatness constraint \eqref{eq:FlatConNew}
\begin{equation}
\eta M T - \eta^{-1} T M = u v^T\,.
\label{eq:FlatnessConstrBis}
\end{equation}
as we take the $R \to 0$ limit. 

Analogously to our analysis for the limit of the XXZ Bethe equations, we can take 
the matrix $M$ to scale as 
\begin{equation}
M=\exp 2\pi R\mathfrak{m}\,,
\end{equation}
with finite $\mathfrak{m}$, and keep $T$ finite. We also scale $\eta=e^{\pi R\epsilon}$. The flatness condition goes to 
\begin{equation}
[\mathfrak{m},T]+\epsilon T = \tilde{u} \tilde{v}^T\,.
\label{eq:constrrRS}
\end{equation}

If we consider a gauge where $\mathfrak{m}$ is diagonal with eigenvalues $m_i$
then the flatness equation makes $T$ into the Lax matrix for a rRS model. We can take the limits of \eqref{eq:lax1} and \eqref{eq:lax2} to obtain
\begin{equation} \label{eq:lax1R}
T_{ij} = \frac{\tilde u_i \tilde v_j}{m_i - m_j+ \epsilon}\,.
\end{equation}
The conjugate momenta are defined through
\begin{equation}\label{eq:lax2R}
\tilde u_i \tilde v_i \prod_{k \neq i} (m_i - m_k) = - p_\mu^i \prod_{k}( m_i - m_k - \epsilon)\,.
\end{equation}

Vice versa, in a gauge where $T$ is diagonal with eigenvalues $\tau_i$ 
the matrix $\mathfrak{m}$ becomes the Lax matrix for the bispectrally dual tCM model.
We have $\tilde{u}_i \tilde{v}_i = \epsilon \tau_i$ and
\begin{equation}
\mathfrak{m}_{ij}=\frac{\tilde{u}_i \tilde{v}_j}{\tau_i-\tau_j}\,, \qquad i \neq j\,.
\end{equation}
The diagonal components of $\mathfrak{m}$ contain the conjugate momenta
\begin{equation} 
\mathfrak{m}_{ii} = P_\tau^i + \frac{\epsilon}{2} \sum_{k \neq i}\frac{\tau_i + \tau_k}{\tau_i - \tau_k}\,,
\label{eq:lax1m}
\end{equation}
where $p^i_\tau = \exp 2 \pi R P_\tau^i$.

The diagonalization of the tCM Lax operator leads to constraints on its momenta which can be solved by parameterization \eqref{eq:PtauXXX}, or, in other words to XXX Bethe equations \eqref{eq:XXXGen}. Similarly, the eigenvalue problem for rRS Lax operator will give us constraints on its momenta \eqref{eq:PtautGaudin} and tGaudin Bethe equations \eqref{eq:BAEGaudinslNGen}. The XXX and tGaudin Bethe equations have isomorphic sets of solutions. 

\subsubsection{rCM/rCM duality}
Finally one can take the $\epsilon \to 0$ limit as prescribed in \figref{fig:tRSduals} from \eqref{eq:constrrRS} with $T = \exp \epsilon\mathfrak{t}$ and obtain
\begin{equation}
[\mathfrak{m},\mathfrak{t}]=-1+ u' v'\,.
\label{eq:constrrCM}
\end{equation}
In the basis where $\mathfrak{m}$ is diagonal we find $u'_i v'_i = 1$ and Lax matrix for the rational CM model with 
\begin{equation}
\mathfrak{t}_{ij}=\frac{u'_i v'_j}{m_i - m_j}\,, \qquad i \neq j \,,
\end{equation}
and diagonal components computed from the limit of \eqref{eq:lax2R}
\begin{equation}\label{eq:lax2Re}
\mathfrak{t}_{ii} = P_\mu^i - \sum_{k \neq i}\frac{1}{m_i - m_k}\,.
\end{equation}

Similarly, in a gauge where $\mathfrak{t}$ is diagonal we find the bispectrally dual rational CM model. 
The diagonalization problem leads to the rational Gaudin Bethe equations.

\section{Open Problems and Future Directions}\label{Sec:FutureWork}
In this paper we have only initiated the systematic analysis of the Lagrangian submanifolds associated to the half-BPS boundary conditions for the $\CN=4$ four-dimensional $U(N)$
gauge theory. Several more possibilities are available and have a known S-dual from 
brane constructions or field theoretic considerations. 

An important example 
are the boundary conditions involved in the engineering of quivers in the shape of 
D-type Dynkin diagrams and of their mirror, linear quivers which end on a node with $Sp$ gauge group. It would be interesting to translate such a mirror symmetry to a
bispectral duality between a XXZ spin chain with $SO$ symmetry and some other unknown 
integrable system.
It would be also natural to consider configurations involving the four-dimensional $U(N)$ gauge theory on a circle, which gives rise to affine $\hat{A}_L$ quivers and their mirror. 
Four dimensional theories with other gauge groups should also prove interesting. 
They give rise, for example, to linear orthosymplectic quivers, with alternating orthogonal and symplectic nodes. 

In this paper we elaborated on trigonometric models and found a nice field theoretical framework to deal with them. The next obvious step is to look at elliptic integrable systems, like the XYZ chain \cite{Sklyanin:1982tf} and elliptic Ruijsenaars-Schneider (eRS) model. 
The Bethe equations for the XYZ chain arise from looking at the vacua of $\CN=2$ linear quiver gauge theories  compactified on a torus. It may be possible to use these four-dimensional theories to define boundary conditions for five-dimensional SYM gauge theory,
which gives rise to the eRS model upon torus compactification. 

We do not know if a S-duality or bispectral duality will be available in this context. We should at least find useful relations between the XYZ spin chain Bethe equations and the eRS model\footnote{Some progress in understanding of the dualities in the elliptic case has been made in \cite{Braden:1999aj,Mironov:1999vj,Mironov:2001du}.}. This direction of inquiry should be related to the work done in \cite{Gaiotto:2012xa}, where the quantum eRS Hamiltonian had a natural action on the superconformal index of $\CN=2$ theories. A semi-classical limit of the index analogous to the $b \to 0$ limit should make contact with the moduli space of vacua on the torus. 

Vice versa, much of our work involving the moduli space of vacua can be extended to 
indices and ellipsoid partition functions. This should be an interesting direction to follow up,
which involves the quantum tRS model. It would be particularly interesting to figure out how much of the structure of the T-Q relations and Hirota \cite{JPSJ.50.3785} survives in this ``quantized'' setup (see also \cite{Gorsky:2012ym, Zabrodin:2012gx} and references therein). This may also provide some interesting information on the BPS line defects of the three and four-dimensional theories. 

 Recently in \cite{Gadde:2013wq} a related type of ``spectral'' duality at classical and quantum levels has been observed. In particular, our main characters, tRS and XXZ models appear in \cite{Gadde:2013wq} as well, however, in a somewhat different context. The authors use a different string(M) theory construction to engineer their gauge theories and defects in them then us; at the moment we are unaware of any simple direct connection between the two constructions. Certainly, it will be an interesting problem to find a connection between the two approaches.

In \secref{Sec:3dN4Quiver} we studied the relationship of vacua of 3d gauge theories and solutions of Bethe ansatz for spin chains with {\it compact} symmetry group. Compactness was imposed by the action of the R-symmetry generators of the $\CN=2^*$ theory. In the recent literature the bispectral duality for noncompact ($GL(N), SL(N)$) chains has been discussed as well \cite{Mironov:2012ba,Mironov:2012uh}. For example, in \cite{MR2409414} a $GL(2)$ XXX chain on two sites was proven to be bispectrally dual to a $GL(2)$ trigonometric Gaudin system on a cylinder with two extra punctures. Also a generic duality between $GL(N)$ XXX on $M$ sites and $GL(M)$ tGaudin with $N$ spins was conjectured. The $N=M=2$ example was extensively used in \cite{Bulycheva:2012ct} to verify the 4d/2d correspondence. Some work on generalizations is now in progress \cite{KorChen}, however, a direct {\it proof} of the bispectral duality in a generic case is still missing. Recent computation by Nekrasov and Pestun \cite{Nekrasov:2012xe} will be of great help.

It will be also interesting to connect our work with the AGT correspondence \cite{Alday:2009aq}. It is a well known conjecture that the ellipsoid partition function of $T[SU(N)]$ should coincide with the S-duality kernel for the one-punctured torus with a minimal puncture. 
Our analysis can be used to show that the S-duality kernel exchanges Verlinde line operators on A- and B-cycles of the torus, i.e. Wilson and 't Hooft operators in the gauge theory. We do not understand, though, the CFT meaning of the restrictions $\rho$ and $\rho^\vee$ impose on the parameters of the problem.

\section*{Acknowledgements}
We are grateful to J. Gomis, B. Le Floch, S. Gukov, I. Yaakov, N. Seiberg, A. Gorsky, N. Nekrasov, S. Shatashvili, V. Pestun, A. Vainshtein, M. Shifman for fruitful discussions.
PK is thankful to Simons Center for Geometry and Physics at Stony Brook University as well as to W. Fine Institute for Theoretical Physics at University of Minnesota, where part of his work was done, for kind hospitality.
The research of DG and PK was supported by the Perimeter Institute for Theoretical Physics. Research at Perimeter Institute is supported by the Government of Canada through Industry Canada and by the Province of Ontario through the Ministry of Economic Development and Innovation.

\appendix

\section{Some Technical Details}\label{sec:AppTech}
Here we give some formulae which were used in the main text to derive some results.

\subsection{$GL(2)$ Flat Connections}
The eigenvalues of $T$ \eqref{eq:2ndEigen} can be computed from an intricate identity involving rational functions, which can be cleanly stated as the residue theorem for the following rational function
\begin{equation}
\frac{1}{z}\frac{1}{\eta^{-1}-\eta}\frac{(\eta z-\eta^{-1} \alpha_1)(\eta z-\eta^{-1}\alpha_2)}{(z-\alpha_1)(z-\alpha_2)}\frac{\eta^{-1} z - \sigma}{z - \eta^{-1} \sigma}\,,
\label{eq:ReFunc}
\end{equation}
which gives (after wrapping the contour around $\alpha_1$ and $\alpha_2$ and shrinking it in the two possible ways)
\begin{equation}
\frac{\alpha_1 \eta-\eta^{-1}\alpha_2}{\alpha_1-\alpha_2}\frac{\eta^{-1}\alpha_1 - \sigma}{\alpha_1 - \eta^{-1} \sigma}+ \frac{\alpha_2 \eta-\eta^{-1}\alpha_1}{\alpha_2-\alpha_1}\frac{\eta^{-1}\alpha_2 - \sigma}{\alpha_2 - \eta^{-1} \sigma}= 1+ \frac{(\sigma-\eta^{-1} \alpha_1)(\sigma-\eta^{-1}\alpha_2)}{(\eta^{-1} \sigma-\alpha_1)(\eta^{-1} \sigma -\alpha_2)}\,,
\end{equation}
thus
\begin{equation}
\text{Tr}\,T = \frac{-\alpha_1 \eta^{-1}+\alpha_2 \eta}{\alpha_1-\alpha_2}p_\alpha^1 + \frac{\alpha_1 \eta-\alpha_2 \eta^{-1}}{\alpha_2-\alpha_1}p_\alpha^2 = -\tau_1-\tau_1 \frac{(\sigma-\eta^{-1} \alpha_1)(\sigma-\eta^{-1}\alpha_2)}{(\eta^{-1} \sigma-\alpha_1)(\eta^{-1} \sigma -\alpha_2)}=-\tau_1 - \tau_2\,.
\label{eq:TrTU2}
\end{equation}

\subsection{$GL(N)$ Flat Connections}\label{sec:GLnFlatApp}
Analogously to the $T[U(2)]$ theory \eqref{eq:TrTU2}, we can look at the trace of the more generic Lax operator \eqref{eq:trLaxtRS}
\begin{equation}
\mathrm{Tr}\, T = \sum_i p_\alpha^i \prod_{k \neq i} \frac{\alpha_i \eta^{-1}-\alpha_k \eta}{\alpha_i - \alpha_k}\,. 
\end{equation}
For the sake of simplicity in this derivation we shall leave all the subtleties related to signs, which we encountered in the body of the paper in $\text{Tr}\, T$ and in XXZ Bethe equations behind and illustrate the main idea. We study the residues of the following function, which is a straightforward generalization of \eqref{eq:ReFunc}
\begin{equation}
\frac{1}{z}\frac{1}{1-\eta^2}\prod_{i=1}^N\frac{z-\eta^2 \alpha_i}{z-\alpha_i}\prod_{k=1}^{N-1}\frac{z - \eta^{-1} \sigma_k}{z - \eta \sigma_k}\,.
\end{equation}
One obtains
\begin{equation}
\sum_i \prod_{j\neq i}\frac{\alpha_i \eta^{-1}-\eta \alpha_j}{\alpha_i-\alpha_j}\prod_{k=1}^{N-1}\frac{\alpha_i - \eta^{-1} \sigma_k}{\eta^{-1} \alpha_i - \sigma_k} = 1+\sum_s \prod_{i=1}^N\frac{\eta^{-1} \sigma_s- \alpha_i}{ \sigma_s-\eta^{-1} \alpha_i}\prod_{k\neq s}\frac{\eta \sigma_s - \eta^{-1} \sigma_k}{\sigma_s - \sigma_k}
\label{eq:recursionRath}
\end{equation}
We can set 
\begin{equation}
p_\alpha^i = \xi_N \prod_{k=1}^{N-1}\frac{\alpha_i - \eta^{-1} \sigma_k}{\eta^{-1} \alpha_i - \sigma_k}\,, 
\end{equation}
so that 
\begin{equation}
\mathrm{Tr}\, T = \xi_N +  \xi_N \sum_s \prod_{i=1}^N\frac{\eta^{-1} \sigma_s- \alpha_i}{ \sigma_s-\eta^{-1} \alpha_i}\prod_{k\neq s}\frac{\eta \sigma_s - \eta^{-1} \sigma_k}{\sigma_s - \sigma_k}
\end{equation}
We can impose $\xi_N = \tau_N$, and define
\begin{equation}
p_\sigma^s = \xi_{N-1}\prod_{i=1}^N\frac{\eta^{-1} \sigma_s- \alpha_i}{ \sigma_s-\eta^{-1} \alpha_i}\prod_{k\neq s}\frac{\eta \sigma_s - \eta^{-1} \sigma_k}{\eta^{-1} \sigma_s - \eta\sigma_k}\,,
\end{equation}
then we get
\begin{equation}
\mathrm{Tr}\, T = \tau_N +  \xi_N \xi_{N-1} \sum_s p_\sigma^s \prod_{k\neq s}\frac{\eta^{-1} \sigma_s - \eta \sigma_k}{\sigma_s - \sigma_k}\,.
\end{equation}
This formula indicates the existence of an inductive reduction, where we introduced the new set of variables $\sigma_s$ in order to enforce the requirement that one of the eigenvalues of $T$ should be $\tau_N$. We are left with the problem of solving a new auxiliary problem of size $N-1$, where the $\sigma_s$ play the role of the $\alpha_i$ and $p_\sigma^s$ the role of the $p_\alpha^i$. With some work one can show that the whole size $N$ linear problem has been reduced to this size $N-1$ auxiliary problem. We can introduce a new set of $N-2$ variables $\sigma_s^{(2)}$ by a parameterization 
\begin{equation}
p_\sigma^i = \xi_{N-1} \prod_{k=1}^{N-2}\frac{\sigma_i - \eta^{-1} \sigma^{(2)}_k}{\eta^{-1} \sigma_i - \sigma^{(2)}_k}\,, 
\end{equation}
We will soon identify $\tau_{N-1}=\xi_N\xi_{N-1}$. This gives us a first set of equations, which allow us to identify the $\sigma_s$ with the Coulomb branch parameters of the $U(N-1)$ node of the quiver, or the first level Bethe roots of an XXZ $SU(N)$ spin chain\footnote{Again, up to signs, which can be fixed}
\begin{equation}
\frac{\tau_N}{\tau_{N-1}}\prod_{i=1}^N\frac{\eta^{-1} \sigma_s- \alpha_i}{ \sigma_s-\eta^{-1} \alpha_i}\prod_{k=1}^{N-2}\frac{\eta^{-1} \sigma_i - \sigma^{(2)}_k}{\sigma_i - \eta^{-1} \sigma^{(2)}_k}  \prod_{k\neq s}\frac{\eta \sigma_s - \eta^{-1} \sigma_k}{\eta^{-1} \sigma_s - \eta\sigma_k} = 1\,. 
\label{eq:XXZtau}
\end{equation}
After that introducing 
\begin{equation}
p^{\sigma,2}_s = \xi_{N}\xi_{N-1}\prod_{i=1}^{N-1}\frac{\eta^{-1} \sigma^{(2)}_s- \sigma_i}{ \sigma^{(2)}_s-\eta^{-1} \sigma_i}\prod_{k\neq s}\frac{\eta \sigma^{(2)}_s - \eta^{-1} \sigma^{(2)}_k}{\eta^{-1} \sigma^{(2)}_s - \eta\sigma^{(2)}_k}\,,
\end{equation}
and by using an obvious analogue of \eqref{eq:recursionRath} for higher nesting levels
\begin{align}
&\sum_{i=1}^{n_l} \prod_{k=1}^{n_{l+1}}\frac{\sigma^{(l)}_i - \eta^{-1} \sigma^{(l+1)}_k}{\eta^{-1} \sigma^{(l)}_i - \sigma^{(l+1)}_k} \prod_{\substack{j=1 \\ j\neq i}}^{n_l}\frac{\eta^{-1}\sigma^{(l)}_i-\eta \sigma^{(l)}_j}{\sigma^{(l)}_i-\sigma^{(l)}_j} \cr
&= 1+\sum_{s=1}^{n_{l+1}} \prod_{i=1}^{n_{l+1}}\frac{\sigma^{(l)}_i-\eta^{-1} \sigma^{(l+1)}_s}{\eta^{-1} \sigma^{(l)}_i-\sigma^{(l+1)}_s}\prod_{\substack{k=1 \\ k\neq s}}^{n_{l+1}}\frac{\eta \sigma^{(l+1)}_s - \eta^{-1} \sigma^{(l+1)}_k}{\sigma^{(l+1)}_s - \sigma^{(l+1)}_k}\,,
\label{eq:RecursionRathGen}
\end{align}
where $\sigma^{(0)}_i=\alpha_i,\,\sigma^{(1)}_i=\sigma_i$ and $n_l = N-l$, where $l=0,\dots,N-1$ we get
\begin{equation}
\mathrm{Tr}\, T = \tau_N + \tau_{N-1}+ \xi_N\xi_{N-1}\xi_{N-2} \sum_s p^{\sigma,2}_s \prod_{k\neq s}\frac{\eta^{-1} \sigma^{(2)}_s - \eta \sigma^{(2)}_k}{\sigma^{(2)}_s - \sigma^{(2)}_k}\,.
\end{equation}
Repeating the inductive steps to get all the way down the last level of nesting where only one Bethe root is left. In the end of the process we will get
\begin{equation}
p^{\sigma, N-1}_i = \xi_1\frac{\sigma^{(N-2)}_i-\eta^{-1} \sigma^{(N-1)}}{\eta^{-1} \sigma^{(N-2)}_i-\sigma^{(N-1)}}\,.
\end{equation}

\bibliography{cpn1}
\bibliographystyle{nb}

\end{document}